\documentclass[aps,prc,twocolumn,superscriptaddress]{revtex4-1}

\usepackage{amssymb,amsmath}
\usepackage{graphicx}
\usepackage{comment}
\usepackage{color}
\usepackage{multirow}
\usepackage{soul}
\usepackage{hyperref}
\usepackage{datetime}
\usdate

\newcommand{\CommentBlock}[1]{}

\definecolor{darkgreen}{RGB}{50,150,50}

\newcommand{\AuAu}{Au+Au}
\newcommand{\PbPb}{Pb+Pb}

\newcommand{\pp}{$pp$}

\newcommand{\aaa}{\ensuremath{A+A}}
\newcommand{\sqrtsNN}{\ensuremath{\sqrt{s_{NN}}}}

\newcommand{\sqrts}{\ensuremath{\sqrt{s}}}
\newcommand{\rr}{\ensuremath{R}}

\newcommand{\gev}{\ensuremath{\mathrm{GeV/}c}}

\newcommand{\TAAavg}{\ensuremath{\left<T_{AA}\right>}}

\newcommand{\kT}{\ensuremath{k_{T}}}
\newcommand{\antikT}{anti-\ensuremath{k_{T}}}

\newcommand{\pT}{\ensuremath{p_{T}}}
\newcommand{\meanpT}{\ensuremath{\left<p_{T}\right>}}
\newcommand{\meanET}{\ensuremath{\left<E_{T}\right>}}

\newcommand{\pTjet}{\ensuremath{p_{T, \mathrm{jet}}}}
\newcommand{\pTjetch}{\ensuremath{p_{T, \mathrm{jet}}^\mathrm{ch}}}

\newcommand{\etajet}{\ensuremath{\eta_{\mathrm{jet}}}}
\newcommand{\etatrack}{\ensuremath{\eta_{\mathrm{track}}}}

\newcommand{\ET}{\ensuremath{E_{T}}}

\newcommand{\AJ}{\ensuremath{A_\mathrm{J}}}

\newcommand{\pizero}{\ensuremath{\pi^0}}

\newcommand{\dNjetdpT}{\ensuremath{\frac{dN^{AA}_\mathrm{{jet}}}{d\pTjet}}}

\newcommand{\dpT}{\ensuremath{\delta{\pT}}}

\newcommand{\pTembed}{\ensuremath{\pT^{\rm{emb}}}}

\newcommand{\qhat}{\ensuremath{\hat{q}}}
\newcommand{\vtwo}{\ensuremath{v_2}}

\newcommand{\RAA}{\ensuremath{R_{AA}}}
\newcommand{\RCP}{\ensuremath{R_\mathrm{CP}}}

\newcommand{\RAApyth}{\ensuremath{R_{AA}^\mathrm{Pythia}}}

\newcommand{\pTconst}{\ensuremath{p_{T}^\mathrm{const}}}

\newcommand{\pTraw}{\ensuremath{p_{T, \mathrm{jet}}^{\mathrm{raw,ch}}}}
\newcommand{\pTrawi}{\ensuremath{p_{T, \mathrm{jet}}^{\mathrm{raw},i}}}

\newcommand{\pTreco}{\ensuremath{p_{T, \mathrm{jet}}^\mathrm{reco,ch}}}
\newcommand{\Ajet}{\ensuremath{A_\mathrm{jet}}}

\newcommand{\pTdet}{\ensuremath{p_{T,\mathrm{jet}}^\mathrm{det}}}

\newcommand{\pTpart}{\ensuremath{p_{T,\mathrm{jet}}^\mathrm{part}}}

\newcommand{\pTrecoi}{\ensuremath{p_{T, \mathrm{jet}}^{\mathrm{reco},i}}}

\newcommand{\Ajeti}{\ensuremath{A_{\mathrm{jet}}^{i}}}

\newcommand{\rhoAi}{\ensuremath{\rho{A_\mathrm{jet}^{i}}}}

\newcommand{\Rtotal}{\ensuremath{R_{\mathrm{total}}}}

\newcommand{\Rdet}{\ensuremath{R_{\mathrm{det}}}}

\newcommand{\Rbkg}{\ensuremath{R_{\mathrm{bkg}}}}

\newcommand{\zvtx}{\ensuremath{z_\mathrm{vtx}}}

\newcommand{\pTlead}{\ensuremath{p_{T, \mathrm{lead}}}}
\newcommand{\pTleadmin}{\ensuremath{p_{T, \mathrm{lead}}^{\mathrm{min}}}}

\newcommand{\pTjetdet}{\ensuremath{p_{T, \mathrm{jet}}^{\mathrm{det}}}}
\newcommand{\pTjetpart}{\ensuremath{p_{T, \mathrm{jet}}^{\mathrm{part}}}}

\newcommand{\drel}{\ensuremath{d_\mathrm{rel}}}

\graphicspath{{./}}

\begin{document}

\title{Measurement of inclusive charged-particle jet production\\ 
in \AuAu\ collisions at \sqrtsNN~=~200 GeV}




\affiliation{Abilene Christian University, Abilene, Texas   79699}
\affiliation{AGH University of Science and Technology, FPACS, Cracow 30-059, Poland}
\affiliation{Alikhanov Institute for Theoretical and Experimental Physics NRC "Kurchatov Institute", Moscow 117218, Russia}
\affiliation{Argonne National Laboratory, Argonne, Illinois 60439}
\affiliation{American University of Cairo, New Cairo 11835, New Cairo, Egypt}
\affiliation{Brookhaven National Laboratory, Upton, New York 11973}
\affiliation{University of California, Berkeley, California 94720}
\affiliation{University of California, Davis, California 95616}
\affiliation{University of California, Los Angeles, California 90095}
\affiliation{University of California, Riverside, California 92521}
\affiliation{Central China Normal University, Wuhan, Hubei 430079 }
\affiliation{University of Illinois at Chicago, Chicago, Illinois 60607}
\affiliation{Creighton University, Omaha, Nebraska 68178}
\affiliation{Czech Technical University in Prague, FNSPE, Prague 115 19, Czech Republic}
\affiliation{Technische Universit\"at Darmstadt, Darmstadt 64289, Germany}
\affiliation{ELTE E\"otv\"os Lor\'and University, Budapest, Hungary H-1117}
\affiliation{Frankfurt Institute for Advanced Studies FIAS, Frankfurt 60438, Germany}
\affiliation{Fudan University, Shanghai, 200433 }
\affiliation{University of Heidelberg, Heidelberg 69120, Germany }
\affiliation{University of Houston, Houston, Texas 77204}
\affiliation{Huzhou University, Huzhou, Zhejiang  313000}
\affiliation{Indian Institute of Science Education and Research (IISER), Berhampur 760010 , India}
\affiliation{Indian Institute of Science Education and Research (IISER) Tirupati, Tirupati 517507, India}
\affiliation{Indian Institute Technology, Patna, Bihar 801106, India}
\affiliation{Indiana University, Bloomington, Indiana 47408}
\affiliation{Institute of Modern Physics, Chinese Academy of Sciences, Lanzhou, Gansu 730000 }
\affiliation{University of Jammu, Jammu 180001, India}
\affiliation{Joint Institute for Nuclear Research, Dubna 141 980, Russia}
\affiliation{Kent State University, Kent, Ohio 44242}
\affiliation{University of Kentucky, Lexington, Kentucky 40506-0055}
\affiliation{Lawrence Berkeley National Laboratory, Berkeley, California 94720}
\affiliation{Lehigh University, Bethlehem, Pennsylvania 18015}
\affiliation{Max-Planck-Institut f\"ur Physik, Munich 80805, Germany}
\affiliation{Michigan State University, East Lansing, Michigan 48824}
\affiliation{National Research Nuclear University MEPhI, Moscow 115409, Russia}
\affiliation{National Institute of Science Education and Research, HBNI, Jatni 752050, India}
\affiliation{National Cheng Kung University, Tainan 70101 }
\affiliation{Nuclear Physics Institute of the CAS, Rez 250 68, Czech Republic}
\affiliation{Ohio State University, Columbus, Ohio 43210}
\affiliation{Institute of Nuclear Physics PAN, Cracow 31-342, Poland}
\affiliation{Panjab University, Chandigarh 160014, India}
\affiliation{Pennsylvania State University, University Park, Pennsylvania 16802}
\affiliation{NRC "Kurchatov Institute", Institute of High Energy Physics, Protvino 142281, Russia}
\affiliation{Purdue University, West Lafayette, Indiana 47907}
\affiliation{Rice University, Houston, Texas 77251}
\affiliation{Rutgers University, Piscataway, New Jersey 08854}
\affiliation{Universidade de S\~ao Paulo, S\~ao Paulo, Brazil 05314-970}
\affiliation{University of Science and Technology of China, Hefei, Anhui 230026}
\affiliation{Shandong University, Qingdao, Shandong 266237}
\affiliation{Shanghai Institute of Applied Physics, Chinese Academy of Sciences, Shanghai 201800}
\affiliation{Southern Connecticut State University, New Haven, Connecticut 06515}
\affiliation{State University of New York, Stony Brook, New York 11794}
\affiliation{Instituto de Alta Investigaci\'on, Universidad de Tarapac\'a, Chile}
\affiliation{Temple University, Philadelphia, Pennsylvania 19122}
\affiliation{Texas A\&M University, College Station, Texas 77843}
\affiliation{University of Texas, Austin, Texas 78712}
\affiliation{Tsinghua University, Beijing 100084}
\affiliation{University of Tsukuba, Tsukuba, Ibaraki 305-8571, Japan}
\affiliation{United States Naval Academy, Annapolis, Maryland 21402}
\affiliation{Valparaiso University, Valparaiso, Indiana 46383}
\affiliation{Variable Energy Cyclotron Centre, Kolkata 700064, India}
\affiliation{Warsaw University of Technology, Warsaw 00-661, Poland}
\affiliation{Wayne State University, Detroit, Michigan 48201}
\affiliation{Yale University, New Haven, Connecticut 06520}

\author{J.~Adam}\affiliation{Brookhaven National Laboratory, Upton, New York 11973}
\author{L.~Adamczyk}\affiliation{AGH University of Science and Technology, FPACS, Cracow 30-059, Poland}
\author{J.~R.~Adams}\affiliation{Ohio State University, Columbus, Ohio 43210}
\author{J.~K.~Adkins}\affiliation{University of Kentucky, Lexington, Kentucky 40506-0055}
\author{G.~Agakishiev}\affiliation{Joint Institute for Nuclear Research, Dubna 141 980, Russia}
\author{M.~M.~Aggarwal}\affiliation{Panjab University, Chandigarh 160014, India}
\author{Z.~Ahammed}\affiliation{Variable Energy Cyclotron Centre, Kolkata 700064, India}
\author{I.~Alekseev}\affiliation{Alikhanov Institute for Theoretical and Experimental Physics NRC "Kurchatov Institute", Moscow 117218, Russia}\affiliation{National Research Nuclear University MEPhI, Moscow 115409, Russia}
\author{D.~M.~Anderson}\affiliation{Texas A\&M University, College Station, Texas 77843}
\author{A.~Aparin}\affiliation{Joint Institute for Nuclear Research, Dubna 141 980, Russia}
% ...

\collaboration{STAR Collaboration}\noaffiliation




\begin{abstract}
The STAR Collaboration at the Relativistic Heavy Ion Collider reports the first 
measurement of inclusive jet production in peripheral and 
central \AuAu\ 
collisions at \sqrtsNN~=~200 GeV.
Jets are reconstructed with the \antikT\ algorithm using 
charged tracks with pseudorapidity $|\eta|<1.0$ and transverse momentum 
$0.2<\pTjetch<30$ \gev, with jet resolution 
parameter \rr~=~0.2, 0.3, and 0.4. The large background yield uncorrelated with 
the jet signal is observed to be dominated by statistical phase space, 
consistent with a
previous coincidence measurement. This background is suppressed by requiring a high-transverse-momentum (high-\pT) leading hadron in accepted jet candidates. The 
bias imposed by this requirement is assessed, and the \pT\ region in which the 
bias is small is identified. 
Inclusive charged-particle jet distributions are reported in peripheral and central \AuAu\ collisions for $5<\pTjetch<25$ \gev\ and
$5<\pTjetch<30$ \gev\ respectively. 
The charged-particle jet inclusive yield is suppressed for central \AuAu\ collisions, 
compared to both the peripheral \AuAu\ yield from this measurement and to the \pp\ yield calculated using the PYTHIA event generator. The magnitude of the 
suppression is consistent with that of inclusive hadron production at 
high \pT, and that of semi-inclusive recoil jet yield when expressed in terms of
energy loss due to medium-induced energy transport. Comparison of inclusive 
charged-particle jet yields for different 
values of \rr\ exhibits no significant evidence for
medium-induced broadening of the transverse jet profile for \rr\ $<0.4$ in central 
\AuAu\ 
collisions. The measured distributions are consistent with theoretical model 
calculations that incorporate jet quenching.  
\end{abstract}

\pacs{}

\maketitle



 


\section{Introduction}
\label{sect:Intro}

Collisions of heavy nuclei at high energy generate a quark-gluon plasma 
(QGP), a state of matter with temperature and energy density similar to those of the 
universe a few microseconds after the Big Bang, and whose dynamics are governed by 
the interactions of
subhadronic quanta (\cite{Busza:2018rrf} and references therein). Extensive measurements of the QGP have 
been carried out with nuclear collisions at the Relativistic Heavy Ion 
Collider (RHIC) and the Large Hadron Collider (LHC). Comparison of these 
measurements with theoretical calculations indicates that the QGP 
is an inviscid fluid exhibiting collective behavior \cite{Heinz:2013th}. The QGP is likewise found to be 
opaque to penetrating probes carrying color charge, a phenomenon known as ``jet quenching"~(Ref.~\cite{Burke:2013yra} and references therein). 

Jets in high-energy collisions are generated by the hard (high momentum-transfer 
$Q^{\mathrm{2}}$) scattering 
of quarks and gluons (collectively, partons) from the incoming projectiles. The scattered 
partons fragment into correlated sprays of stable 
hadrons that are 
observed in the detector. Jet production has been measured 
extensively in \pp\ collisions, with theoretical 
calculations based on high-order perturbative quantum chromodynamics (pQCD)
describing such measurements accurately over a wide kinematic range 
\cite{Abelev:2006uq,Adamczyk:2016okk,Abelev:2013fn,Aad:2014vwa,Khachatryan:2016mlc}.

Jets are likewise generated in high-energy nuclear collisions, with production
rates that are accurately calculable using pQCD methods~\cite{Majumder:2010qh}. Because high-$Q^{\mathrm{2}}$ 
processes occur early in the evolution of a nuclear collision, jets probe the 
QGP at its highest temperature and energy density. Jet quenching, which arises from the 
interaction of energetic partons with the QGP via elastic and radiative processes, is expected to 
generate modifications in
observed jet production rates and internal structure~\cite{Cao:2020wlm}.

Measurement of reconstructed jets in heavy-ion collisions is challenging: A 
jet, which comprises $\approx$10 correlated particles at RHIC energies, must be distinguished from the many hundreds of 
particles generated by uncorrelated processes~\cite{Adams:2005dq}.
High transverse-momentum (high-\pT) hadrons, which are the leading fragments 
of jets, can be more readily distinguished from this background than fully 
reconstructed jets. The production rate of high-\pT\ hadrons was also predicted to be suppressed due 
to jet 
quenching~\cite{Wang:1991xy}, and suppression of inclusive production and 
correlations of high-\pT\ hadrons due to jet quenching has indeed been observed at 
RHIC~\cite{Adler:2002xw,Adler:2002tq,Adams:2003kv,Adams:2006yt,Adamczyk:2013jei,Adcox:2001jp,Adare:2012wg,Adare:2010ry} 
and the 
LHC~\cite{Aamodt:2011vg,Abelev:2012hxa,Adam:2016jp,CMS:2012aa,Chatrchyan:2012wg}.
The comparison of inclusive hadron suppression measurements 
with theoretical calculations has been used to constrain the QGP transport 
parameter \qhat~\cite{Burke:2013yra}, which characterizes the momentum transfer 
between a jet probe and the QGP medium. 

High-\pT\ hadron suppression provides limited insight into the 
mechanisms and dynamics of jet 
quenching, however. Observed high-\pT\ hadrons arise 
predominantly from jets that have lost relatively little energy  
in-medium, due to the interplay of the shape of the jet-\pT\ distribution, jet 
fragmentation, and jet energy 
loss~\cite{Baier:2002tc,Drees:2003zh,Dainese:2004te,Eskola:2004cr,Renk:2006nd,Loizides:2006cs,Zhang:2007ja,Renk:2012cb}. 
The contribution to the inclusive high-\pT\  hadron yield arising from jets undergoing significant 
modification due to quenching is thereby suppressed.

Broader exploration of jet 
quenching requires measurements with reconstructed jets. At the LHC, 
reconstructed-jet 
measurements in \PbPb\ collisions have been reported for inclusive 
production~\cite{Abelev:2013kqa,Adam:2015ewa,Acharya:2019jyg,Aad:2014bxa,Khachatryan:2016jfl}, 
correlations~\cite{Adam:2015doa,Aad:2010bu,Chatrchyan:2012nia, 
Chatrchyan:2012gt,Sirunyan:2018qec}, and jet 
substructure~\cite{Acharya:2017goa,Acharya:2019djg,Sirunyan:2017bsd}.
At RHIC, reconstructed-jet 
measurements in \AuAu\ collisions have been reported for 
correlations~\cite{Adamczyk:2016fqm,Adamczyk:2017yhe}. While the inclusive jet 
and dijet production cross sections 
have been reported for \pp\ collisions at 
RHIC~\cite{Abelev:2006uq,Adamczyk:2016okk}, the measurement of inclusive jet 
production in \AuAu\ collisions at RHIC has not been reported to date.

This paper presents the first measurement of inclusive jet 
production in \AuAu\ collisions at \sqrtsNN~=~200 GeV.  
Jets are reconstructed in central (0--10 percentile bin of the inelastic cross 
section) and peripheral (60--80 percentile bin) 
\AuAu\ collisions using charged tracks with transverse momentum $\pTconst>0.2$ \gev\ and pseudo--rapidity $|\etatrack|<1.0$, 
using the \antikT\ algorithm~\cite{FastJetAntikt} with resolution parameter 
\rr~=~0.2, 0.3, and 0.4. 
Uncorrelated background yield is suppressed by a cut on the 
leading (highest \pT) hadron of 
each jet candidate, $\pTlead>\pTleadmin$, which imposes a bias on 
the fragmentation pattern of the reported jet population; we label the resulting jet population 
``quasi-inclusive". The effect of the bias is 
determined by varying the value of \pTleadmin. 
The distribution of the jet population arising from the large 
uncorrelated 
background is well-described by a 
model calculation based on statistical phase space, without taking into account any multi-particle correlations whatsoever. This observation is
consistent with the accurate 
description of the background to semi-inclusive recoil jet yields by event 
mixing~\cite{Adamczyk:2017yhe}. 

Quasi-inclusive charged-particle jet distributions are reported in the range 
$5<\pTjetch<30$ \gev\ for central \AuAu\ collisions. Charged-particle jet yield suppression is 
quantified by comparing the quasi-inclusive distribution measured in central 
\AuAu\ collisions to that measured in peripheral \AuAu\ collisions, and 
to the inclusive charged-particle jet distribution for \pp\ collisions generated using 
the PYTHIA Monte Carlo generator~\cite{Skands:2010ak}, which has been
validated by comparison to inclusive measurements of 
pions and 
fully reconstructed jets at RHIC~\cite{Adam:2019aml}. These measurements are also compared 
to similar 
inclusive jet measurements 
at the LHC, to semi-inclusive hadron+jet measurements at RHIC, 
and to theoretical calculations of jet quenching.

The paper is organized as follows: Section~\ref{sect:DataSet} describes the 
experiment and data selection;
Sec. \ref{sect:Bkgd} presents considerations for heavy-ion jet analysis and the measurement approach;
Sec. \ref{Sect:JetReco} 
presents the jet reconstruction; Sec. \ref{Sect:RawData} presents raw jet spectra;
Sec. \ref{sect:Corrections} presents the corrections due to background 
fluctuations and detector effects; Sec. \ref{sect:SysUncert} presents the 
systematic uncertainties; Sec.~\ref{sect:ClosureTest} presents the parametrized model (PM) and closure test; Sec. \ref{sect:ppRef} presents the reference spectrum 
from \pp\ collisions calculated using PYTHIA; 
Sec.~\ref{sect:Theory} describes the theoretical calculations used for comparison;
Sec. \ref{sect:results} presents the results; and Sec. 
\ref{sect:Summary} presents the summary. 

\section{Detector and dataset}
\label{sect:DataSet}

The STAR detector is described in Ref.~\cite{Ackermann:2002ad}. STAR is a large, general-purpose collider detector with high-precision tracking, 
particle identification, electromagnetic calorimetry, and forward detectors. The central region is immersed in a 0.5 T solenoidal magnetic field.
The data for this 
analysis were recorded during the 2011 RHIC
run with \AuAu\ collisions at \sqrtsNN~=~200 GeV. Events were selected online 
using a minimum bias (MB) trigger that requires signals in both
forward scintillator Vertex Position Detectors (VPD), with a timing cut to 
constrain
the primary vertex position within $|\zvtx|<30$ cm of the
nominal center of STAR along the beamline, and with the requirement of 
at least one 
neutron in each Zero Degree Calorimeter (ZDC), to bias toward the hadronic 
interaction of both Au ions. The MB trigger minimizes pileup by requiring that no 
additional interactions occur in a time interval of 40 $\mu\mathrm{s}$ before or 
after the triggered collision, consistent with the drift time of the Time Projection Chamber (TPC)~\cite{Anderson:2003ur}.

Charged-particle tracks are reconstructed offline using the 
TPC, which has an inner radius 
of 50 cm and an outer radius of 200 cm, and covers the full 
azimuth within $|\etatrack|<1$. TPC tracks have a maximum number of 
45 space points. 

Global tracks, which do not include the primary event vertex in 
the momentum fit, are accepted if they have more than 14 space points, with the 
ratio of the number of space points to the number of potential space points 
greater than 0.52. The location of the primary vertex is determined using global tracks. The primary vertex position resolution 
along the beam direction is 350~$\mu$m for the most central \AuAu\ events used in the 
analysis. 

Jet reconstruction utilizes 
primary tracks, which are global tracks whose momenta have been refit with 
inclusion of the primary event vertex. Primary 
tracks with $0.2<\pTconst<30$ \gev\ and which have distance of closest 
approach (DCA) to the primary vertex in the transverse plane ${\mathrm{DCA}}_{xy}<1$~cm
are accepted for further analysis.

Events are accepted for the analysis if their reconstructed vertex lies 
within $|z_\mathrm{vtx}|<30$ cm of the
nominal center of STAR along the beamline, and within 2 cm of the beam axis in the 
transverse
plane.
After offline event
selection cuts, a total of $\approx$400~M \AuAu\ events were accepted, corresponding 
to an integrated luminosity of $\approx$6 $\mu {\mathrm{b}}^{-1}$.

Events are classified offline in percentile bins of centrality,
based on charged-particle multiplicity measured in $|\etatrack|<0.5$. The
accepted event population has $\approx$47~M central collision events 
and $\approx$94~M peripheral collision events. The online
trigger efficiency is consistent with 100\% for central \AuAu\ collisions
and is approximately 70\% for peripheral \AuAu\ collisions.

Simulated events for \pp\ collisions at \sqrts~=~200 GeV were generated using 
PYTHIA 6.428, tune Perugia 2012~\cite{Skands:2010ak}.
Simulated events without 
instrumental effects are 
denoted ``particle level," whereas events incorporating instrumental effects are denoted ``detector level"; see Sec.~\ref{sect:Corrections}.  
The largest 
instrumental effects in the measurement of charged-particle jets are tracking 
efficiency and track momentum resolution. Fast simulation events are 
generated by applying a \pT-dependent parametrization of these effects to 
PYTHIA-generated events. 

Tracking efficiency is determined by embedding single tracks simulated at the detector level into real \AuAu\ events. Tracking efficiency depends on particle species; tracking efficiency for nonidentified charged tracks therefore depends on the 
relative population of different species. In order to assess the magnitude of this dependence, two different assumptions are made for the 
relative yield of charged pions, charged kaons, protons, and antiprotons comprising the charged track 
population: the relative yields measured in \pp\ 
collisions~\cite{Adams:2003qm,Agakishiev:2011dc}, and those measured in
\AuAu\ collisions~\cite{Abelev:2006jr,Agakishiev:2011dc,Agakishiev:2011ar}. 
The relative yields for \AuAu\ collisions are used in the principal analysis, 
giving tracking efficiency for primary charged tracks of 68\% at \pT~=~0.5 \gev\ 
and 72\% for $\pT>1$ \gev\ in central \AuAu\ collisions; and 85\% at \pT~=~0.5 \gev\ and 88\% 
for $\pT>1$ \gev\  in peripheral \AuAu\ collisions.
The relative yields from \pp\ collisions give tracking efficiency that is 1\% lower for 
$\pT<1~\gev$, with negligible differences for $\pT>1~\gev$. This variation is 
smaller than the overall systematic uncertainty assigned to the tracking 
efficiency, which is discussed below.

Primary track momentum resolution, which is also determined by embedding 
simulated tracks into real \AuAu\ events, is parametrized for $\pT>1.2$ \gev\ as 
$\sigma_{\pT} = -0.026 + 0.020\pT + 0.003(\pT)^2$ (\pT\ in units of \gev),
with a variation $\sigma_{\pT} = 0.003(\pT)^2$ used for systematic uncertainty.

Comparison of inclusive jet spectra at different centralities requires the 
scaling of yields by the centrality-dependent nuclear thickness factor \TAAavg, which is 
calculated using Glauber modeling~\cite{Miller:2007ri}. In this 
analysis, \TAAavg\ has the value $22.8\pm{1.6}$ mb$^{-1}$ for central \AuAu\ collisions and 
$0.49\pm{0.14}$ mb$^{-1}$ for peripheral \AuAu\ collisions. 

\section{Analysis strategy}
\label{sect:Bkgd}

Jet reconstruction algorithms provide a systematically well-controlled 
approach to jet measurements and corresponding theoretical calculations in \pp\ 
collisions at collider 
energies \cite{Adamczyk:2016okk,Aad:2014vwa,Khachatryan:2016mlc}. Jet 
measurements in 
heavy-ion  collisions are significantly more complex, 
however, due to the large uncorrelated background in such events. In 
this section 
we discuss the main considerations for a theoretically 
interpretable measurement of the inclusive jet distribution
in the large-background environment of heavy-ion collisions and the 
consequent strategy for this analysis.

The constituents of a jet reconstructed in a high-energy nuclear collision arise 
from multiple different
sources, which we classify qualitatively as due to hard processes (${Q^2}>$~few 
GeV$^2$) or to soft processes (all others). Multiple hard processes can occur in a 
single nuclear collision; in the framework of QCD 
factorization they are considered to be incoherent. These
processes can  
generate multiple energetic 
jets that overlap in $(\eta,\phi)$ space, whose hadronic fragments are thereby 
clustered by a jet reconstruction 
algorithm into a single jet candidate. Each such jet candidate will also contain 
copiously produced hadrons 
from soft processes. Jet candidates in central high-energy nuclear collisions 
therefore have a significant contribution from hadrons due to soft processes, and may also 
contain hadronic fragments of one or more primordial jets arising from hard processes. 

For an inclusive jet measurement in central high-energy nuclear collisions to be 
theoretically interpretable, it 
must report the distribution of a unique, well-defined jet population arising 
from hard processes. The measurement must therefore 
exclude the yield of purely 
combinatorial jet candidates arising solely due to contributions from soft 
processes, and disentangle the effects of multiple overlapping primordial 
jets arising from hard processes. It should also correct for the 
shift and smearing of 
the jet \pT-scale due to the large number of hadrons arising from soft processes in 
each identified hard-jet candidate. 

In semi-inclusive hadron jet analyses~\cite{Adam:2015doa,Adamczyk:2017yhe} these
corrections are implemented in three distinct steps: (i) approximate 
adjustment event-by-event of jet candidate \pTjet\ for the uncorrelated 
background contribution; (ii) rejection of 
background yield not correlated with the trigger, giving the raw trigger-correlated jet yield; and 
(iii) final correction via unfolding of the jet \pTjet\ for shift and 
fluctuations in the background energy density. Steps (ii) and (iii) are carried out at the level 
of ensemble-averaged distributions (``statistical correction"). This 
approach enables the measurement of trigger-normalized recoil jet distributions 
for large jet radius \rr\ and low \pTjet\ in the most central \aaa\ collisions, 
without imposing fragmentation bias on the reported jet 
population~\cite{Adam:2015doa,Adamczyk:2017yhe}. 

The inclusive jet distribution that is the goal of this analysis is not defined with respect to a 
trigger, however, and a different approach is needed for step (ii) to identify jet candidates that arise 
from hard processes. We therefore accept 
for analysis only those jet 
candidates whose highest-\pT\ hadronic 
constituent (``leading hadron") has $\pTlead>\pTleadmin$ 
\cite{deBarros:2012ws,Adam:2015ewa}. No cut is made on \pTjet\ in this analysis, in contrast to other 
current measurements of inclusive jet distributions in heavy-ion 
collisions~\cite{Abelev:2013kqa,Adam:2015ewa, Aad:2014bxa, Khachatryan:2016jfl}.

There are competing considerations for the value of \pTleadmin~\cite{deBarros:2012ws}:

\begin{itemize}

\item The value of \pTleadmin\ must be sufficiently high that the probability for such a hadron to arise from 
purely combinatorial jet is negligible; i.e. with high probability it is the fragment of a hard process.

\item The value of \pTleadmin\ must be sufficiently high that the probability for 
multiple hadrons to satisfy this cut in a central \AuAu\ collision is 
negligible. The probability of two hard jets in an event 
passing this acceptance cut is therefore also negligible; with high 
probability there will be at most one such jet candidate in an event. This 
selection thereby identifies a unique, well-defined jet population arising from 
a specified hard process, as required.

\item The value of \pTleadmin\ should be as low as possible, to 
minimize the bias imposed on the accepted jet population.

\end{itemize}

The second consideration, that the value of \pTleadmin\ is sufficiently high that the probability to find two such hadrons in an event is negligible, is required to ensure applicability of the correction scheme based on unfolding (Sec.~\ref{sect:Corrections}), which is a linear transformation of a distribution that is a function of jet \pT.

The bias relative to the inclusive jet population imposed by the \pTleadmin\ cut must be determined
experimentally, for the 
measurement to be theoretically interpretable. 
The value of \pTleadmin\ is consequently varied in the analysis, and the \pTjet\ range in which the corrected inclusive 
jet distribution does not depend significantly on \pTleadmin\ is found. This is identified as the range 
where the bias is small.

\section{Jet reconstruction}
\label{Sect:JetReco}

Jet reconstruction utilizes the \kT~\cite{Cacciari:2011ma} and \antikT~\cite{FastJetAntikt} algorithms with the boost-invariant
\pT-recombination scheme~\cite{Cacciari:2011ma}, 
applied to all accepted charged tracks. The jet area is
calculated by the Fastjet algorithm~\cite{FastJetArea} with a ghost particle area of 0.01. The jet centroid is calculated as the sum of the four-vectors of its constituents~\cite{Cacciari:2011ma}. 

This analysis employs several types of charged-particle jet, which are referred to 
using the notation defined in Ref.~\cite{Adamczyk:2017yhe}: 
The raw transverse momentum 
of reconstructed jets is denoted \pTraw, jet 
transverse momentum after the eventwise adjustment for uncorrelated
 background density is denoted \pTreco, and jet transverse momentum after
full correction for instrumental effects and background fluctuations is denoted 
\pTjetch.

Jet reconstruction is carried out twice for each event. The first jet reconstruction pass
applies the \kT\ algorithm with \rr~=~0.3 to calculate $\rho$, the estimated
transverse-momentum density of background in the event~\cite{FastJetPileup},

\begin{equation}
\rho=\mathrm{median}\left\{ \frac{\pTrawi}{\Ajeti} \right\},
\label{eq:rho}
\end{equation}

\noindent
where index $i$ labels the charged-particle jet candidates in the event from this reconstruction pass, and \pTrawi\
and \Ajeti\ are the transverse momentum and area of the
$i$th jet. For central \AuAu\ collisions, the two jets 
with largest \pTrawi\ are excluded from the median calculation, while for 
peripheral collisions the single jet with largest \pTrawi\ is excluded. 
Different choices for the number of excluded jets are used for systematic 
variation (Sec.~\ref{subsect:SysUncRho}).

The second reconstruction pass, which generates jet candidates
for the measured distributions, applies the \antikT\
algorithm with \rr~=~0.2, 0.3, or 0.4. Jet candidates are accepted for further analysis if their centroid lies within $|\etajet|<1-\rr$, due to the TPC acceptance. 

The value of \pTrawi\ is adjusted according to \cite{FastJetPileup}

\begin{equation}
\pTrecoi=\pTrawi - \rhoAi,
\label{eq:pTraw}
\end{equation}

\noindent
where $i$ in this case labels the jet candidates from the second reconstruction pass and $\rho$ is determined from Eq.~(\ref{eq:rho}). The value of $\rho$ varies 
event to event: For central \AuAu\ collisions in this analysis, its
most probable value is 31 GeV/($c$-sr), with 
RMS~=~3 GeV/($c$-sr); for peripheral \AuAu\ collisions its most probable 
value is 0, with RMS~=~1 GeV/($c$-sr). 

The definition of $\rho$ in 
Eq.~(\ref{eq:rho}) requires algorithmic choices that are not unique, including 
reconstruction algorithm, jet-resolution parameter \rr, and the 
number of jet candidates excluded from the median calculation. 
The adjustment to \pTraw\ in 
Eq.~(\ref{eq:pTraw}) is therefore only an estimate of the 
eventwise pedestal due to uncorrelated background.
The absolute jet energy scale is imposed in the 
unfolding step described below (see also Refs.~\cite{Adam:2015doa,Adamczyk:2017yhe}). 

\begin{figure*}[htbp]
\includegraphics[width=0.49\textwidth]{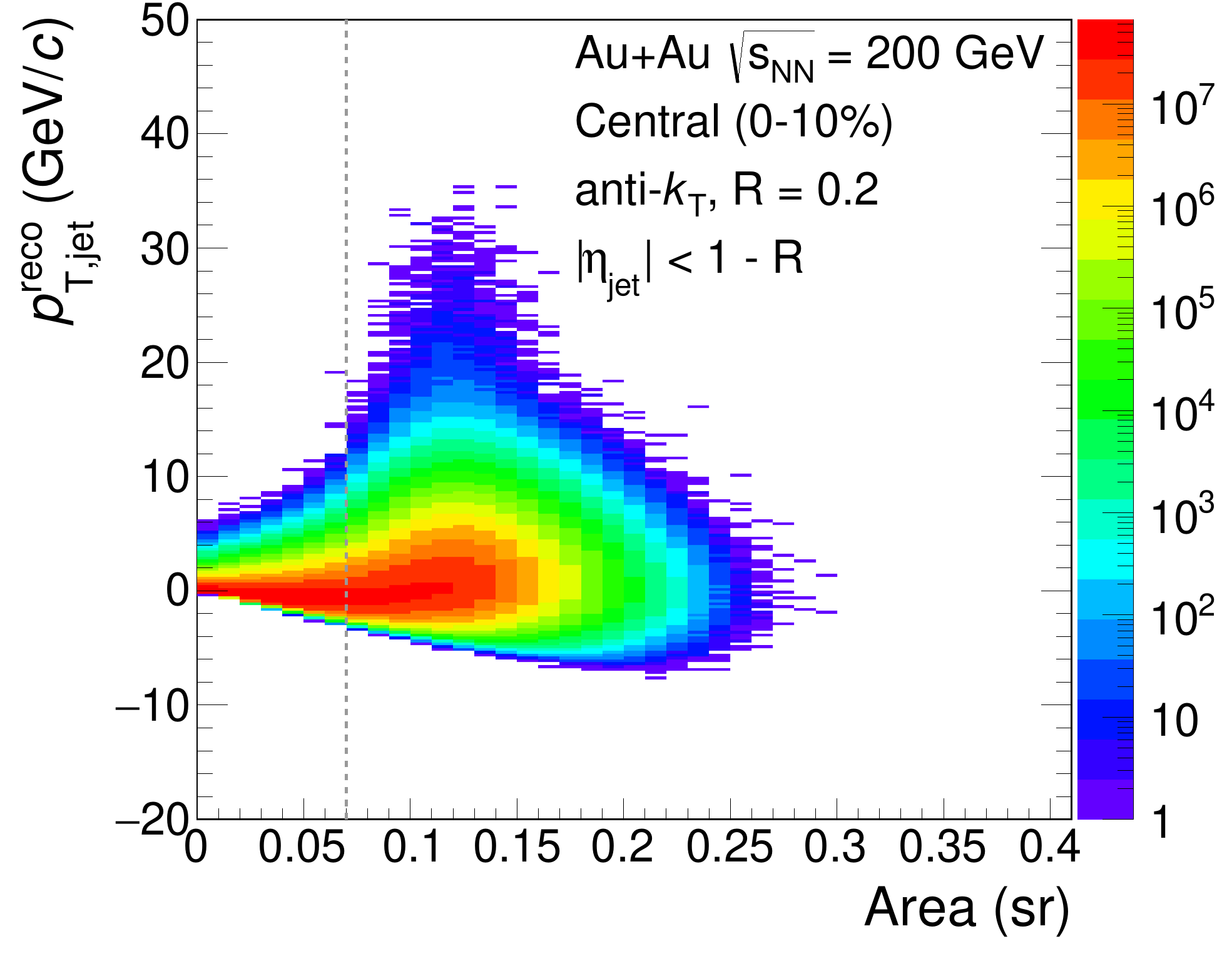}
\includegraphics[width=0.49\textwidth]{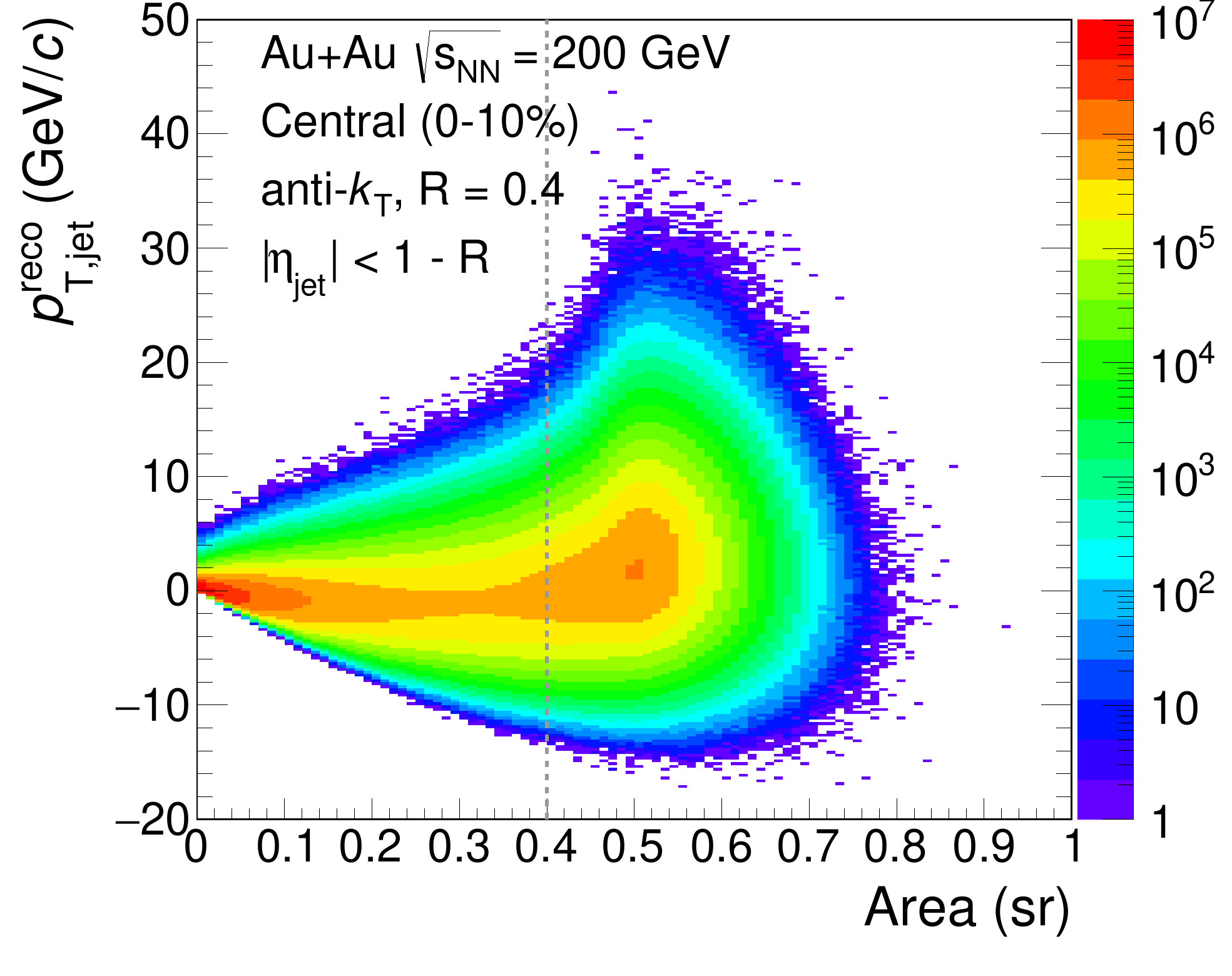}\\
\includegraphics[width=0.49\textwidth]{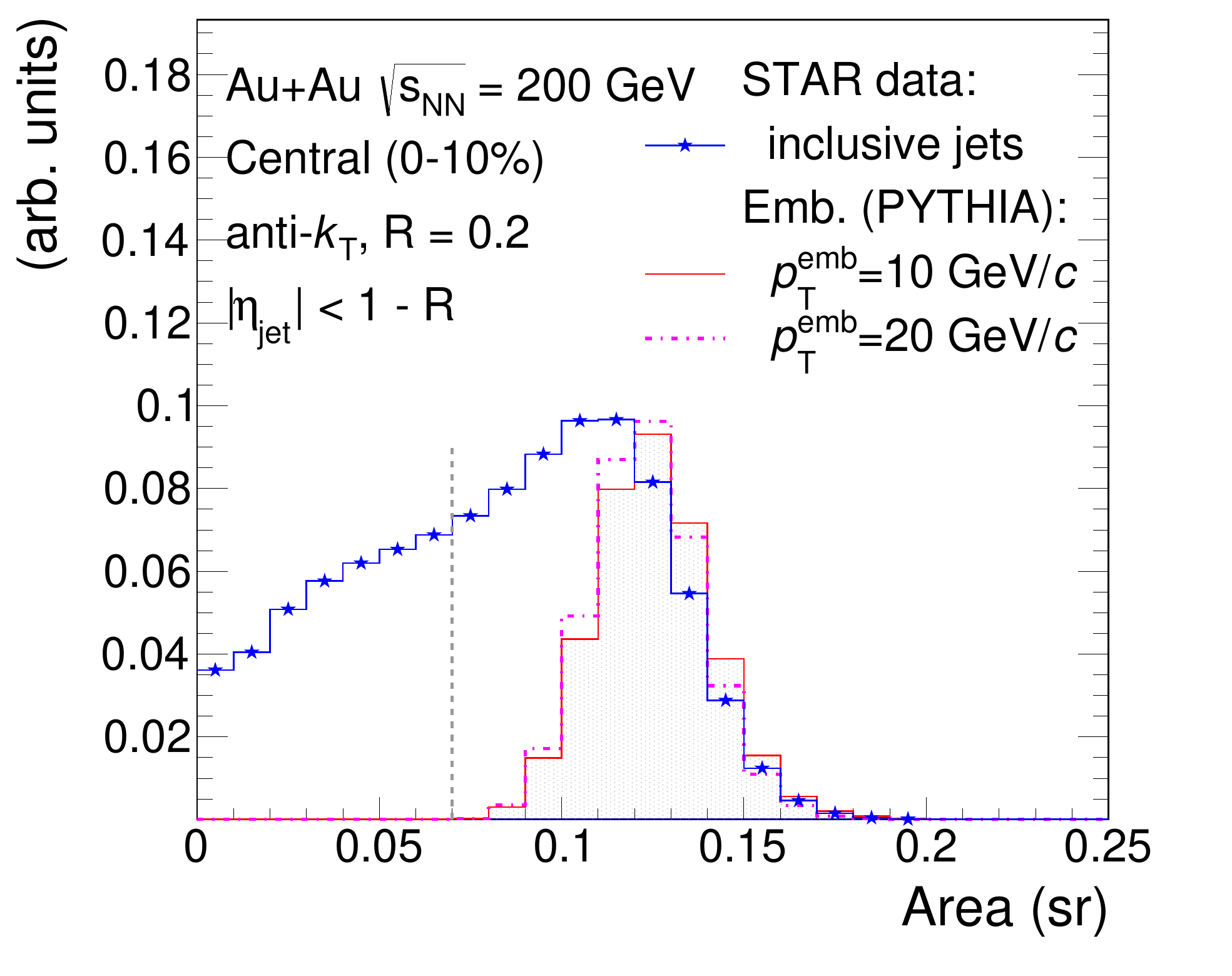}
\includegraphics[width=0.49\textwidth]{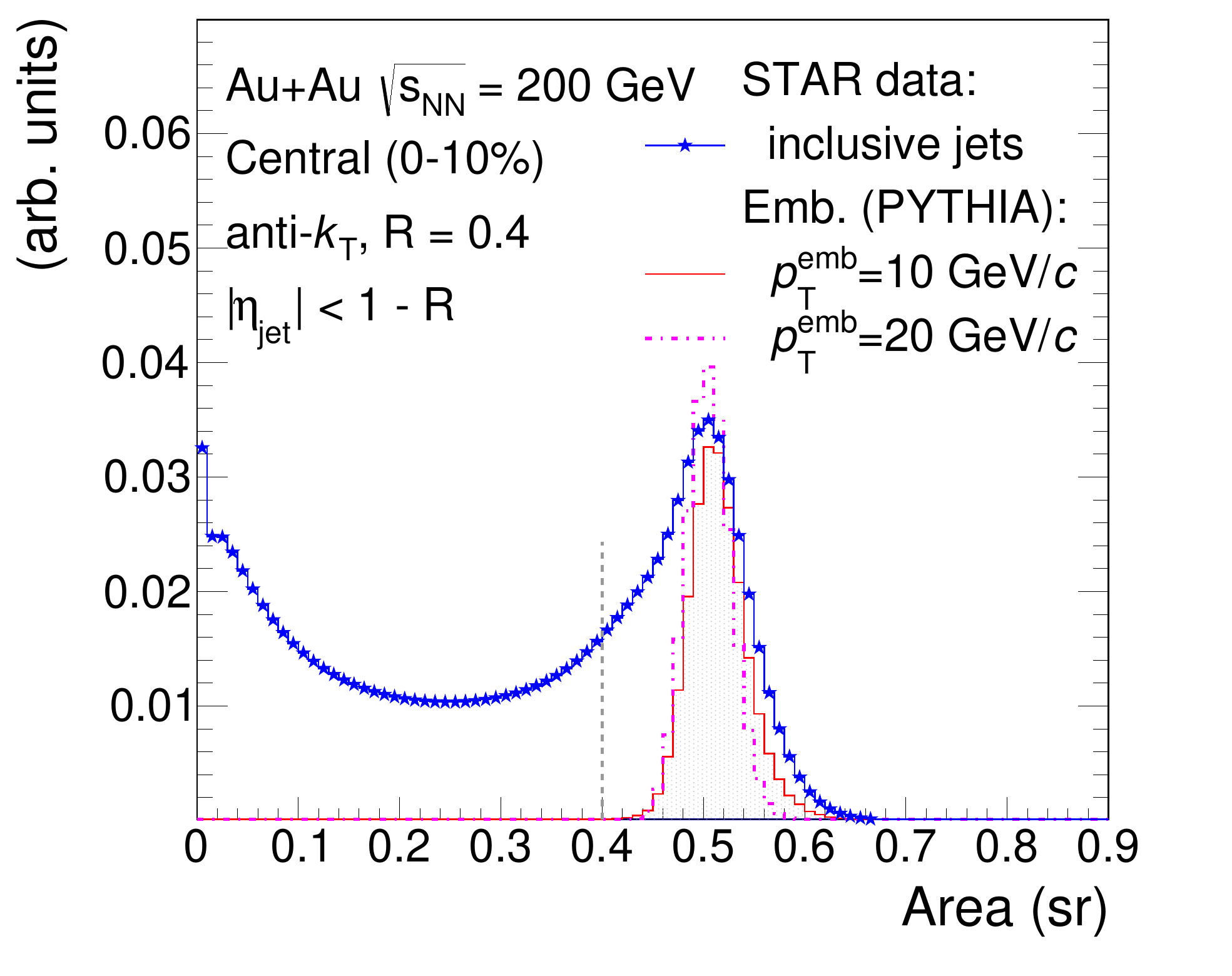}\\
\includegraphics[width=0.49\textwidth]{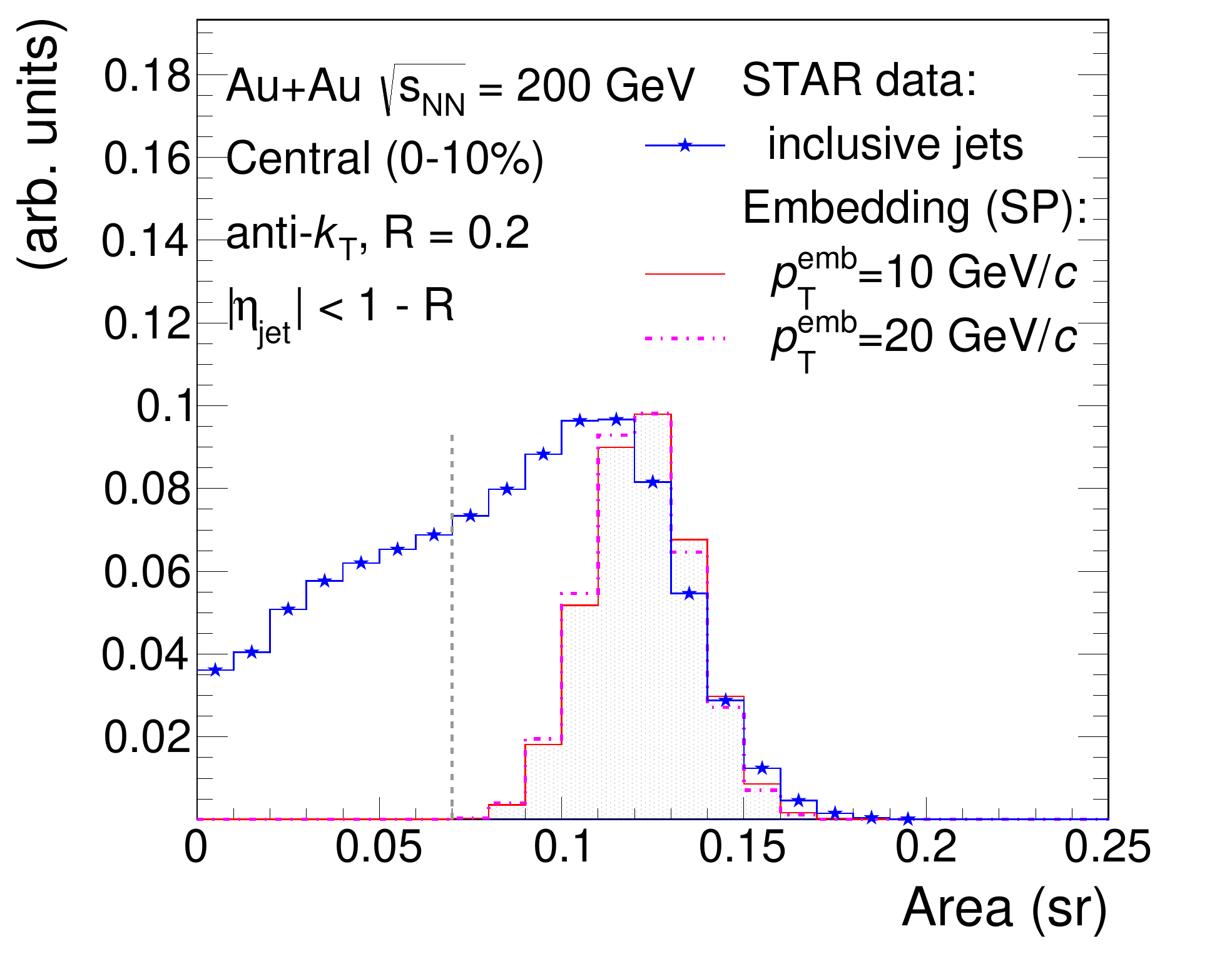}
\includegraphics[width=0.49\textwidth]{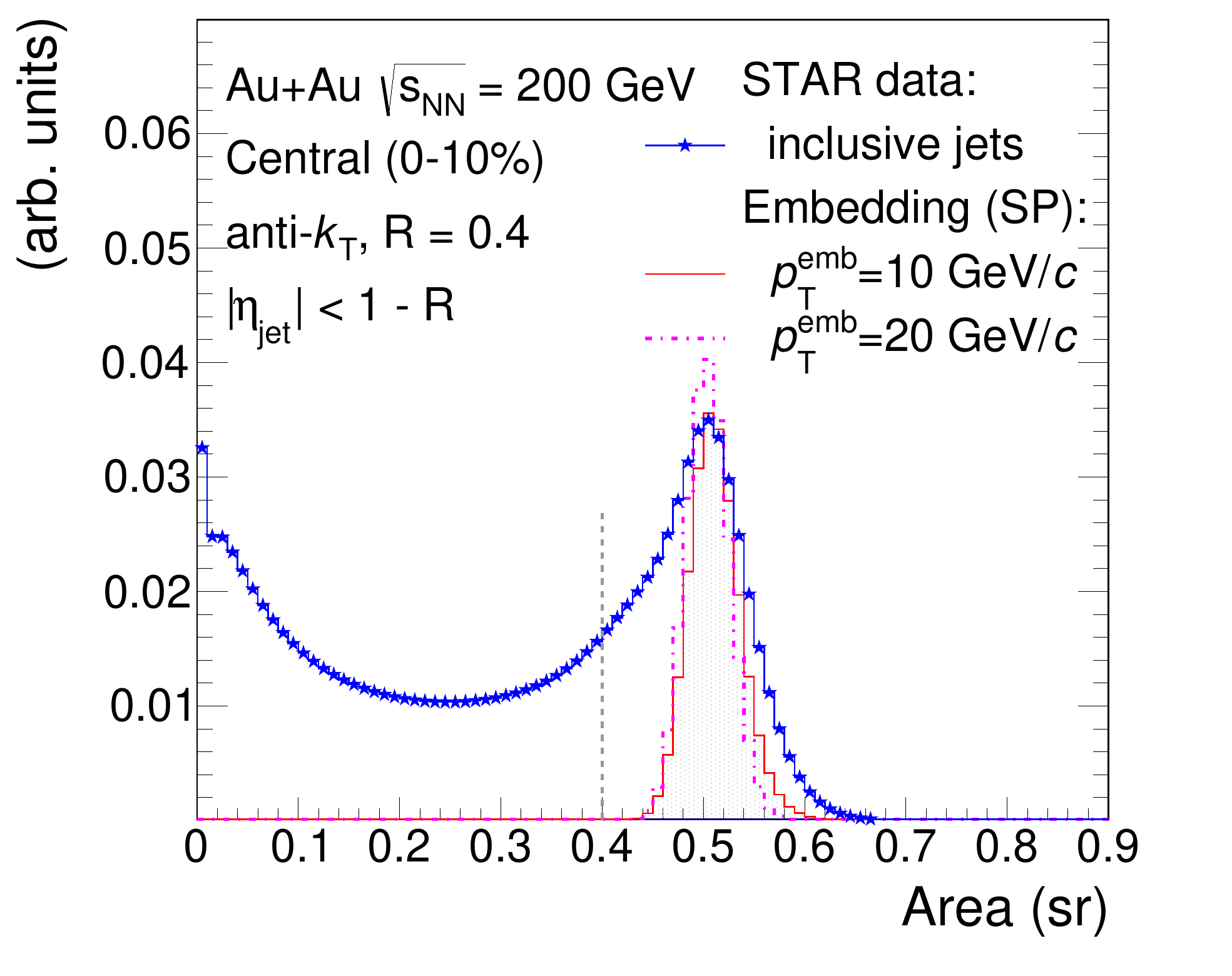}\\
\caption{(Color online) Distribution of \pTreco\ and jet area for the inclusive 
charged-particle jet 
population (\pTleadmin~=~0) in central \AuAu\ collisions. Upper panels: \pTreco\ 
vs. jet area for 
\rr~=~0.2 
(left) and \rr~=~0.4 (right). Middle and lower panels: Projection onto 
the jet area axis. Also shown are area distributions for PYTHIA-generated 
(middle) and SP jets (lower) with \pTembed~=~10 and 20 \gev, embedded into real 
\AuAu\ data for central collisions. The vertical dashed lines show the jet area 
cut. 
}
\label{fig:area}
\end{figure*}

Figure~\ref{fig:area} (upper panels) show distributions of \pTreco\ vs. jet 
area in central \AuAu\ collisions for the 
inclusive charged-particle jet population without a leading particle cut (indicated by $\pTleadmin~=~0$; note that tracks have $\pTconst>0.2$ \gev) with \rr~=~0.2 and 0.4. Jets 
with small area predominantly have $\pTreco\approx0$. The middle and lower panels 
show area 
projections of these distributions, together with 
those for jets in \pp\ collisions simulated using PYTHIA with $\pTjetch~=~10$ and 
20 \gev\ that have been 
embedded into real events, and for single-particle ``jets'' (SP, 
Sec.~\ref{sect:dpT}). 
The area distributions for 
PYTHIA-generated and SP jets in central \AuAu\ collisions are similar, with negligible dependence on 
\pTjetch. The area distributions for 
PYTHIA-generated and SP jets are similar in peripheral \AuAu\ collisions (not shown). 

Figure~\ref{fig:area} shows that, for jets with $\pTjetch>10$ \gev, the jet area 
is largely a geometric quantity, with little dependence on the pattern of jet 
fragmentation into hadrons. The area distribution for embedded jets 
is peaked at $\Ajet\approx\pi\rr^2$, while the 
inclusive jet population exhibits a tail toward small area, which arises from 
purely combinatorial jets without a hard component. A cut on jet area is therefore applied to suppress 
purely combinatorial jet candidates, while preserving high efficiency for jets 
that include a hard component~\cite{
Adamczyk:2017yhe}. Jet candidates are 
rejected if $\Ajet<0.07$ sr for \rr~=~0.2, $\Ajet<0.2$ sr for \rr~=~0.3, and
$\Ajet<0.4$  sr for \rr~=~0.4.

\section{Uncorrected jet distributions}
\label{Sect:RawData}
 
\begin{figure}[htbp]
\includegraphics[width=0.49\textwidth]{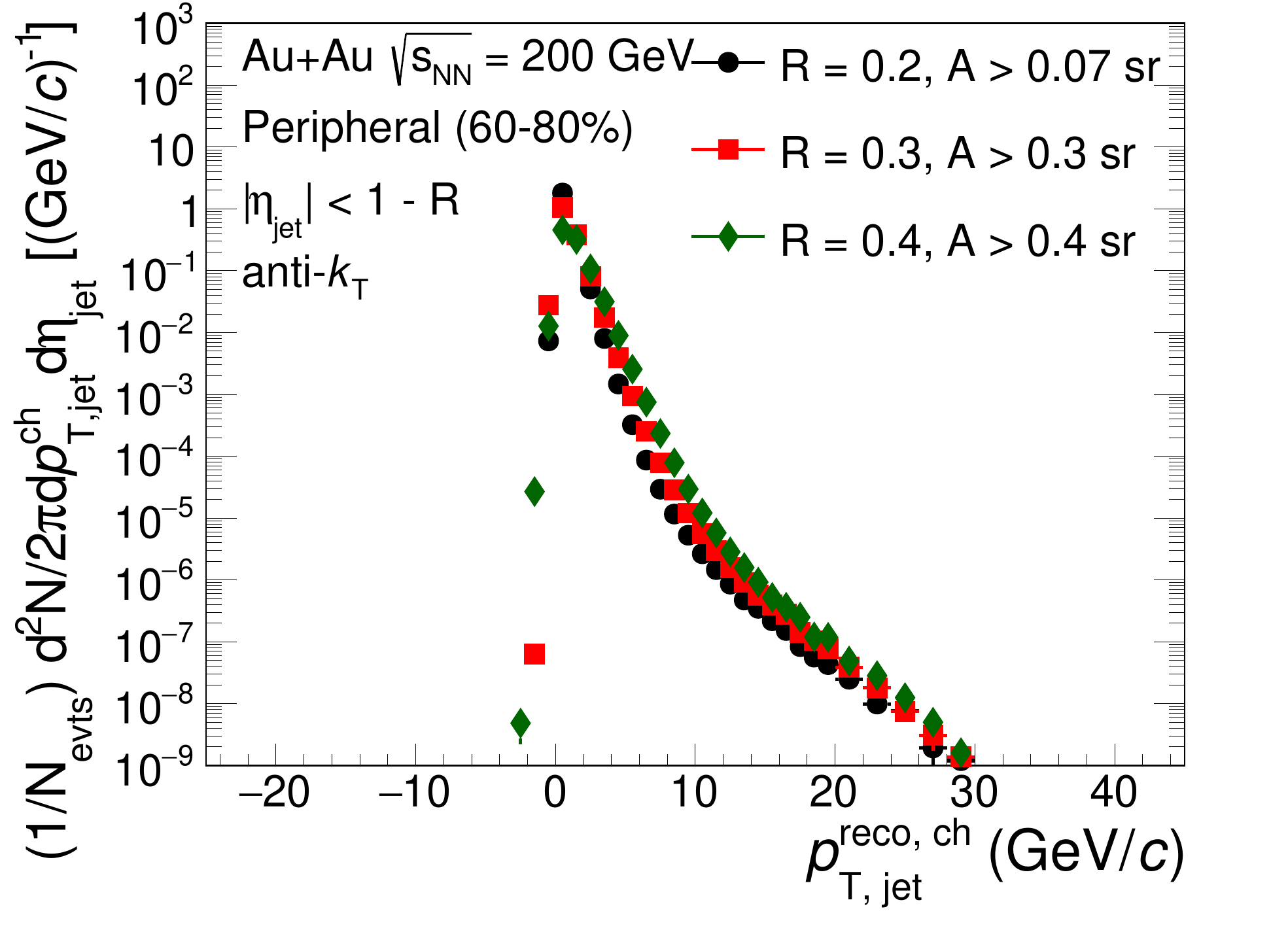}
\includegraphics[width=0.49\textwidth]{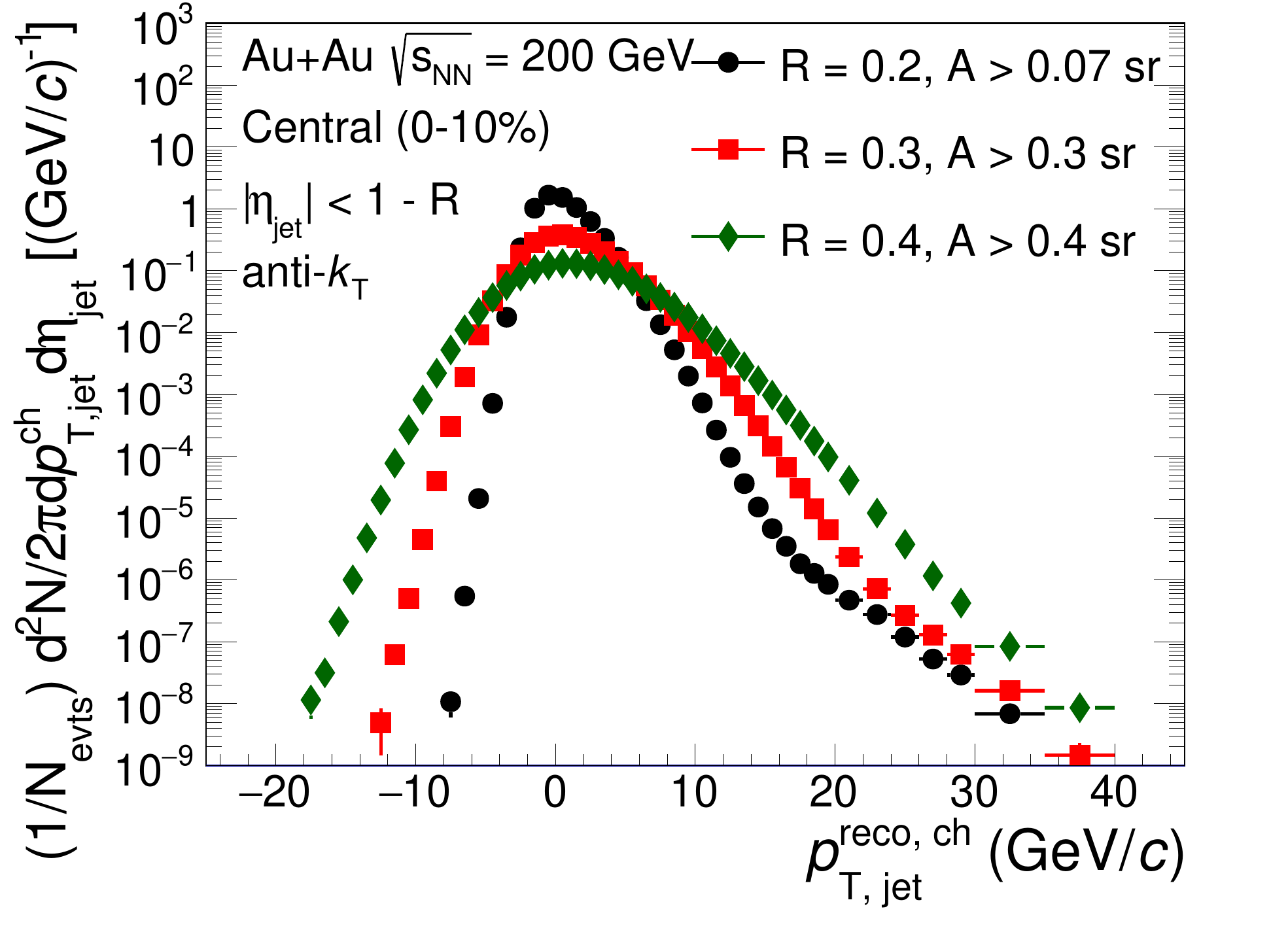}
\caption{(Color online) Distribution of inclusive charged-particle jet candidates passing the jet area cut as a 
function of \pTreco\ for \rr~=~0.2, 0.3, and 0.4, in peripheral (upper) and central (lower) \AuAu\ collisions.
}
\label{fig:ptreco_R}
\end{figure}

Figure~\ref{fig:ptreco_R} shows measured \pTreco\ distributions for inclusive 
jet candidates with \rr~=~0.2, 0.3, and 0.4 which pass the jet area cut, in 
peripheral and
central \AuAu\ collisions. The distributions for central \AuAu\ collisions have 
significant 
yield in the region $\pTreco<0$. This feature is also observed in 
hadron-triggered semi-inclusive 
analyses~\cite{Adam:2015doa,Adamczyk:2017yhe}, where it is attributed 
predominantly to combinatorial jet candidates generated by soft processes that are
uncorrelated with the trigger. 

The distributions exhibit a change in slope at $\pTreco\approx10$ \gev\ for 
all \rr\ in peripheral \AuAu\ collisions, and at $\pTreco\approx15$ \gev\ for 
\rr~=~0.2 in central \AuAu\ collisions, suggesting two distinct 
contributions to the spectrum that are visible for the configurations with 
smallest background. In this picture the yield at low \pTreco\ is dominated by 
combinatorial jet candidates, similarly to distributions in the hadron+jet analysis, 
while the yield at large 
\pTreco\ is dominated by jets arising from hard processes.

\begin{figure*}[htbp]
\includegraphics[width=0.48\textwidth]{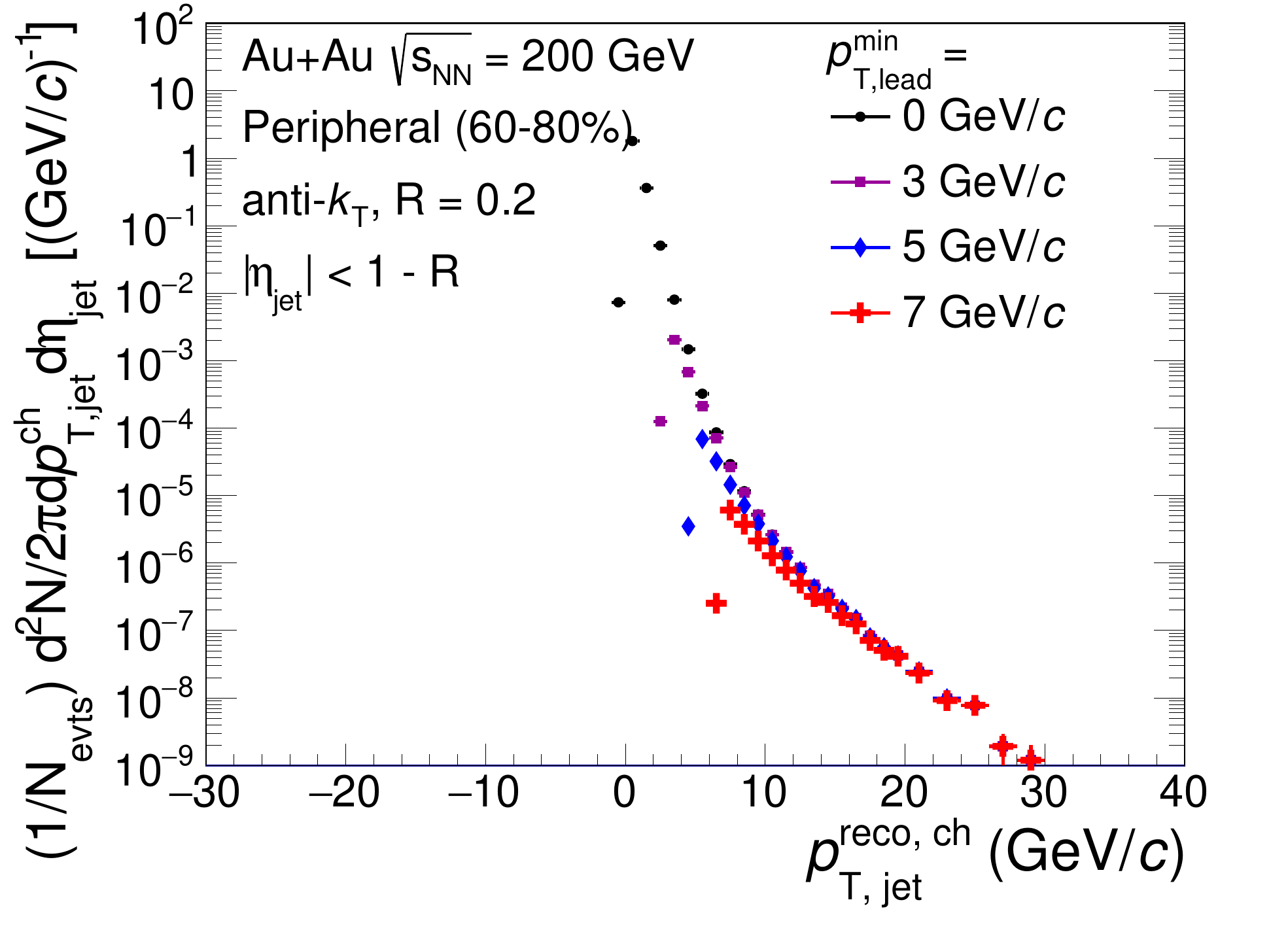}
\includegraphics[width=0.48\textwidth]{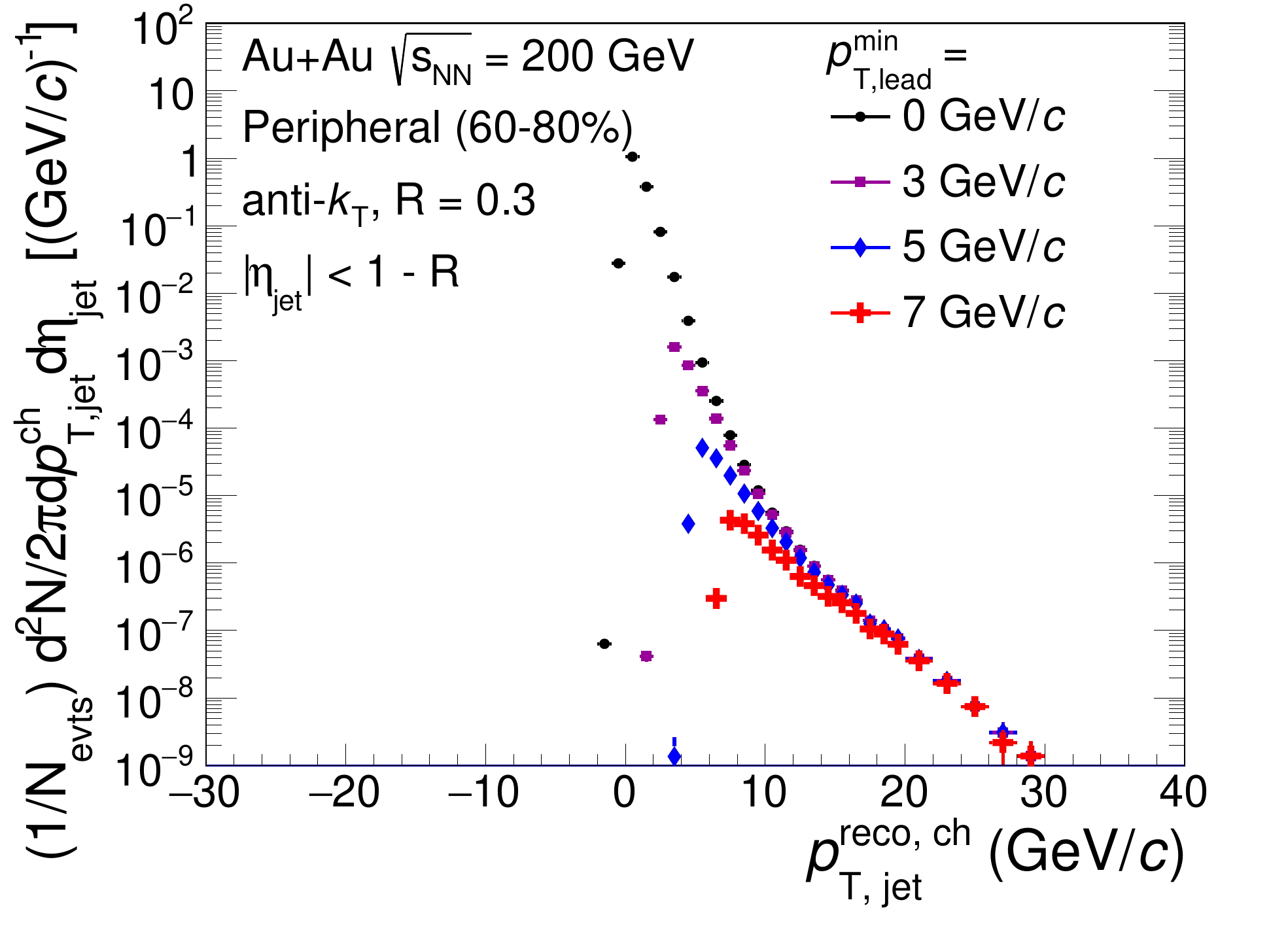}
\includegraphics[width=0.48\textwidth]{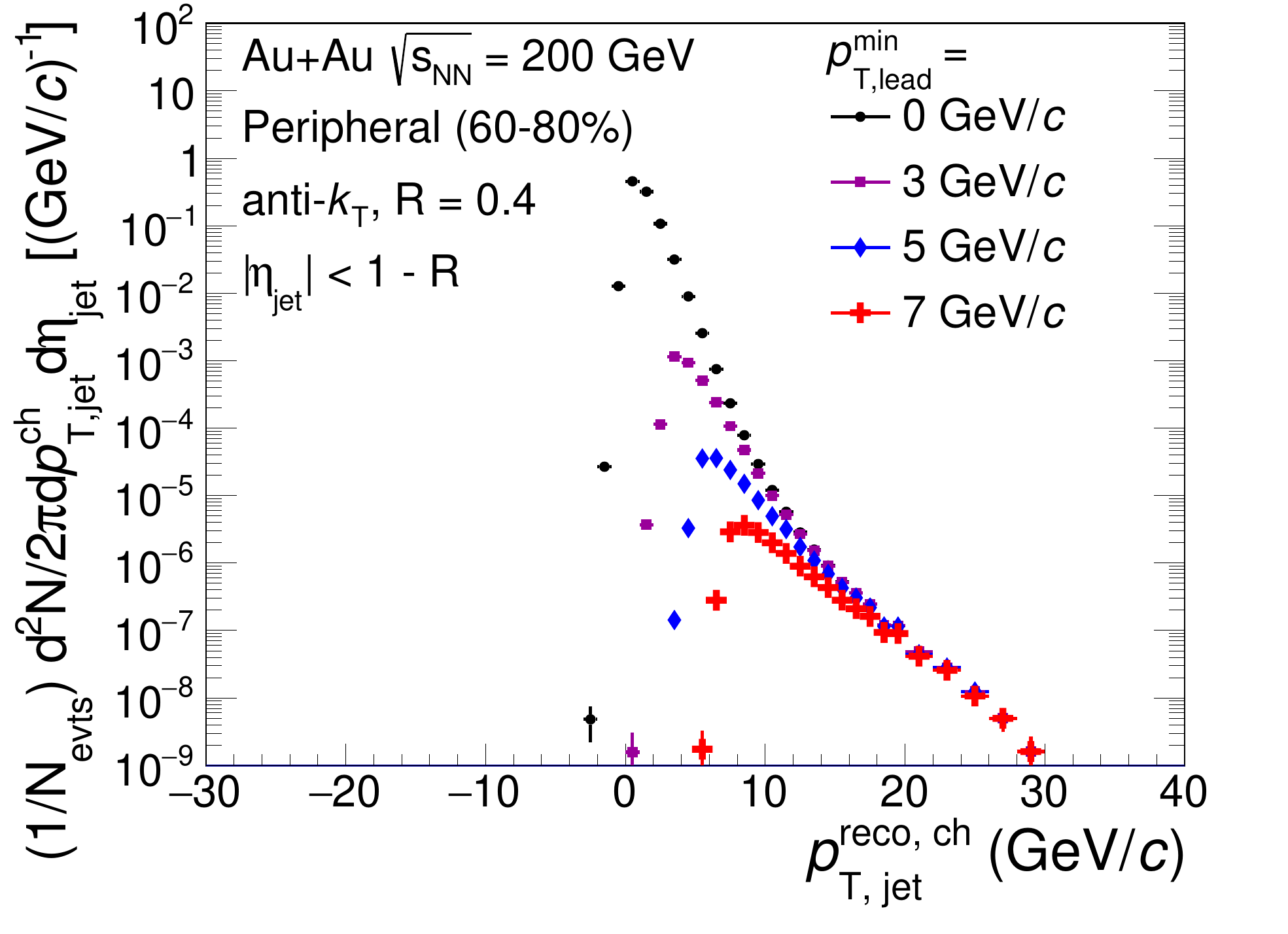}
\includegraphics[width=0.48\textwidth]{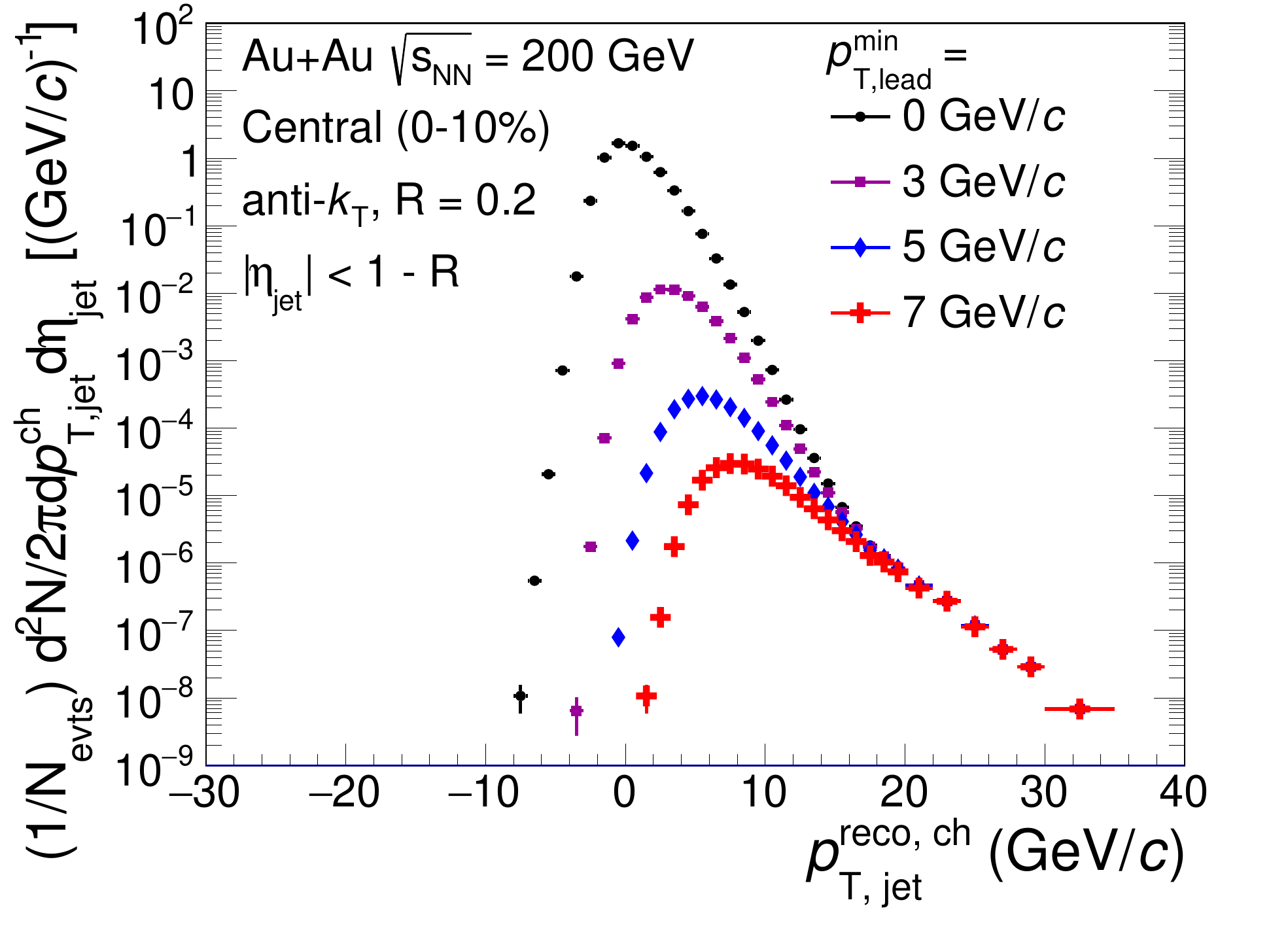}
\includegraphics[width=0.48\textwidth]{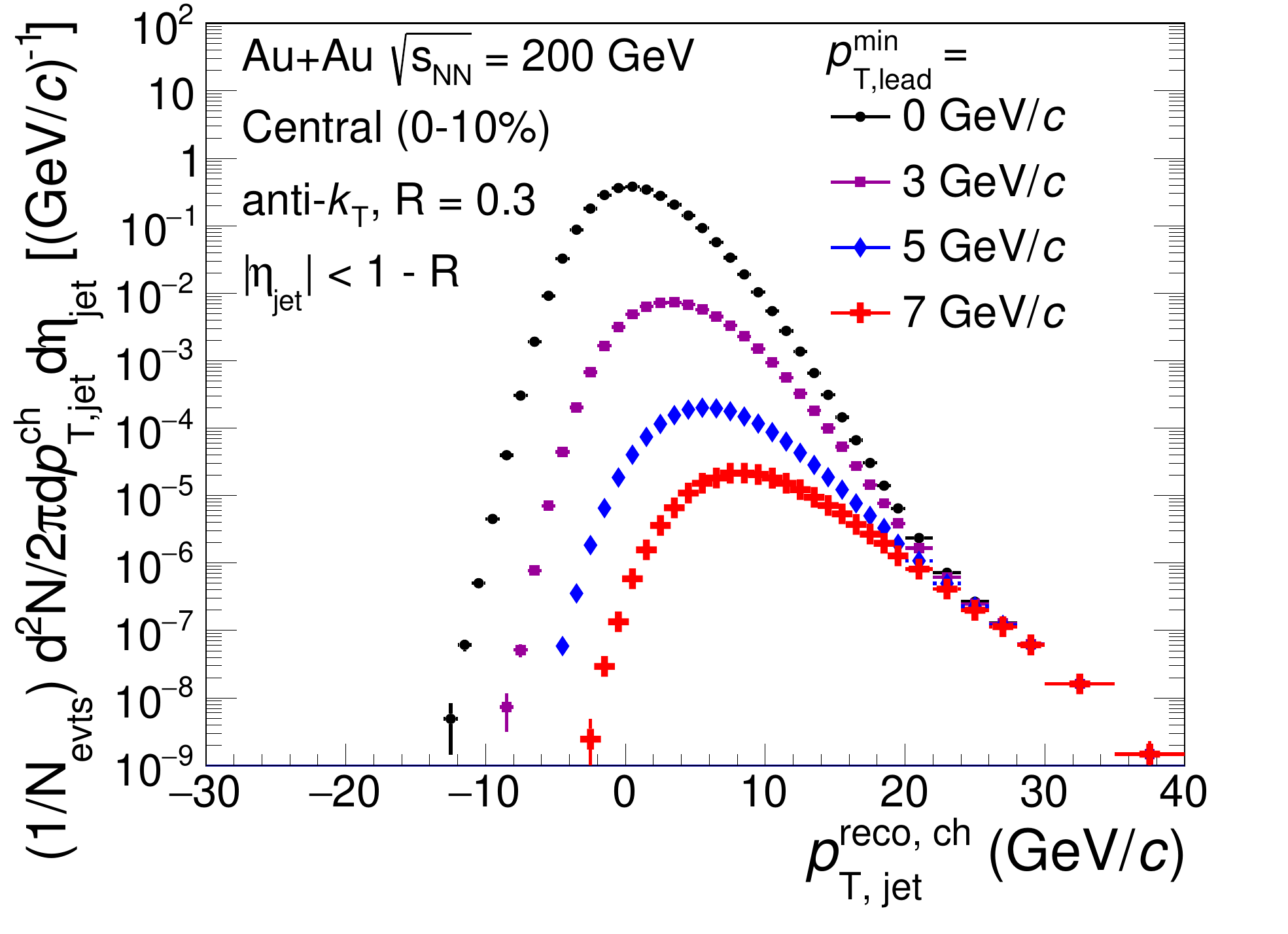}
\includegraphics[width=0.48\textwidth]{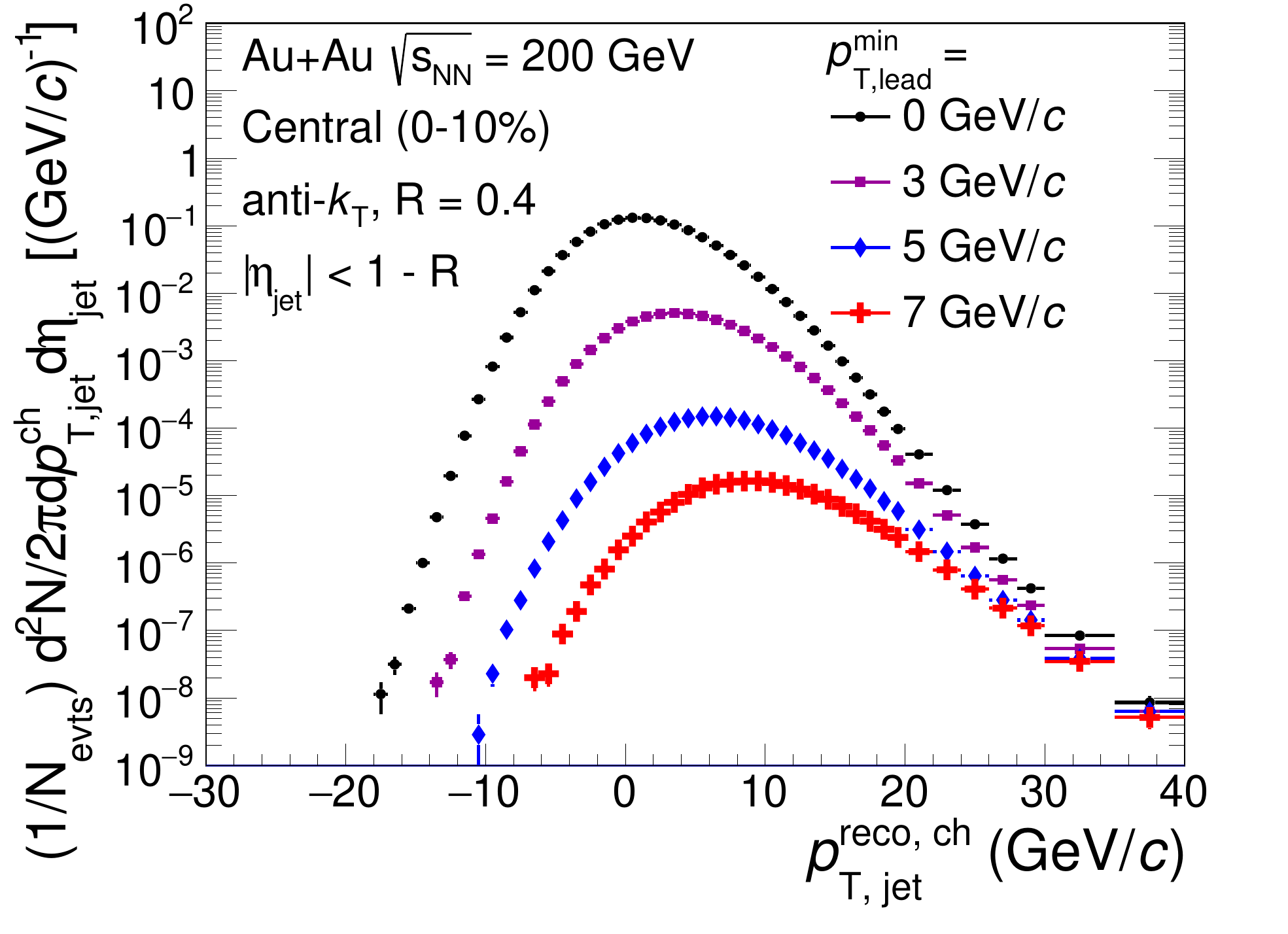}
\caption{(Color online) Distribution of \pTreco\ measured in peripheral (left) and central (right) \AuAu\ collisions at \sqrtsNN~=~200 GeV, for \pTleadmin~=~0, 3, 5 and 7 \gev. Upper: \rr~=~0.2; middle: \rr~=~0.3; lower: \rr~=~0.4.  The distributions for 
\pTleadmin~=~0 are the same as those in Fig.~\ref{fig:ptreco_R}. 
}
\label{fig:pTreco_pTlead}
\end{figure*}

Figure~\ref{fig:pTreco_pTlead} shows the effect of the cut $\pTlead>\pTleadmin$ on 
\pTreco\ distributions in peripheral and central \AuAu\ collisions. 
The \pTleadmin\ cut suppresses the 
yield most strongly for large negative values of \pTreco, with much reduced
suppression at large positive values of \pTreco. 

Larger values of \pTleadmin\ generate larger suppression, with correspondingly larger bias expected in the fully corrected distributions. Section~\ref{sect:Bkgd} specifies the competing criteria for optimizing the value of \pTleadmin. The optimum value of \pTleadmin\ for this analysis is found to be \pTleadmin~=~5 \gev, which is the lowest value that gives stable unfolding results (Sec.~\ref{sect:Unfolding}) and successful closure (Sec.~\ref{sect:ClosureTest}). The value \pTleadmin~=~7 \gev\ is used for systematic variation, to determine the range in \pTjetch\ over which the bias is small (Sec.~\ref{sect:results}).

\section{Corrections}
\label{sect:Corrections}

The raw distributions are corrected for the effects of 
instrumental response and background fluctuations, using regularized 
unfolding~\cite{Cowan:2002in,Hocker:1995kb,dAgostini2010}. We utilize the approach and notation described in Ref.~\cite{Adamczyk:2017yhe}.

\subsection{Instrumental response matrix \Rdet}
\label{sect:Rdet}

The instrumental response matrix \Rdet\ is constructed using PYTHIA-generated 
events for \pp\ collisions at \sqrts~=~200 GeV. A detector-level event is 
generated by applying the fast 
simulator to each particle-level event.
Jet reconstruction is then carried 
out with the \antikT\ algorithm at both the particle and detector levels, and jets are selected by applying the 
fiducial acceptance ($|\etajet|<1-\rr$) and \pTleadmin\ cuts. Jets at the particle and 
detector levels are matched following the procedure 
in Ref.~\cite{Adamczyk:2017yhe}: Tracks are first matched at the detector and particle levels, and then the detector-level jet with the largest fraction of the energy of a given particle-level jet is matched to it, if that fraction is greater than 50\%.

\begin{figure}[htbp]
\includegraphics[width=0.49\textwidth]{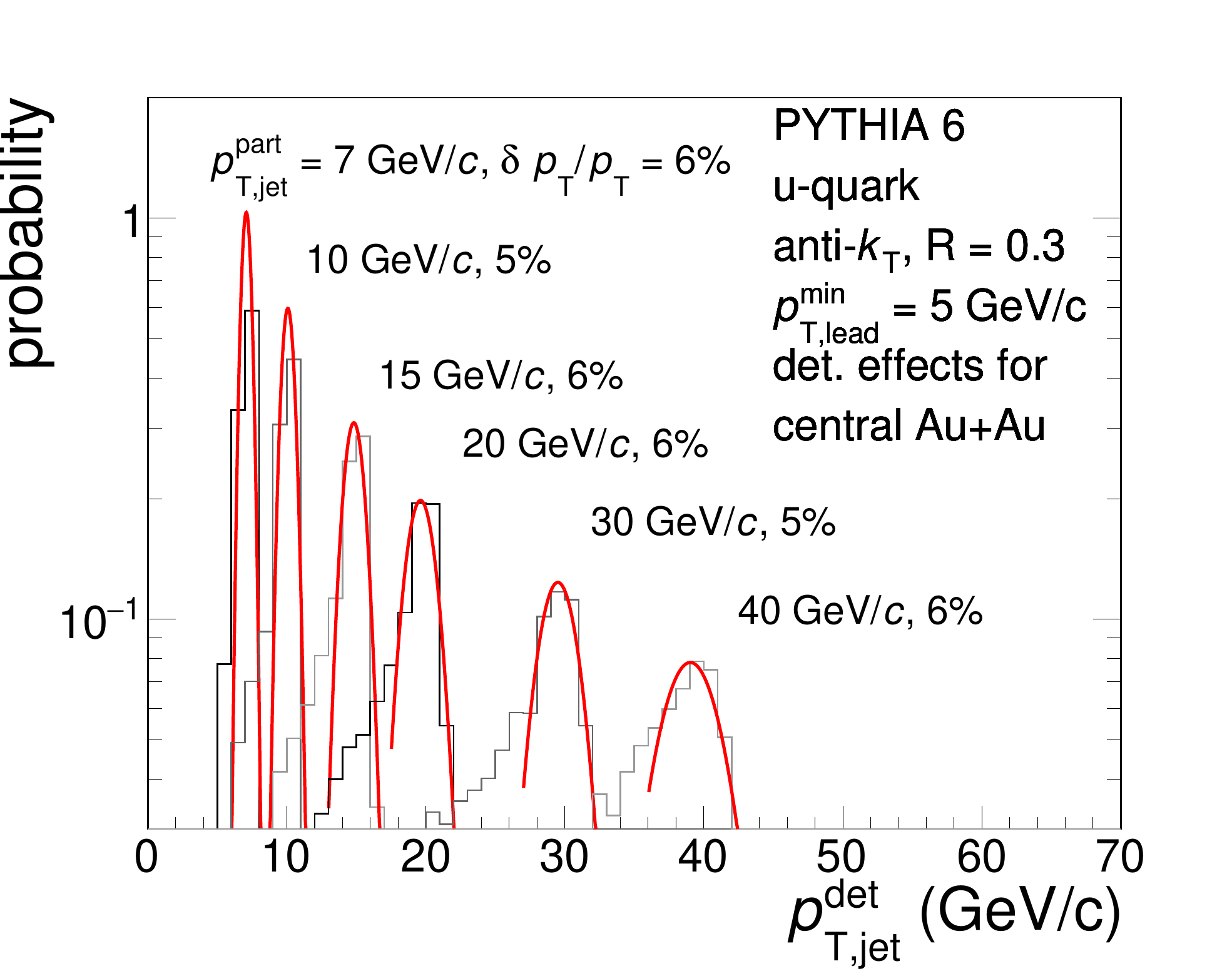}
\caption{(Color online) Instrumental jet response: Distribution of \pTdet\ 
for u-quark jets  generated by PYTHIA with various values of \pTpart, with detector effects corresponding to those in central \AuAu\ collisions. Jets have \rr~=~0.3 and are selected with the requirement \pTleadmin~=~5 \gev\ at both the particle and detector levels.
The red distributions show a Gaussian functional fit to the peak region of each distribution, 
with relative width of the fit as shown. 
}
\label{fig:JetResponse}
\end{figure}

The instrumental response is determined by comparing
matched
jets at the particle and detector levels. 
Figure~\ref{fig:JetResponse} shows the distribution of \pTdet\ for u-quark-initiated jets 
with several values 
of \pTpart, with detector-level effects corresponding to those 
for central \AuAu\ collisions. The cut $\pTleadmin~=~5$ 
\gev\ is applied both at the particle  and detector levels
for the primary analysis, with $\pTleadmin~=~7$ \gev\ used to correct the corresponding analysis used for systematic variation (not shown).
The distributions 
for 
gluon-initiated jets are very similar, suggesting that the instrumental 
response does not depend significantly on the specific mixture of light quark- 
and gluon-initiated jets in the population.

The instrumental response in Fig.~\ref{fig:JetResponse} is 
asymmetric, with a tail for $\pTdet<\pTpart$ that arises 
predominantly from the loss of a single charged hadron with high 
momentum-fraction (high-$z$)  
due to tracking inefficiency~\cite{Adamczyk:2017yhe}. This asymmetric response 
cannot be characterized fully by a Jet 
Energy Resolution (JER) figure, 
and so the entire distribution shown in Fig.~\ref{fig:JetResponse}  is used to correct the spectrum for instrumental effects. Nevertheless, as an approximation to the 
JER, we fit the main peak of these 
distributions with a Gaussian function and report its relative width, as shown 
in the figure.
For jets with $7<\pTpart<40$ \gev, the relative width has values between 4 and 
10\%, with negligible dependence on fragmentation model or jet resolution 
parameter \rr.

A detector-level jet corresponding to a particle-level jet in the experimental 
acceptance can be lost due to fiducial cuts and 
instrumental response. The most significant contribution to this loss is tracking inefficiency, 
especially for low-\pT\ jets containing few tracks. The jet area cut has negligible inefficiency for 
$\pTjetpart>4$ \gev.

\begin{figure}[htbp]
\includegraphics[width=0.49\textwidth]{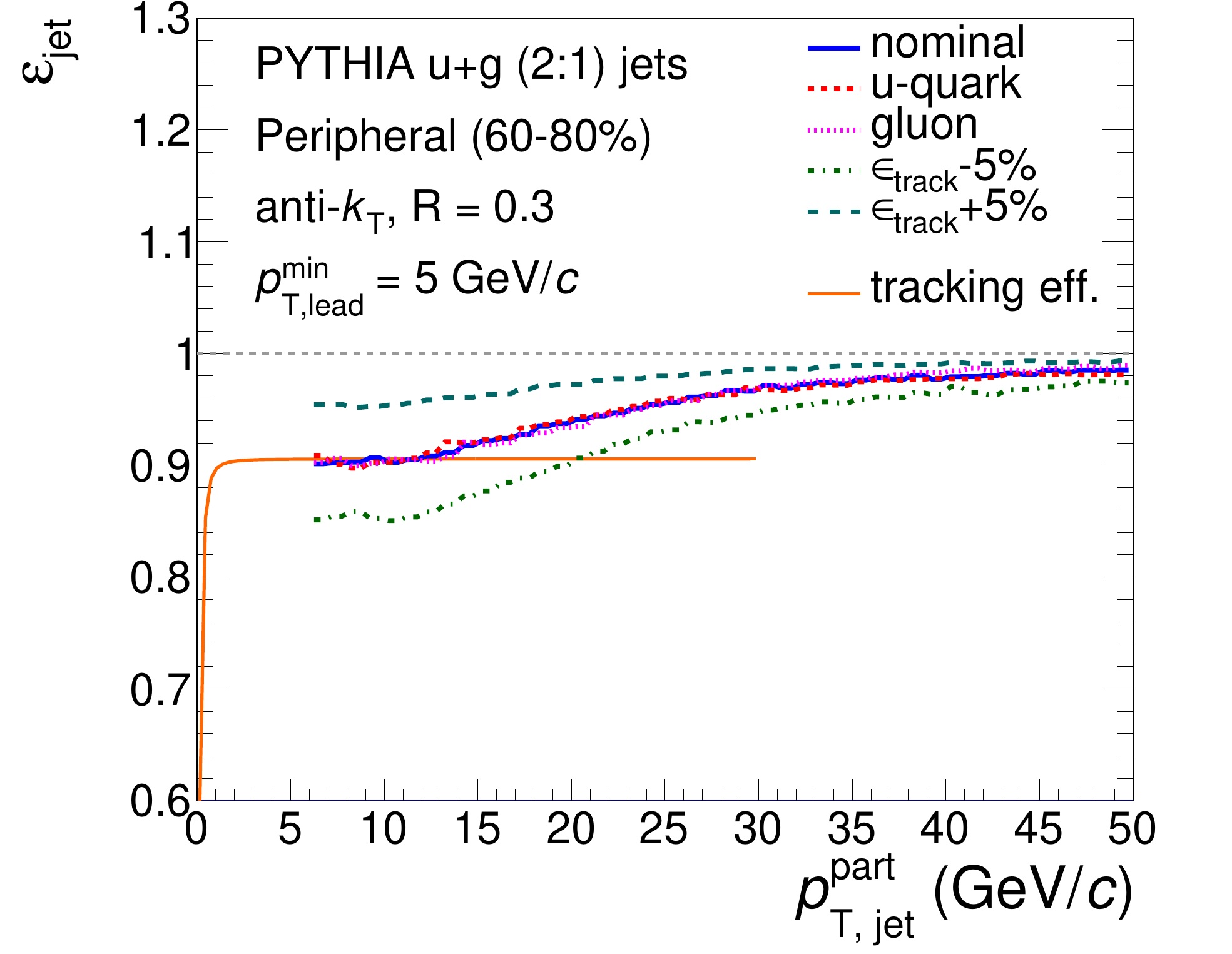}
\includegraphics[width=0.49\textwidth]{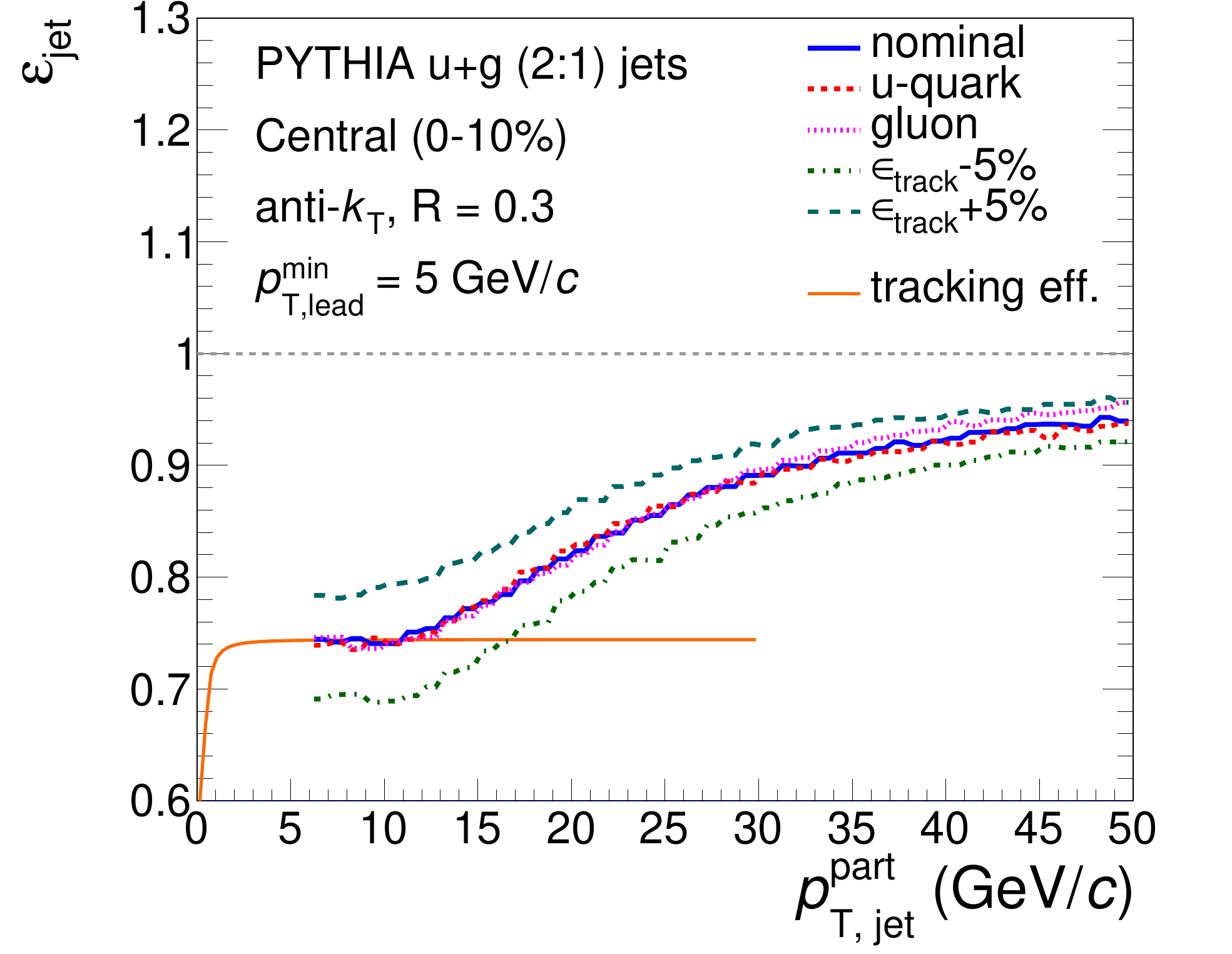}
\caption{(Color online) Jet reconstruction efficiency in peripheral (upper) and 
central (lower) \AuAu\ collisions, for \rr~=~0.3 and \pTleadmin~=~5 \gev, for 
u quarks and gluons in the ratio 2:1 (labeled ``nominal"), pure u, pure g, and 
variation of the relative tracking efficiency by $\pm5\%$ for the nominal 
population. The orange line shows the single-track efficiency.
}
\label{fig:effi}
\end{figure}

For transparency in understanding the effects of unfolding, corrections for \pT-smearing and jet finding efficiency are applied in separate steps. This is implemented by normalizing the elements of \Rdet\ such that, for each bin in \pTpart, the integral over \pTdet\ is unity; \Rdet\ thereby only corrects for \pT-smearing. The effects of jet finding efficiency are then corrected by multiplying the unfolded solution with the efficiency as a function of \pTpart.

Figure~\ref{fig:effi} shows the jet reconstruction 
efficiency. The nominal calculation is carried out for a mixture of u-quark and 
gluon 
jets with yield ratio 2:1, and the nominal tracking efficiency. The efficiencies 
for pure u-quark or gluon populations are also shown, as is the jet-finding 
efficiency for $\pm$ 5\% relative variation 
in tracking efficiency, corresponding to its systematic uncertainty. The single-track efficiency is also shown, which corresponds to the jet reconstruction efficiency for \pTpart~=~\pTleadmin.

A particle-level jet without a sufficiently hard leading track may be accepted 
at the detector-level due to track momentum smearing. This jet feed-up increases 
the jet finding efficiency for the lowest \pTjet\ values by 1--2\% (absolute). 
Figure~\ref{fig:effi} includes this effect.

A track from a displaced vertex arising from a weak decay may be assigned an 
incorrect momentum that situates it inside or outside of a jet cone differently 
than its parent particle. However, such effects are found to be 
negligible~\cite{Adamczyk:2017yhe} and no correction for them is applied.

\begin{figure*}[htbp]
\includegraphics[width=0.32\textwidth]{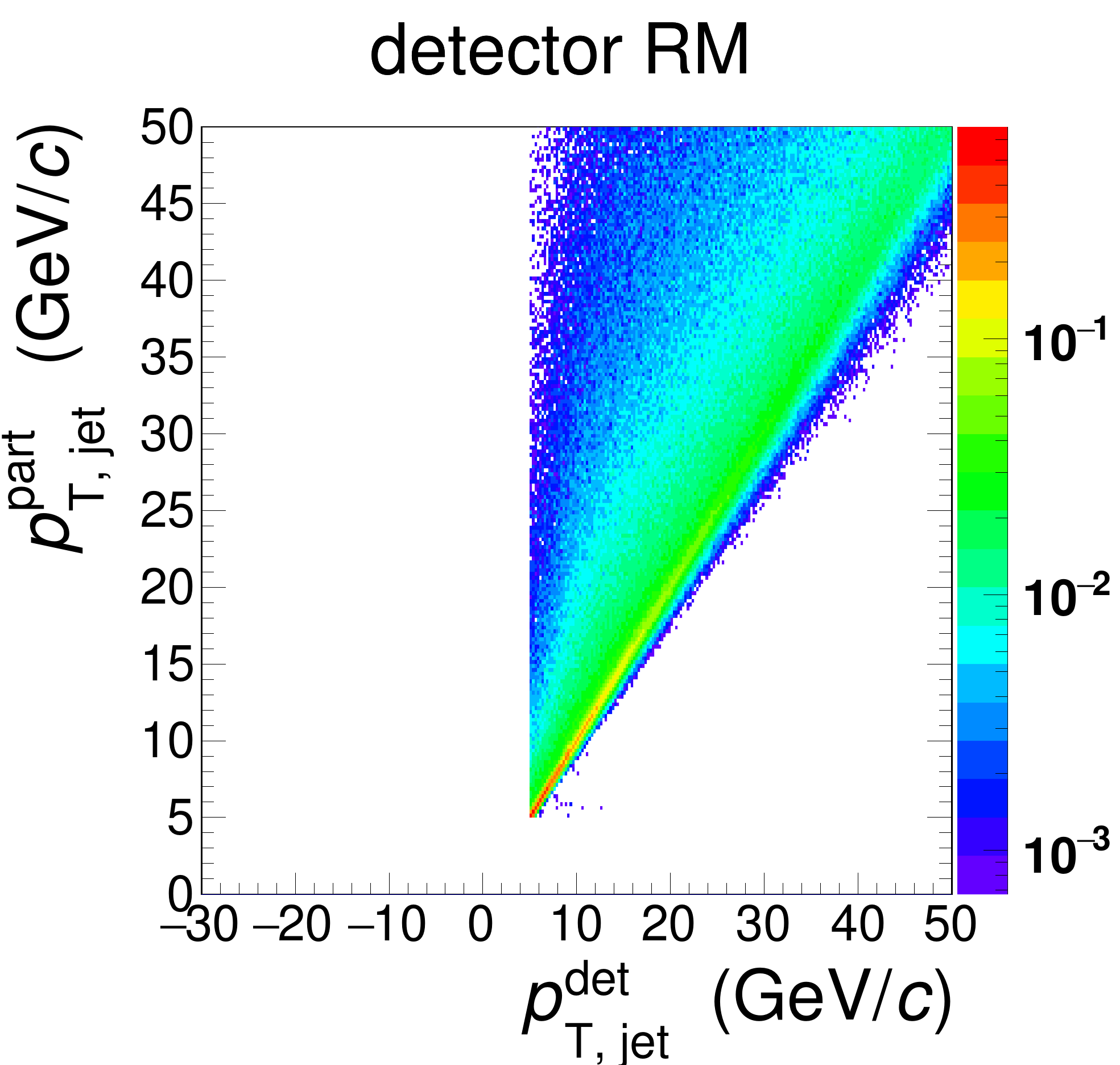}
\includegraphics[width=0.32\textwidth]{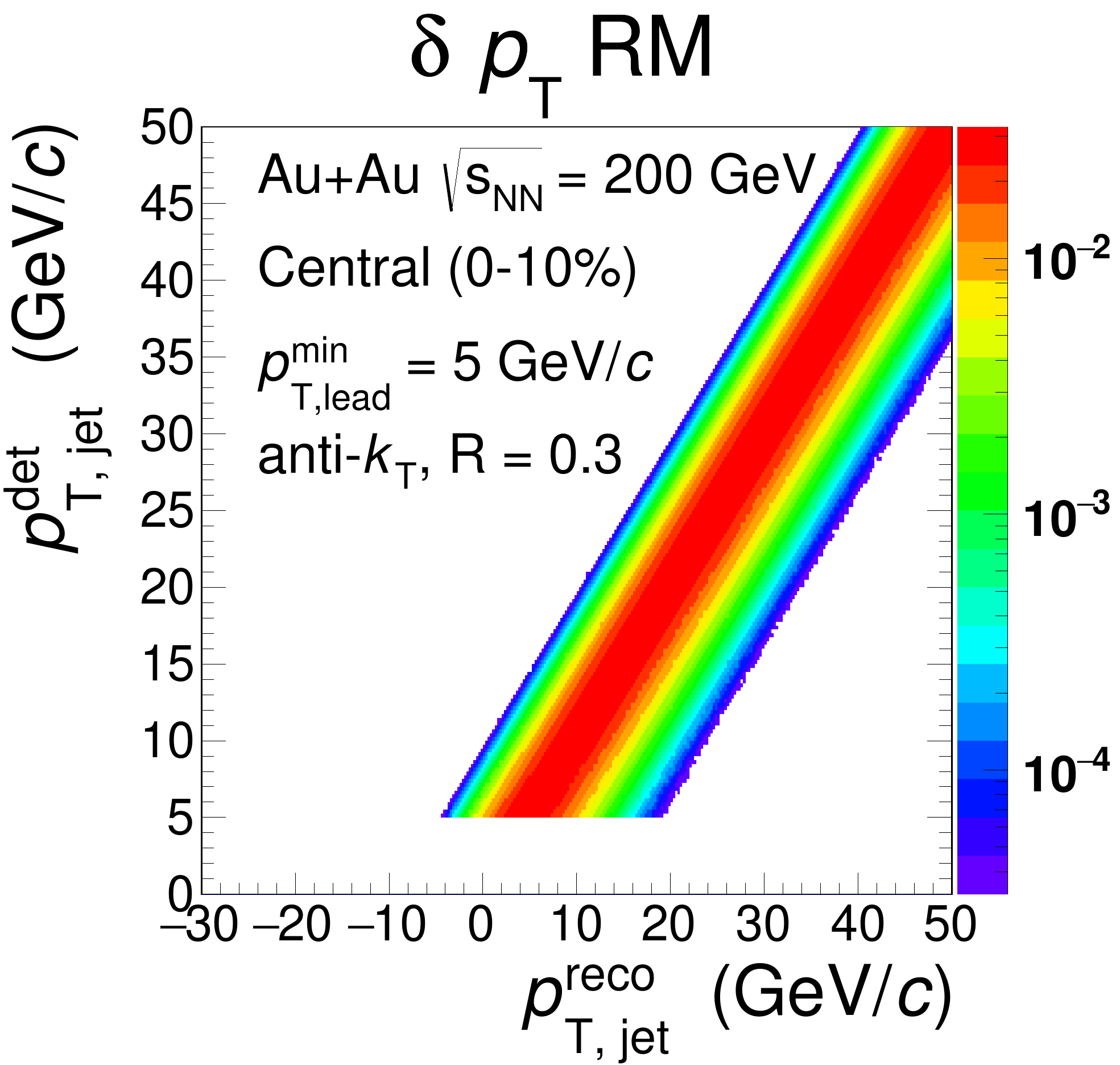}
\includegraphics[width=0.32\textwidth]{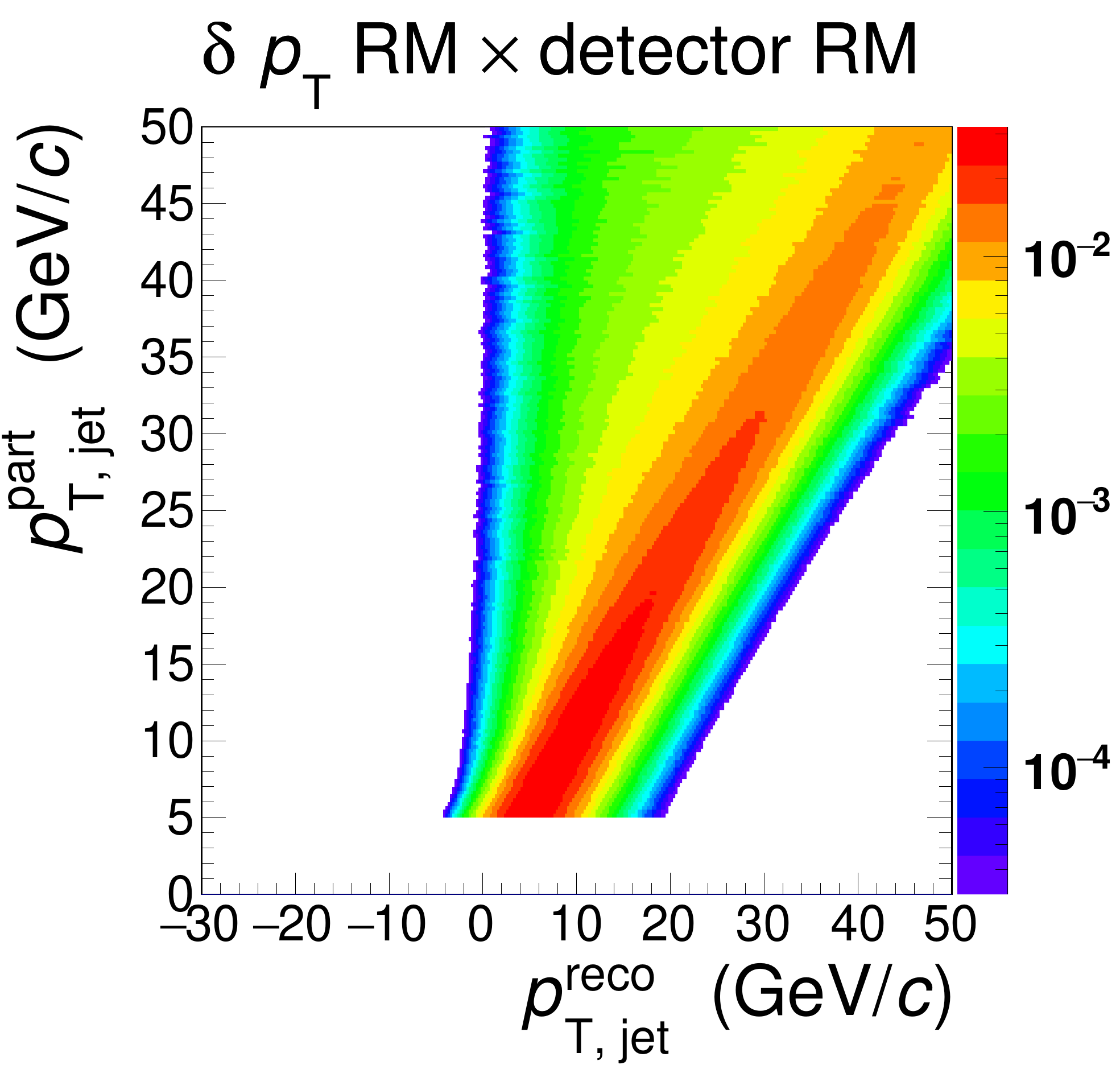}\\
\caption{(Color online) Response matrices (RM) for charged jets with \pTleadmin~=~5 \gev\ and \rr~=~0.3. Left: Detector effects  \Rdet; center: background fluctuations \Rbkg\ (SP embedding); right: $\Rtotal = \Rbkg \times \Rdet$. 
} 
\label{fig:RM}
\end{figure*}

Figure \ref{fig:RM} (left panel) shows the matrix \Rdet\ for 
central \AuAu\ collisions. Contributions in the region
$\pTjetdet<\pTjetpart$ are due primarily to tracking efficiency, which causes 
tracks to be lost from the jet. Contributions in the region 
$\pTjetdet>\pTjetpart$, which are less probable, arise primarily from the 
effect of momentum resolution for cases in which the fraction of \pTjet\ lost due to tracking 
inefficiency is small.




The Jet Energy Scale (JES) uncertainty due to instrumental effects, which is 
dominated by the uncertainty of 
the tracking efficiency, is $\approx$5\% for \rr~=~0.2 and 0.3, and 7\% for 
\rr~=~0.4, in 
central \AuAu\ collisions; and 3\% for \rr~=~0.2, 0.3, and 0.4 in peripheral 
\AuAu\ collisions.
The dependence of JES on \pTdet\ is negligible.

\subsection{Uncorrelated background response matrix \Rbkg}
\label{sect:dpT}

The response matrix representing fluctuations in energy density 
uncorrelated with a jet arising from a hard process is calculated by embedding 
detector-level simulated jets into real events, reconstructing the hybrid 
events, and then matching each embedded jet with 
a reconstructed jet. The matching of particle- and detector-level jets likewise 
follows the procedure in Ref.~\cite{Adamczyk:2017yhe}. The response matrix 
corresponds to the probability distribution for \dpT, where

\begin{equation}
\dpT=\pTreco-\pTembed.
\label{eq:dpT}
\end{equation}

Jet reconstruction algorithms are infrared and collinear-safe (IRC-safe) in 
elementary collision systems, i.e., they measure energy flow and are
insensitive to the specific pattern of jet fragmentation into 
hadrons. In this analysis we likewise seek to measure energy flow for 
charged-particle jets in heavy-ion collisions, without bias 
toward specific patterns of jet fragmentation. That goal requires the 
\dpT\ distribution not to have significant dependence on the jet fragmentation 
pattern.

\begin{figure*}[htbp]
\includegraphics[width=0.49\textwidth]{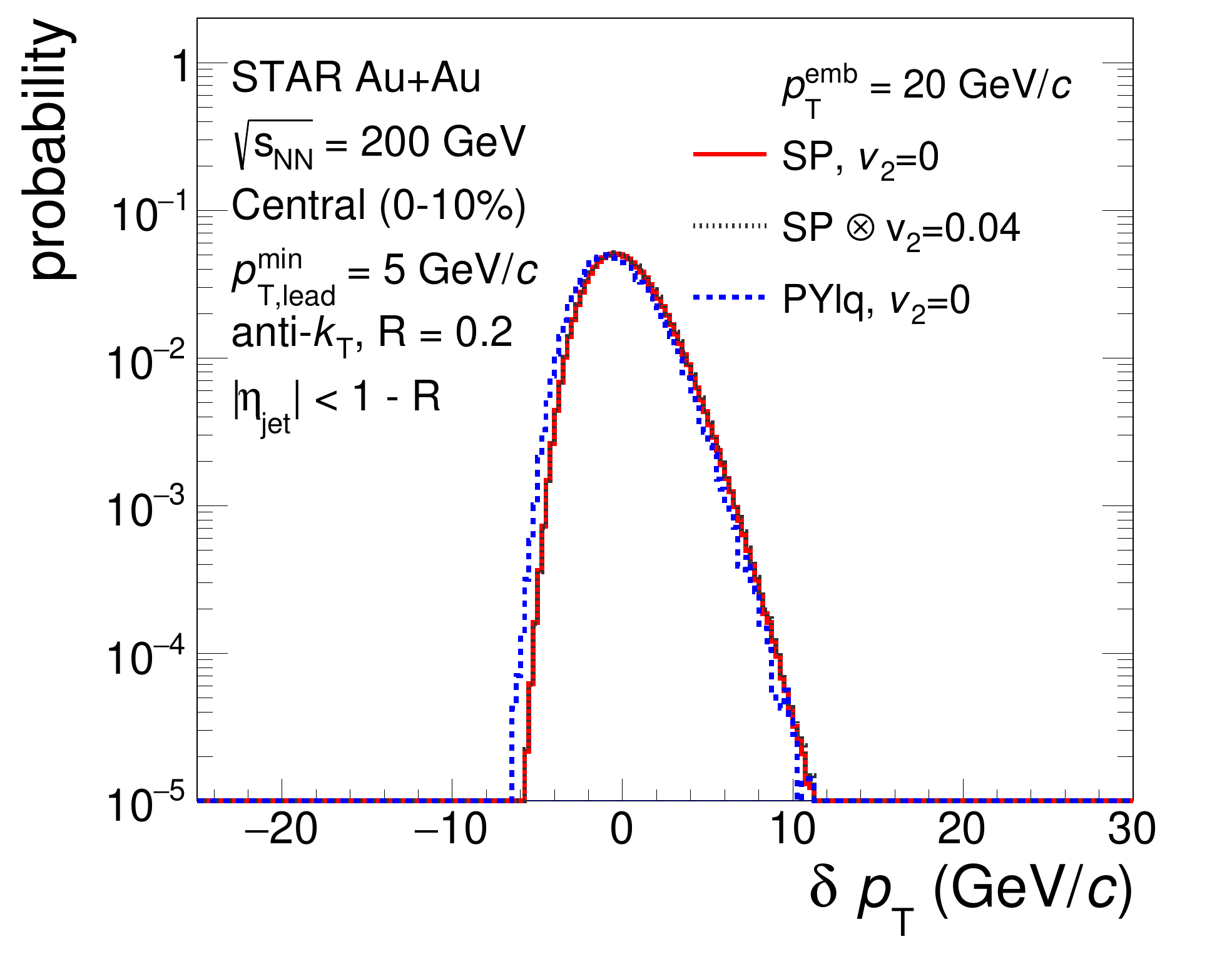}
\includegraphics[width=0.49\textwidth]{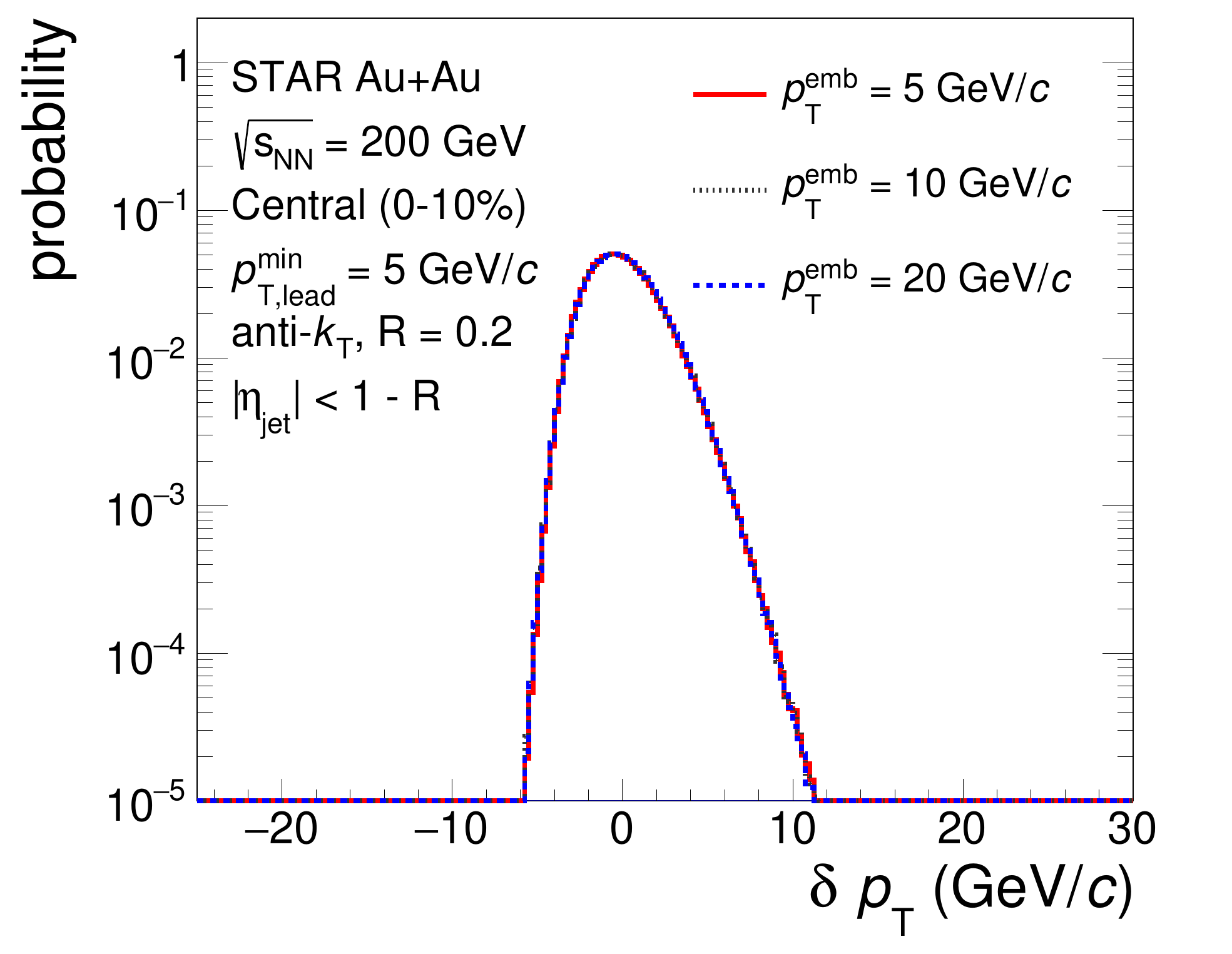}
\includegraphics[width=0.49\textwidth]{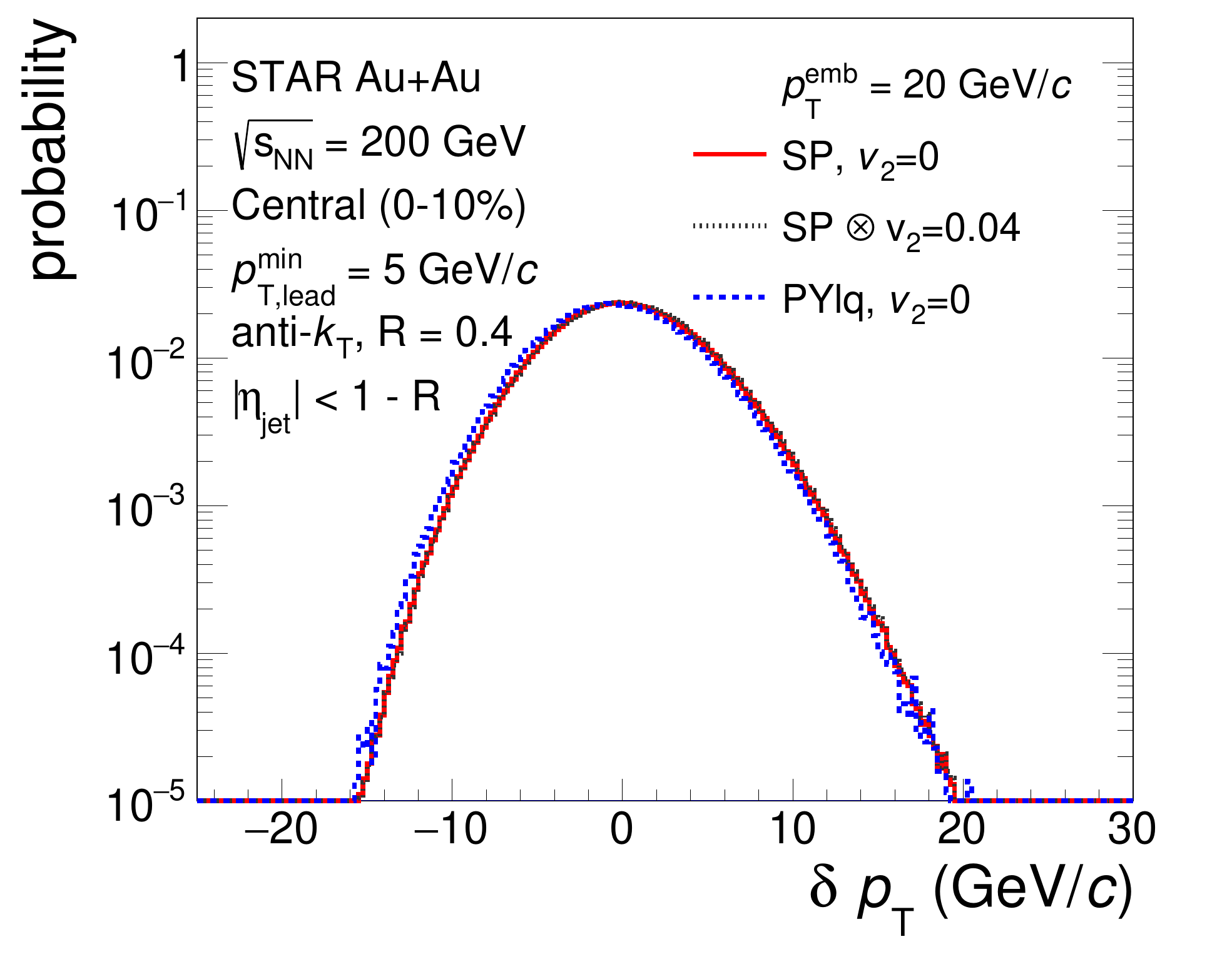}
\includegraphics[width=0.49\textwidth]{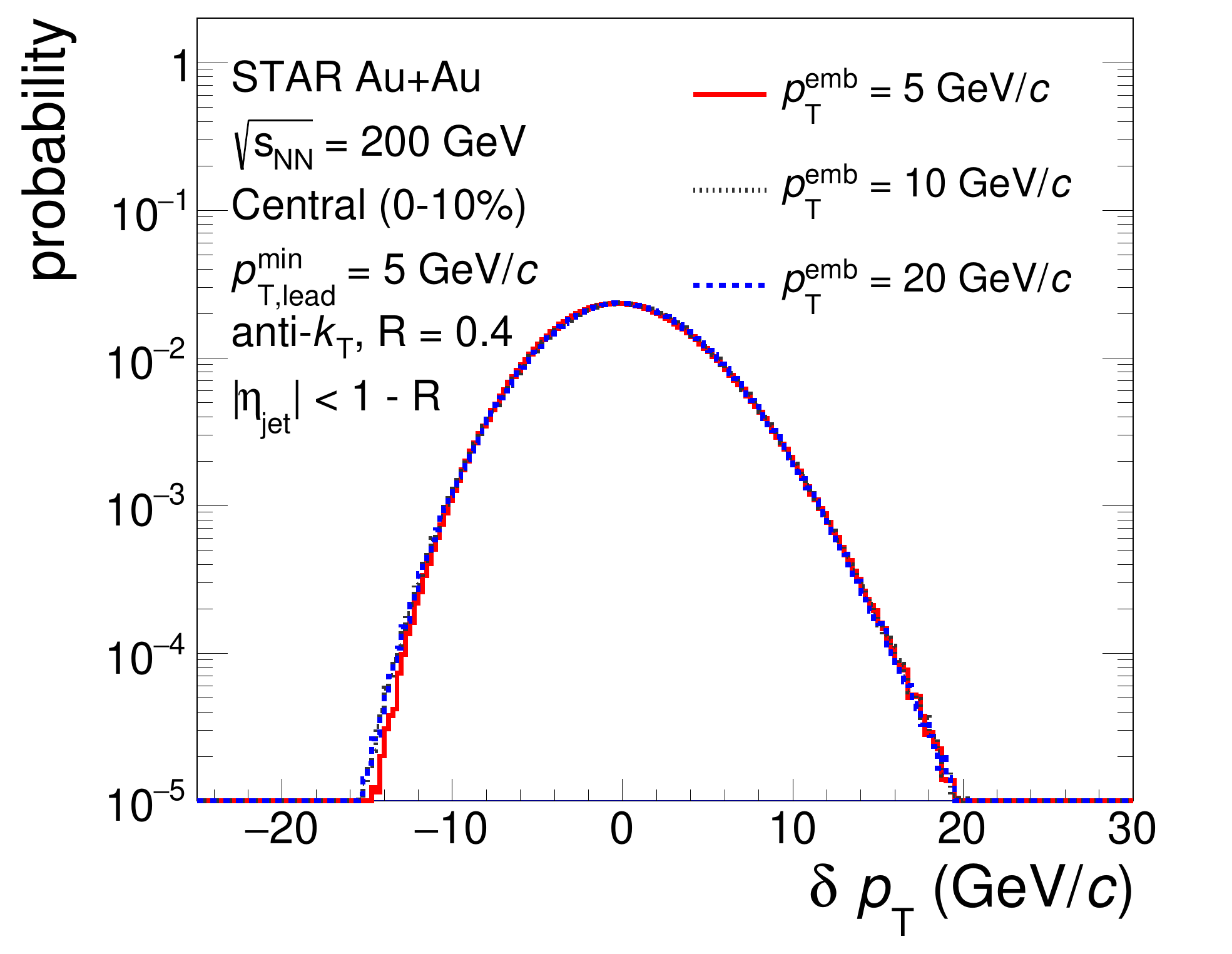}
\caption{(Color online) \dpT\ distributions calculated by embedding various types of simulated jet in central \AuAu\ 
collisions at \sqrtsNN~=~200 GeV, for \rr~=~0.2 (left) and \rr~=~0.4 (right). Upper panels: \pTembed~=~20 \gev\ for SP jets, 
SP jets with \vtwo-modulated background, and light quark jets generated by PYTHIA. 
 Lower panels: SP jets for several values 
of \pTembed. See text for details. 
}
\label{fig:dpT}
\end{figure*}

In order to test this dependence, we calculate the \dpT\ 
distribution in Eq.~(\ref{eq:dpT}) with two significantly different jet 
fragmentation models: light quark jets generated by PYTHIA (PYlq), utilizing the PYTHIA fragmentation routines for a quark of specified flavor and momentum;  
and ``single-particle" jets (SP), in which the entire jet \pT\ is carried by a single hard 
particle~\cite{Jacobs:2010wq}. Figure~\ref{fig:dpT} (upper panels) compare the
\dpT\ probability distributions for the SP and PYlq 
fragmentation models for \rr~=~0.2 and 0.4 jets with \pTjetch~=~20 \gev\ embedded into central \AuAu\ collisions; the cut \pTleadmin~=~5 \gev\ is applied to the PYlq jets.
The two fragmentation models generate similar \dpT\ distributions, having similar shape and differing by a shift of $\approx$500 MeV. This variation is accounted for in the systematic uncertainty, discussed below.  
Figure~\ref{fig:area} shows that the jet area distributions for these two fragmentation models are also similar.

Figure~\ref{fig:RM} (middle panel) shows the background response matrix \Rbkg, 
whose  elements are the \dpT\ probability distribution as a function of \pTjet, 
calculated by SP embedding. 

High-\pT\ hadrons may be correlated in azimuth with the event plane (EP)
orientation~\cite{Adare:2014bga}. The strength of this correlation is characterized by \vtwo, the 
second-order Fourier coefficient of 
the azimuthal distribution between the hadron and the EP. Nonzero \vtwo\ for hadrons with $\pT>\pTleadmin$ will bias the orientation of the accepted jet population relative to the EP, thereby biasing the level of uncorrelated 
background. This bias is taken into account in the calculation of the \dpT\ probability distribution by weighting each embedded jet with a weight $w$,

\begin{equation}
 w = 1+ 2 \vtwo \cos (2 \Delta \phi),
 \label{eq:v2weight}
\end{equation}
\noindent
where $\Delta\phi$ is the azimuthal angle of the leading hadron relative to the EP 
axis. Figure \ref{fig:dpT} (upper panels) show \dpT\ probability distributions 
with SP embedding for \pTjet~=~20 \gev, for \vtwo~=~0 and for \vtwo~=~0.04,
with the latter value consistent with hadron \vtwo\ measured in the region $\pT>\pTleadmin$~\cite{Adare:2014bga}. This 
variation in \vtwo\ is seen to generate negligible variation in the \dpT\ 
distributions, and its effect is likewise negligible in the final corrected spectra. 
This is the only contribution of azimuthal asymmetry effects to the analysis.

Figure~\ref{fig:dpT} (lower panels) show \dpT\ probability distributions for SP 
jets with embedded \pTjet~=~5, 10 and 20 \gev. These distributions exhibit negligible 
dependence on \pTjet\ for \rr~=~0.2, and minor dependence for \rr~=~0.4.

Figure~\ref{fig:dpT} shows that the response matrix for 
background fluctuations in central \AuAu\ collisions is largely independent of both
\pTjet\ and the fragmentation model used in the calculation. A similar lack of 
dependence on fragmentation model is found for 
peripheral \AuAu\ collisions. This indicates that 
jet reconstruction in this analysis indeed measures energy flow within jets, as 
required. The 
small residual variations seen in Fig.~\ref{fig:dpT} are taken into account in the 
systematic uncertainty of the corrected spectra. 

\subsection{Unfolding}
\label{sect:Unfolding}

The unfolding procedure utilizes the cumulative response matrix 
[Fig.~\ref{fig:RM}, right panel], which is the product of \Rbkg\ and \Rdet. Two 
different unfolding methods are used: an iterative method 
based on Bayes's theorem~\cite{D'Agostini:1994zf}, and a method based on 
singular value decomposition (SVD)~\cite{Hocker:1995kb}. 

Several different functional forms are used for the prior distribution: a 
power-law distribution, $p_{T}^{-n}$, with $n=$~4.5, 5.0, and 5.5; the 
inclusive jet distribution generated by PYTHIA for \pp\ collisions at \sqrts~=~200 
GeV, with \pTleadmin~=~5 \gev; and the Tsallis 
function~\cite{Adamczyk:2017yhe,Adare:2010fe}, with $n$ varying between 4 and 20 
and $T$ varying between 0.6 and 1.2.

The unfolding procedure is regularized, which imposes a smoothness 
constraint on the solution~\cite{Cowan:2002in,Hocker:1995kb,dAgostini2010}. 
The backfolded distribution, which is 
the unfolded distribution smeared by the response matrix, is used to optimize 
the regularization.
For iterative Bayesian 
unfolding, regularization corresponds to limiting the number of iterations $i$; 
optimization of the regularization is based on comparison of unfolded 
distributions for two successive iterations, and comparison of 
the backfolded and uncorrected distributions. For SVD unfolding, 
regularization corresponds to truncation of the expansion at $k$ terms; 
optimization of the value of $k$ is determined by comparing the backfolded and 
uncorrected distributions.

Values of $i$ or $k$ are accepted if the distance between the unfolded and 
backfolded histogram (or between successive iterations in the case of $i$)
is sufficiently small. The histogram distance is quantified using the average relative distance 
between the central values of two distributions $a$ and $b$,

\begin{equation}
\drel =\frac{1}{n} \sum_{i=1}^{n} \frac{ |a_{i} - b_{i}| 
}{\mathrm{min}(a_{i},b_{i})},
\label{eq:drel} 
\end{equation}
\noindent
where $n$ is the number of bins, and $a_i$ and $b_i$ denote the central values in 
bin $i$.

This approach is based on PM simulations (Sec.~\ref{sect:ClosureTest}) which show that a small distance between backfolded 
and unfolded solutions, or between successive iterations for Bayesian unfolding,
corresponds to a small distance between the unfolded and generated spectra.
The $\drel$ metric is found to provide better discrimination than 
$\chi^2$ and Kolmogorov-Smirnov metrics. 

\subsection{Magnitude of corrections}
\label{sect:EstCorrections}

In this section we estimate the magnitude 
of corrections to the quasi-inclusive jet spectrum, to provide context for the 
systematic uncertainties discussed below. This estimate utilizes PYTHIA-generated events 
for \pp\ collisions at \sqrts~=~200 GeV, with instrumental effects corresponding 
to central \AuAu\ 
collisions. The detector-level spectrum is smeared to account for background 
fluctuations and is scaled by \TAAavg, likewise for central \AuAu\ 
collisions.

\begin{figure}[htbp]
\includegraphics[width=0.49\textwidth]{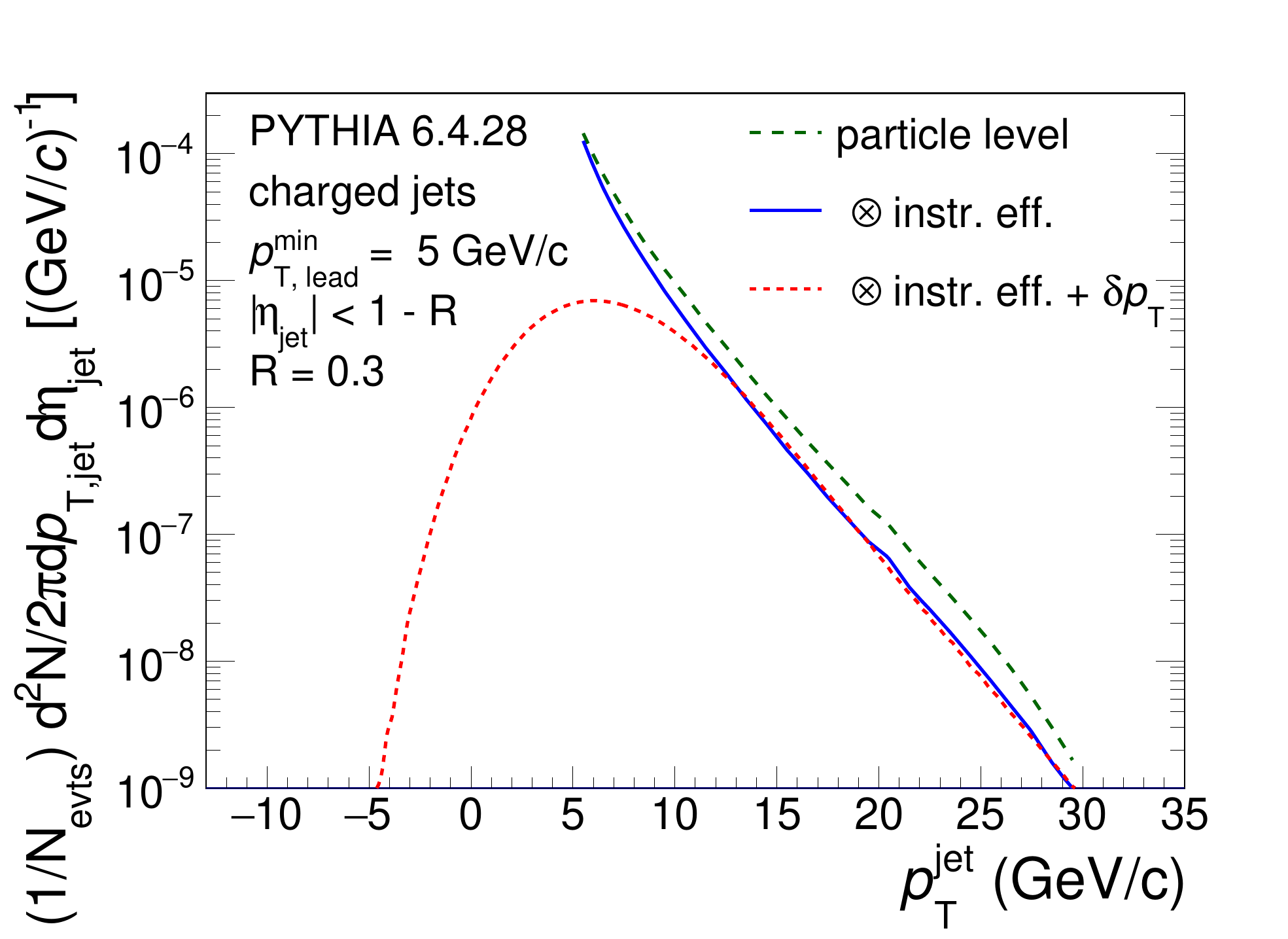}
\caption{(Color online) Estimated magnitude of corrections for charged jets with \rr~=~0.3 and 
\pTleadmin~=~5.0 \gev, for central \AuAu\ collisions. See text for details.
}
\label{fig:mag_corr}
\end{figure}

Figure \ref{fig:mag_corr} shows the results of this calculation: the 
distribution of charged jets with \rr~=~0.3 and \pTleadmin~=~5.0 \gev\ at the 
particle level (green dashed line), which is modified cumulatively by instrumental effects
(blue solid line) and 
background fluctuations ($\delta_{\pT}$, red dashed). Correction by unfolding for this case 
corresponds to transforming the red-dashed to the green-dashed distribution. 
At fixed values of \pTjet, the effect of the unfolding correction for $\pTjetch>15$ 
\gev\ is a factor $\approx$2 in yield, while the yield correction at lower \pTjetch\ is significantly larger, due predominantly to the effect of background fluctuations that transport yield to 
the region $\pTjetch<0$.

\section{Systematic Uncertainties}
\label{sect:SysUncert}

Systematic uncertainties arise from corrections 
for instrumental response and background fluctuations, and from the 
unfolding procedure. We distinguish two categories of systematic uncertainty: 
correlated 
uncertainties, which do not change the shape of the distribution, 
and shape uncertainties. 

Table~\ref{Tab:SysUncert} shows the significant contributions to the systematic 
uncertainty. For each component, the corresponding contribution to the response 
matrix is varied and 
the full correction procedure was carried out. The 
resulting variation in the corrected spectrum gives the systematic uncertainty 
due to that component. 

\subsection{Tracking}
\label{subsect:SysUncTrack}

The largest instrumental uncertainty is due to tracking efficiency (``tracking 
efficiency" Table~\ref{Tab:SysUncert}), whose 
relative uncertainty is $\pm$5\%~\cite{Adamczyk:2015lme}. 

\subsection{Fragmentation model for \Rdet}
\label{subsect:SysUncFrag}

The calculation of \Rdet\ incorporates a fragmentation 
model to determine the instrumental response to a jet. The primary analysis 
utilizes a relative population of light quarks and gluons in the ratio 2:1 at all \pTdet. 
Systematic variations utilize 100\% light 
quark or 100\% gluon fragmentation. The corresponding entry in 
Table~\ref{Tab:SysUncert} is labeled ``fragmentation for \Rdet."

\subsection{\dpT\ for \Rbkg}
\label{subsect:SysUncdpT}

The primary analysis utilizes SP jets to calculate \dpT. For 
systematic variation, \dpT\ distributions are calculated utilizing 
PYTHIA-generated fragmentation for light-quark jets. 
The requirement that accepted jets have \pTleadmin~=~5 \gev\ biases the background 
distribution, since hadrons with $\pT>5$ \gev\ may be correlated in azimuth 
with the EP [Eq.~(\ref{eq:v2weight})]. The primary analysis utilizes \vtwo~=~0.04, 
which is the maximum value compatible with the current 
measurement~\cite{Adare:2014bga}, while \vtwo~=~0 is used for systematic variation.
The corresponding entry in Table~\ref{Tab:SysUncert} is labeled ``\dpT."

\subsection{Median background density $\rho$}
\label{subsect:SysUncRho}

The calculation of $\rho$ [Eq.~(\ref{eq:rho})] is varied relative to that for 
the primary analysis by using \rr~=~0.2 or 0.4 for the first jet reconstruction 
pass, or by excluding 
only the single most energetic jet for central \AuAu\ collisions and no jets for 
peripheral \AuAu\ collisions. The corresponding entry in 
Table~\ref{Tab:SysUncert} is labeled ``$\rho$."

\subsection{Unfolding}
\label{subsect:SysUncUnfold}

Systematic variation of the unfolding procedure corresponds to 
variation of its components: algorithm,
prior distribution, and regularization criterion. The components are varied independently and 
the
unfolding procedure is carried out for each such variant. The unfolded solution 
from a variant is accepted  
if it satisfies the same quality criteria as those used in the primary analysis 
(see Sec.~\ref{sect:Unfolding}).

The algorithm is varied by using the Bayesian 
and SVD approaches. Variation of the prior distribution is discussed 
in Sec.~\ref{sect:Unfolding}. Variation of the regularization parameter 
corresponds to variation of 
the number of iterations 
$i$ for Bayesian unfolding and the number of terms $k$ in the series 
expansion for SVD unfolding: Both $i$ 
and $k$ were increased by 1 relative to their optimum values found in the 
primary analysis.

For each bin in \pTjetch,
the central value of the reported distribution is the mean of all 
accepted unfolded distributions from this variation procedure. The systematic 
uncertainty due to unfolding is the corresponding RMS, calculated separately for 
positive and negative excursions relative to the mean; the resulting uncertainty is therefore asymmetric. The corresponding entry in Table~\ref{Tab:SysUncert} is 
labeled ``unfolding."

\subsection{\TAAavg}
\label{subsect:SysUncTAA}

The uncertainties of the nuclear thickness factor \TAAavg\ are specified in 
Sec.~\ref{sect:DataSet}.

\subsection{Cumulative uncertainty}
\label{sect:CumulUncert} 

\begin{center}
\begin{table*}
\caption{Components of the systematic 
uncertainty for jets with \rr~=~0.2, 0.3 and 0.4 in central and peripheral \AuAu\ collisions.  
See text for details.}
\label{Tab:SysUncert}
\begin{tabular}{ |l|l||c|c|c|c|c|c||c|c|c|c|c|c|  }
\hline
\multicolumn{14}{|c|}{Systematic uncertainty (\%)} \\ \hline 
\multicolumn{2}{|c||}{ } & \multicolumn{6}{|c||}{central \AuAu\ collisions, \sqrtsNN~=~200 GeV } & \multicolumn{6}{|c|}{peripheral \AuAu\ collisions, \sqrtsNN~=~200 GeV } \\ \hline
 & \rr & \multicolumn{2}{|c|}{0.2} & \multicolumn{2}{|c|}{0.3} & \multicolumn{2}{|c||}{0.4} & \multicolumn{2}{|c|}{0.2} & \multicolumn{2}{|c|}{0.3} & \multicolumn{2}{|c|}{0.4}  \\ \cline{2-14}
 & \pTjetch\ [\gev] & [14,16] & [20,25] & [14,16] & [20,25] & [14,16] & [20,25] & [14,16] & [18,20] & [14,16] & [18,20] & [14,16] & [18,20] \\ \hline \hline
\multirow{5}{*}{\bf correlated} & 
tracking efficiency      & $_{+15}\atop^{-12}$ &  $_{+16}\atop^{-10}$     & $_{+16}\atop^{-13}$ &  $_{+12}\atop^{-22}$     & $_{+14}\atop^{-11}$ & $_{+18}\atop^{-12}$
			 & $_{+6}\atop^{-8}$ & $_{+10}\atop^{-12}$      & $_{+12}\atop^{-11}$ & $_{+14}\atop^{-12}$      & $_{+13}\atop^{-12}$ & $_{+16}\atop^{-12}$\\ 
 
 &fragmentation for \Rdet\ & $_{+1}\atop^{-3}$ &  $_{+3}\atop^{-1}$       & $_{+3}\atop^{-1}$   &  $_{+4}\atop^{-5}$       & $_{+4}\atop^{-1}$   & $_{+12}\atop^{-2}$
			 & $_{+0}\atop^{-5}$ & $_{+0}\atop^{-5}$      & $_{+0}\atop^{-1}$ & $_{+2}\atop^{-2}$      & $_{+2}\atop^{-1}$ & $_{+3}\atop^{-1}$\\
 
 &$\dpT$                 & $_{+8}\atop^{-3}$   &  $_{+16}\atop^{-1}$      & $_{+10}\atop^{-2}$  &  $_{+17}\atop^{-2}$      & $_{+7}\atop^{-5}$   & $_{+14}\atop^{-3}$
			 & $_{+10}\atop^{-1}$ & $_{+15}\atop^{-1}$      & $_{+9}\atop^{-1}$ & $_{+11}\atop^{-1}$      & $_{+8}\atop^{-1}$ & $_{+11}\atop^{-1}$\\ 
 
 &$\rho$                 & $_{+1}\atop^{-1}$   &  $_{+1}\atop^{-1}$       & $_{+1}\atop^{-0}$   &  $_{+0}\atop^{-1}$       & $_{+1}\atop^{-1}$   & $_{+1}\atop^{-1}$
			 & $_{+1}\atop^{-3}$ & $_{+4}\atop^{-1}$      & $_{+1}\atop^{-3}$ & $_{+2}\atop^{-4}$      & $_{+1}\atop^{-3}$ & $_{+1}\atop^{-4}$\\
\cline{2-14} 
& {\bf total correlated} & ${+17}\atop{-13}$ &  ${+24}\atop{-10}$     & ${+19}\atop{-13}$ &  ${+21}\atop{-23}$     & ${+17}\atop{-11}$ & ${+26}\atop{-13}$
			 & ${+12}\atop{-10}$ & ${+18}\atop{-14}$      & ${+15}\atop{-11}$ & ${+18}\atop{-13}$      & ${+15}\atop{-12}$ & ${+20}\atop{-13}$\\
\hline 
\hline
{\bf shape} & unfolding  & ${+17}\atop{-14}$ & ${+12}\atop{-10}$      & ${+24}\atop{-19}$ &${+25}\atop{-18}$       & ${+46}\atop{-29}$ & ${+51}\atop{-31}$
			 & ${+14}\atop{-11}$ & ${+8}\atop{-7}$        & ${+8}\atop{-6}$   & ${+17}\atop{-12}$      & ${+4}\atop{-3}$   & ${+11}\atop{-9}$\\ \hline 
\end{tabular}
\end{table*}
\end{center}
The total correlated systematic 
uncertainty in Table~\ref{Tab:SysUncert} is the quadrature 
sum of the individual component contributions for each bin in \pTjetch. 
The most significant sources of systematic uncertainty in both 
peripheral and central collisions are the unfolding procedure, tracking 
efficiency, and the choice of \dpT\ probe. Other uncertainty sources generate 
smaller contributions.

\section{Parametrized Model and Closure Test}
\label{sect:ClosureTest}

The contribution of uncorrelated background to semi-inclusive hadron+jet 
distributions in central \AuAu\ collisions at 
\sqrtsNN~=~200 GeV is well-described by a mixed-event 
population~\cite{Adamczyk:2017yhe}. This indicates that such background 
distributions are largely 
statistical in nature, with dynamically generated correlations having small or 
negligible influence. In this paper we explore a related approach to 
describe 
the uncorrelated background to the inclusive jet distribution, utilizing a  
PM calculation that accurately describes the 
eventwise distributions of mean-\pT\ (\meanpT) and mean transverse energy 
(\meanET) in high-energy nuclear 
collisions~\cite{Adcox:2002pa,Adams:2003uw,Adamczyk:2013up} (see 
also Refs.~\cite{Appelshauser:1999ft,Adamova:2003pz,Adler:2003xq,Abelev:2014ckr}). We 
apply this model in a closure test of this analysis, which assesses the 
precision with which a known signal is reproduced by the full measurement 
procedure. 

For trigger-normalized coincidence measurements, a closure test can be carried out by 
embedding 
simulated signal pairs into real events, reconstructing the hybrid 
event, and executing the full analysis chain~\cite{Adamczyk:2017yhe}. 
If the rate per real event of the process of interest is
much less than unity, identification of the embedded signal trigger 
can be made without significant ambiguity in such a procedure. In contrast,
for an inclusive jet analysis, the jet distribution is normalized per event, not 
per trigger, and such an embedding procedure effectively 
modifies the inclusive jet distribution found in real events. 
The closure test in this approach then corresponds 
to measuring this  modification. The modification is, however, not well-defined, since the 
intrinsic jet spectrum 
of real events is unknown in central \AuAu\ 
collisions; indeed, measuring it is the goal of the analysis. A different 
approach to the closure test is therefore required for inclusive jet distributions.

The inclusive jet measurement closure test therefore requires the 
analysis of fully simulated events, whose global properties mimic those of 
\AuAu\ 
collisions and whose inclusive jet distribution is known by construction.
One approach for the closure test is to generate events using 
established Monte Carlo event generators such as HIJING~\cite{Wang:1991hta} or 
PYQUEN~\cite{Lokhtin:2011qq}, which reproduce the global features of heavy-ion 
collisions at RHIC and the LHC. However, the statistical precision of a 
meaningful closure test must be similar to that of the real data analysis, which is 
difficult to achieve with such MC calculations. We therefore utilize events 
generated by the PM, which is computationally more efficient than MC generators, 
and which likewise reproduces the global properties of 
\AuAu\ collisions at \sqrtsNN~=~200 GeV and has a specified inclusive 
jet distribution. Comparison of the PM calculation with data has the additional 
benefit of providing insight into the nature of the backgrounds in this 
measurement. 

The following considerations motivate a statistical approach to modeling the
background in this
analysis. Eventwise distributions of 
\meanpT\ and \meanET\ in limited acceptance have been measured 
in high-energy nuclear 
collisions~\cite{Appelshauser:1999ft,Adcox:2002pa,Adamova:2003pz,Adler:2003xq,Adams:2003uw,Adamczyk:2013up,Abelev:2014ckr}. 
These distributions are well-described by mixed-event 
analyses~\cite{Appelshauser:1999ft,Adcox:2002pa,Adamova:2003pz,Adler:2003xq,Adamczyk:2013up}, 
and by calculations based on uncorrelated particle 
emission~\cite{Adcox:2002pa,Adams:2003uw,Adamczyk:2013up,Tannenbaum:2001gs}. 
The uncorrelated background in semi-inclusive hadron+jet distributions at 
\sqrtsNN~=~200 GeV is likewise well-reproduced by a mixed-event 
approach~\cite{Adamczyk:2017yhe}, showing that the background distribution in 
heavy-ion jet measurements is predominantly  statistical, with
dynamically generated correlations on the scale of the resolution parameter \rr, due to jets and other QCD 
mechanisms, playing a smaller, even negligible, role. 

In the PM, hadrons are generated from two sources~\cite{deBarros:2012ws}: 
a soft physics 
component based on uncorrelated particle emission, and the production and
fragmentation of hard jets based
on a PYTHIA calculation for \pp\ collisions. All generated ``hadrons'' are 
identical, with zero mass and charge.

The soft hadronic component comprises $M$ independent 
particles distributed uniformly in azimuth ($0<\varphi<2\pi$) and 
pseudo-rapidity ($|\eta|<1$), and distributed in \pT\ according to a Boltzmann 
function,

\begin{equation}
\frac{dN^{AA}_\mathrm{soft}}{d\pT}\propto\frac{4\pT}{{\meanpT}^2}{e}^{-2\pT/\meanpT},
\label{eq:Boltzmann}
\end{equation}

\noindent
where the parameters \meanpT\  and $M$ are constants. This
approach provides an accurate description of the eventwise distribution of 
transverse energy \ET\ in high-energy nuclear collisions
~\cite{Adcox:2002pa,Adams:2003uw,Adamczyk:2013up}. 

The hard jet yield per \AuAu\ collision is

\begin{equation}
\dNjetdpT=\frac{d\sigma^\mathrm{jet}_\mathrm{pp}}{d\pTjet}\TAAavg\RAA C(\pTjet),
\label{eq:PMjet}
\end{equation}

\noindent
where $\frac{d\sigma^\mathrm{jet}_\mathrm{pp}}{d\pTjet}$ is the inclusive 
charged-particle jet cross 
section within $|\etajet|<1$ for \pp\ collisions at \sqrts~=~200 GeV, calculated by 
PYTHIA; \TAAavg\ has value 22.8 mb$^{-1}$ for central \AuAu\ 
collisions; \RAA\ is the jet yield suppression due to quenching, with value chosen such that the hard tail of the reconstructed 
jet distributions matches the data at high-\pTjet; and $C(\pTjet)$ is a function that cuts 
the \dNjetdpT\ 
distribution off smoothly for $\pTjetch\lesssim4$ \gev, in order not to double-count 
soft particle production. 

\begin{center}
\begin{table}
\caption{Model parameters for central \AuAu\ collisions. Figure~\ref{fig:PMvsSTAR} shows the comparison of PM distributions using these parameters to measured STAR data.}
\label{Tab:PMparams}
\begin{tabular}{|c|c|}
\hline
\multicolumn{2}{|c|}{PM parameters, \AuAu\ collisions, \sqrtsNN~=~200 GeV } \\ \hline \hline
\meanpT & 0.6 \gev \\ \hline
$M$ & 600 \\ \hline
\RAA, \rr~=~0.2 & 0.2 \\ \hline
\RAA, \rr~=~0.4 & 0.2-0.5 \\ \hline
\end{tabular}
\end{table}
\end{center}

Table~\ref{Tab:PMparams} shows the PM parameters used to model central \AuAu\ 
collisions at \sqrtsNN~=~200 GeV.  The values for \meanpT\ and $M$ are similar to those observed in STAR data~\cite{Adams:2005dq,Abelev:2008ab}.
\RAA\ is constant for \rr~=~0.2 and a linear 
function of \pTreco\ for \rr~=~0.4, to provide model variation that spans inclusive hadron measurements at RHIC and the LHC and jet measurements at the LHC (see Fig.~\ref{fig:RCP_ALICESTAR}).
 For these 
parameters, the integral of Eq.~(\ref{eq:PMjet}) for $\pTjet>4$ \gev\ is 0.126, 
which is the 
average rate of such hard jets per central \AuAu\ collision. For PM event 
generation, 
the number of hard jets in each event is Poisson-distributed 
about this average, with \pTjet\ distributed 
according to Eq.~(\ref{eq:PMjet}), and with uniform distribution
over the full azimuth and $|\eta|<1$. 
PYTHIA fragmentation is then run 
for either a light quark or a gluon jet, chosen in ratio 
2:1, with transverse momentum equal to \pTjet. The charged particles generated 
by this procedure are the ``hadrons'' of the PM, comprising the hard jet 
component 
of PM events.

\begin{figure*}[htbp]
\includegraphics[width=0.95\textwidth]{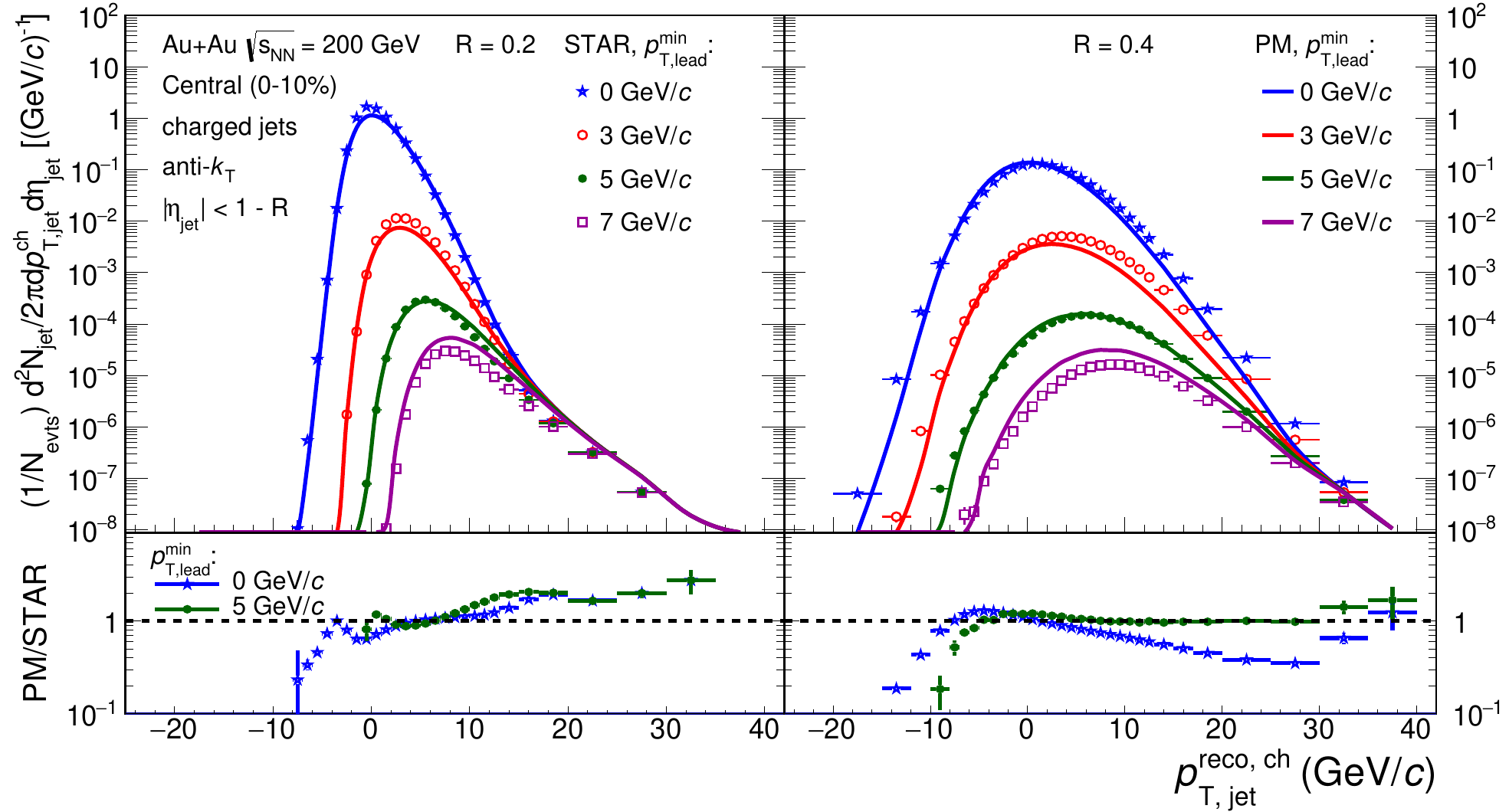}
\caption{(Color online) (Quasi-)inclusive jet \pTreco\ distributions for 
various values of \pTleadmin for \rr~=~0.2 (left) and \rr~=~0.4 (right), for PM-generated events and for the STAR measurements of central \AuAu\ collisions (data from Fig.~\ref{fig:pTreco_pTlead}). Lower panels show the ratio of the PM
and data distributions, for $\pTleadmin~=~0$ and $\pTleadmin~=~5$ \gev. 
}
\label{fig:PMvsSTAR}
\end{figure*}

Figure~\ref{fig:PMvsSTAR} shows (quasi-)inclusive jet \pTreco\ distributions  for 
various values of \pTleadmin, for PM-generated events and for the STAR measurements in central \AuAu\ collisions shown in 
Fig.~\ref{fig:pTreco_pTlead}. 
The good level of agreement of the PM-generated distributions with data is 
notable, in light of the very simple nature of the model. For \pTleadmin~=~5 
\gev, 
the PM-generated distributions agree with data within 
10\%, except in 
the extreme tails, over three 
orders of magnitude variation in yield. 
For \pTleadmin~=~0, the level of agreement 
is poorer, though the yields in this case vary by six orders of 
magnitude over the range of comparison. While the agreement of the model with data could be improved 
further by introducing additional parameters, the focus of this 
analysis is on \pTleadmin~=~5 \gev, where the agreement is already good, and we 
therefore choose not to do so.

Figure~\ref{fig:PMvsSTAR} shows that the background
distribution in this analysis is driven predominantly by gross features of 
the collisions and measurement --- acceptance, track multiplicity $M$, and \meanpT\ --- 
with dynamical correlations due to both soft and hard QCD processes
playing a secondary or even negligible role. This picture, in which the 
background distribution is determined largely by statistical phase space, is 
consistent with that derived from the mixed event 
background analysis in Ref.~\cite{Adamczyk:2017yhe}.

\begin{figure}[htbp]
\includegraphics[width=0.49\textwidth]{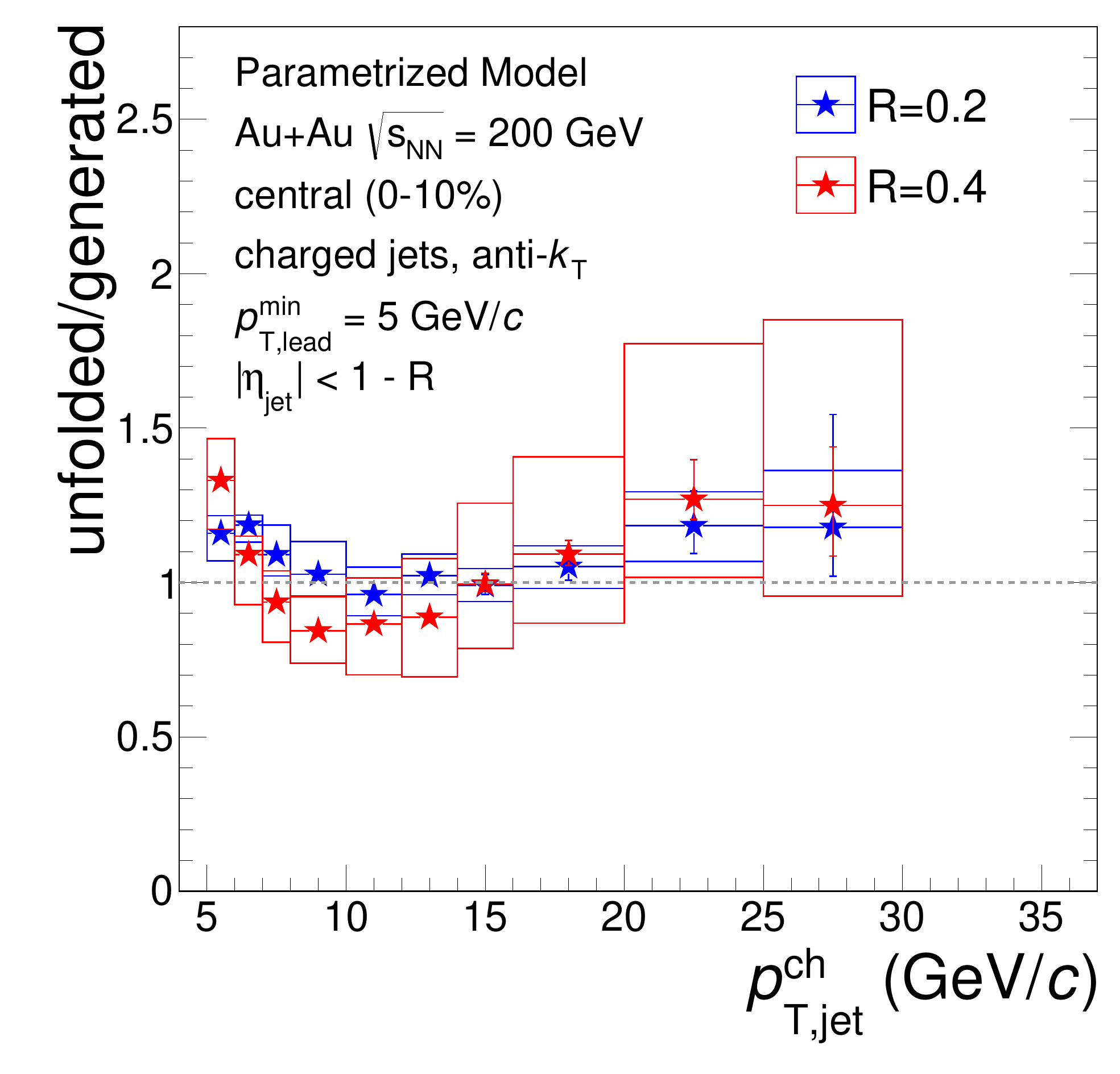}
\caption{(Color online) Closure test for PM-generated events corresponding to central \AuAu\ 
collisions at \sqrtsNN~=~200 GeV. }
\label{fig:Closure}
\end{figure}

We turn now to the closure test, to assess the validity of the 
correction procedure described above. Closure of this correction procedure for 
instrumental effects was shown 
in Ref.~\cite{Adamczyk:2017yhe}. The focus of this closure test is therefore the 
large smearing of the jet spectrum due to fluctuations of uncorrelated 
background, 
which are well represented by the PM generator 
(Fig.~\ref{fig:PMvsSTAR}).

The closure test utilizes 20M PM-generated events modeling central 
\AuAu\ collisions, which has similar statistical precision to the real 
dataset. The cut \pTleadmin~=~5 \gev\ is imposed on all jet candidates.
The full analysis to generate the \pTjetch\ distribution and to 
correct background fluctuations 
was then run, including generation of \dpT\ distributions, 
unfolding, and the determination of systematic uncertainties. 

Figure~\ref{fig:Closure} shows the ratio of the 
corrected distributions from this procedure to the 
reconstructed hard jet distribution without background or detector effects (``Truth''),
for \rr~=~0.2 and \rr~=~0.4.   
The ratio in the range $\pTjetch>15$ \gev\ is consistent with unity within 
uncertainties for both values of \rr. The ratio 
is, however, significantly above unity in the first 
bin at threshold, \pTleadmin~=~5 \gev. This feature is expected, since by 
construction the generated distribution has magnitude zero for 
$\pTjet<\pTleadmin$ and its magnitude is small and changing rapidly for \pTjet\ just above 
\pTleadmin, while the output of a regularized unfolding procedure cannot vary 
arbitrarily rapidly. In 
Sec.~\ref{sect:results}, the first bin at \pTjet~=~5 \gev\ in the corrected 
distributions is therefore not shown. For larger values of \pTjet,
Fig.~\ref{fig:Closure} validates the correction procedure for background 
fluctuations in this analysis.

\section{Reference spectrum from \pp\ collisions}
\label{sect:ppRef}

\begin{figure}[htbp]
\includegraphics[width=0.49\textwidth]{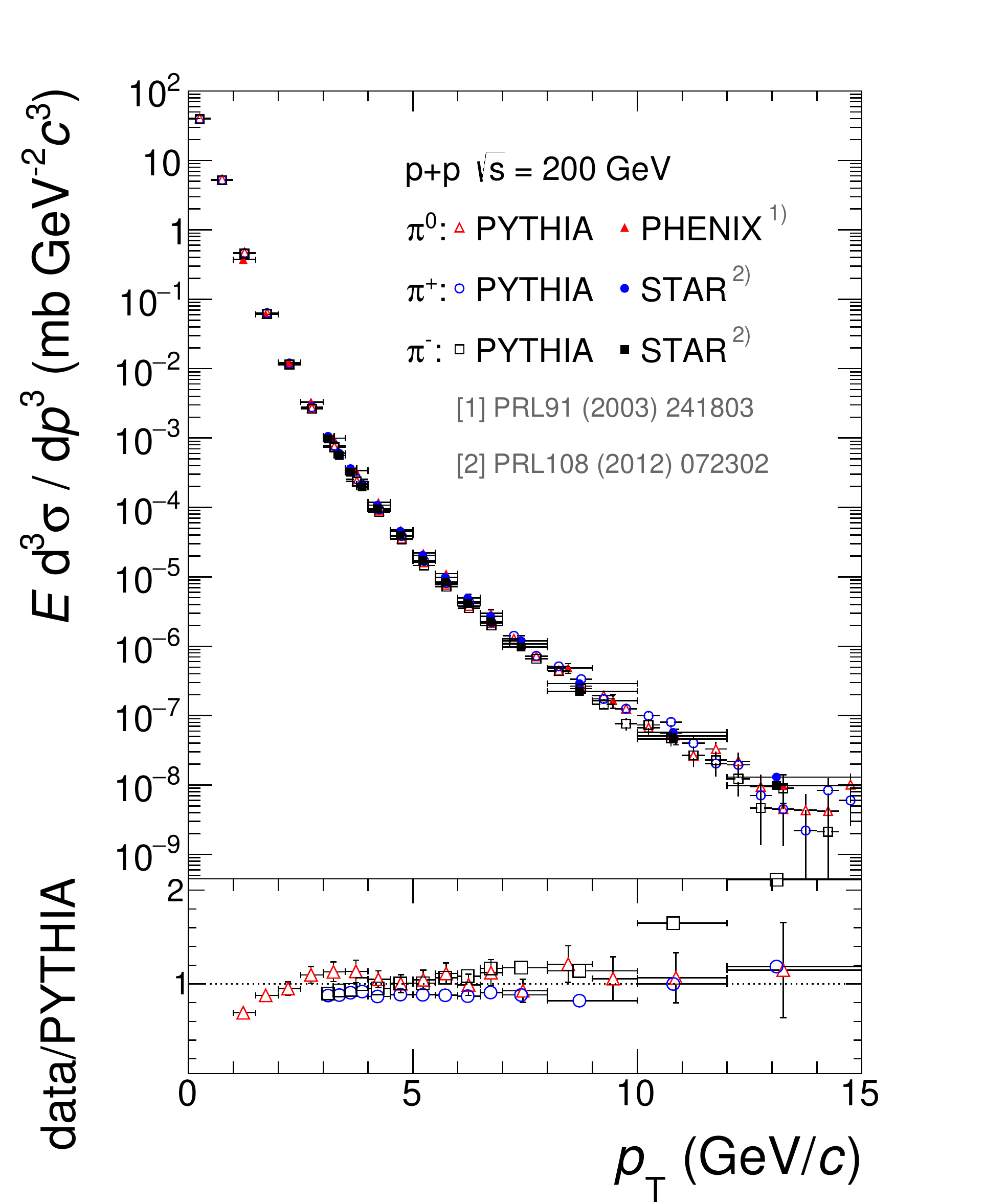}  
\caption{(Color online) Upper panel: Inclusive pion cross section in \pp\ collisions at \sqrts~=~200 GeV from measurements~\cite{Agakishiev:2011dc,Adler:2003pb,Abelev:2006uq} and from a 
PYTHIA simulation (see text for details). Lower panel: ratio of 
data and PYTHIA.}
\label{fig:ppBase}
\end{figure}

The modification  of inclusive jet production due to quenching is 
quantified by 
comparing measurements in central \AuAu\ collisions to those in smaller 
systems, specifically peripheral \AuAu\ and \pp\ collisions. 
\RAA\ is the ratio of inclusive yields in 
$A$+$A$ and \pp\ 
collisions, with the latter scaled by \TAAavg\ to account for the effects of nuclear geometry. \RCP\ is a similar ratio, in which the yield for peripheral \AuAu\ collisions is used as normalization.

The inclusive charged-particle jet spectrum in \pp\ collisions at \sqrts~=~200 GeV is 
not currently available with statistical precision comparable to the inclusive 
charged-particle jet spectrum in central \AuAu\ collisions reported here. We therefore 
simulate the 
charged-particle jet distribution for \pp\ collisions at \sqrts~=~200 GeV using PYTHIA 
Monte Carlo generator version 6.428~\cite{Sjostrand:2006za}, with the Perugia 2012 tune (370) and 
CTEQ6L1 LO parton distribution functions~\cite{Skands:2010ak}. 
However, a calculation of charged-pion yields using this PYTHIA tune 
overestimates the measured pion distribution by up to 
30\%. 
It was found that changing the PYTHIA parameter that 
controls the energy dependence of the low momentum cut-off for
underlying event generation [PARP(90)] from its default value of 0.24 to 0.213 
improves the agreement of the calculated inclusive pion yields with data, for 
both charged and neutral 
pions~\cite{Adam:2019aml}.
 
Figure~\ref{fig:ppBase} shows the comparison of PYTHIA-generated distributions 
using this tune with modified PARP(90) to inclusive pion 
measurements~\cite{Agakishiev:2011dc,Adler:2003pb}; agreement of model and data 
is seen to be within 10\% for $\pT>3$ \gev. This configuration of PYTHIA is also in good 
agreement with measurements of inclusive jet yields, hadron distributions within jets, 
electromagnetic jet energy fraction, and dijet properties measured in \pp\ collisions 
at \sqrts~=~510 GeV~\cite{Adam:2019aml}, and the underlying event measured in \pp\ collisions at \sqrts~=~200 
GeV~\cite{Adam:2019xpp}. These comparisons validate 
this PYTHIA-based calculation with modified tune for calculating inclusive jet \RAA\ in the \AuAu\ analysis 
presented here.

The systematic uncertainty of the inclusive jet cross section generated by 
PYTHIA was estimated using several alternative PYTHIA 
tunes~\cite{Sjostrand:2006za}: 
tune pairs 371 and 372 with $\alpha_s$($\frac{1}{2}p_{\perp}$) and 
$\alpha_s$(2$p_{\perp}$) to vary the magnitude of initial- and final-state 
radiation;
tune 374 with reduced color re-connection; tunes 376 and 377 with modified 
longitudinal and transverse fragmentation; and tune 383 with Innsbruck 
hadronization parameters. The tune pair 371 and 372, which bracket the 
distribution generated by the default PYTHIA tune and those of the other tunes, are used 
as the systematic uncertainty of the reference jet \pp\ spectrum, corresponding 
to 22\% for $R$~=~0.2; 20\% for $R$~=~0.3; and 18\% for $R$~=~0.4, with negligible 
dependence on \pTjet.

\section{Theoretical calculations of jet quenching}
\label{sect:Theory}

We compare our results to several theoretical calculations incorporating jet quenching, which are labeled as follows:

\begin{itemize}

\item {\bf NLO}~\cite{Vitev:2009rd}: a next-to-leading-order (NLO) 
pQCD calculation that accounts for initial-state nuclear 
modification (EMC 
effect, initial-state energy loss)~\cite{Vitev:2008vk,Sharma:2009hn}, with 
collisional partonic energy loss in the QGP calculated using a weak-coupling 
approach. This calculation provides a good description of the inclusive jet cross 
section for \rr~=~0.4 in \pp\ collisions at \sqrts~=~200 GeV~\cite{Abelev:2006uq}
and predicts that inclusive jet \RAA\ for \rr~=~0.2 in \AuAu\ collisions at 
\sqrtsNN~=~200 GeV should be similar to \RAA\ for neutral pions~\cite{Adare:2012wg}.

\item {\bf SCET}~\cite{Chien:2015hda,Chien:2015vja}: soft-collinear effective 
theory extended to describe jet propagation in matter 
(SCET$_G$)~\cite{Idilbi:2008vm,DEramo:2010wup,Ovanesyan:2011xy}; initial-state 
effects include dynamical nuclear shadowing, Cronin effect, and initial-state 
partonic energy loss. This approach describes well the measurement of 
charged-hadron \RAA\ in \PbPb\ collisions at \sqrtsNN~=~2.76~TeV~\cite{Abelev:2012hxa,Abelev:2014ypa,Aad:2015wga,CMS:2012aa}, though a 
similar level of agreement can be achieved with different parameter choices for initial state 
energy loss and Cronin effect, which are anticorrelated with \RAA\ in the model. 
From the two SCET implementations available we use the one with slightly larger Cronin effect and smaller energy loss (SCET1). The error band for this model reflects two values of coupling constant $g$ between the jet and the medium; the lower edge of the band corresponds to $g$~=~2.2, while the upper edge corresponds to $g$~=~2.0.

\item {\bf Hybrid model}~\cite{Casalderrey-Solana:2016jvj}: 
combines several processes governing the evolution and interaction of jet 
showers in the medium. The production and evolution of the jet shower uses a weakly coupled approach based on PYTHIA, while the interaction of shower partons with 
the QGP uses a strongly coupled holographic approach based on $N$~=~4 supersymmetric 
Yang-Mills theory. The model includes \pT-broadening of the shower in the QGP and 
back-reaction of the medium due to passage of the jet. 
The value of $\kappa_{\mathrm{sc}}$, the free parameter in the model, was fixed by using 
LHC hadron and jet data as described in Ref.~\cite{Casalderrey-Solana:2018wrw}.  We note 
that calculations based on this global fit to LHC data disagree with measurements 
of high-\pT\ hadron suppression at RHIC at the 3$\sigma$ level, suggesting 
stronger jet-medium interaction at RHIC.

\item~{\bf LBT model}~\cite{He:2015pra,He:2018xjv}: The Linear Boltzmann Transport model utilizes pQCD for elastic and inelastic scattering between jet shower and thermal medium partons. Dynamic evolution of the QGP is calculated using the 3+1D ``CLVisc'' hydrodynamic model~\cite{Pang:2014ipa}, with initial conditions fluctuating event by event. The recoil of thermal partons is accounted for, enabling the calculation of medium response. LBT model calculations agree well with measurements of inclusive jet yield suppression in \PbPb\ collisions at the LHC~\cite{He:2018xjv}.

\item~{\bf LIDO model}~\cite{Ke:2018jem,Ke:2020nsm}: The LIDO model is based on a modified formulation of semiclassical Boltzmann transport using pQCD cross sections with running coupling and an approximate treatment of in-medium multiple-scattering coherence (Landau-Pomeranchuk-Migdal or LPM effect). Medium excitation is accounted for using a linearized approximation to the hydrodynamic equations. LIDO model calculations reproduce inclusive jet and hadron suppression measurements  in \PbPb\ collisions at the LHC. The LIDO calculations presented here~\cite{Ke:2020nsm} are shown as a band, corresponding to variation of the temperature-dependent coupling constant scale parameter between $1.5\pi T$ and $2\pi T$. 

\end{itemize}

\section{Results}
\label{sect:results}

\begin{figure*}[htbp]
\includegraphics[width=0.45\textwidth]{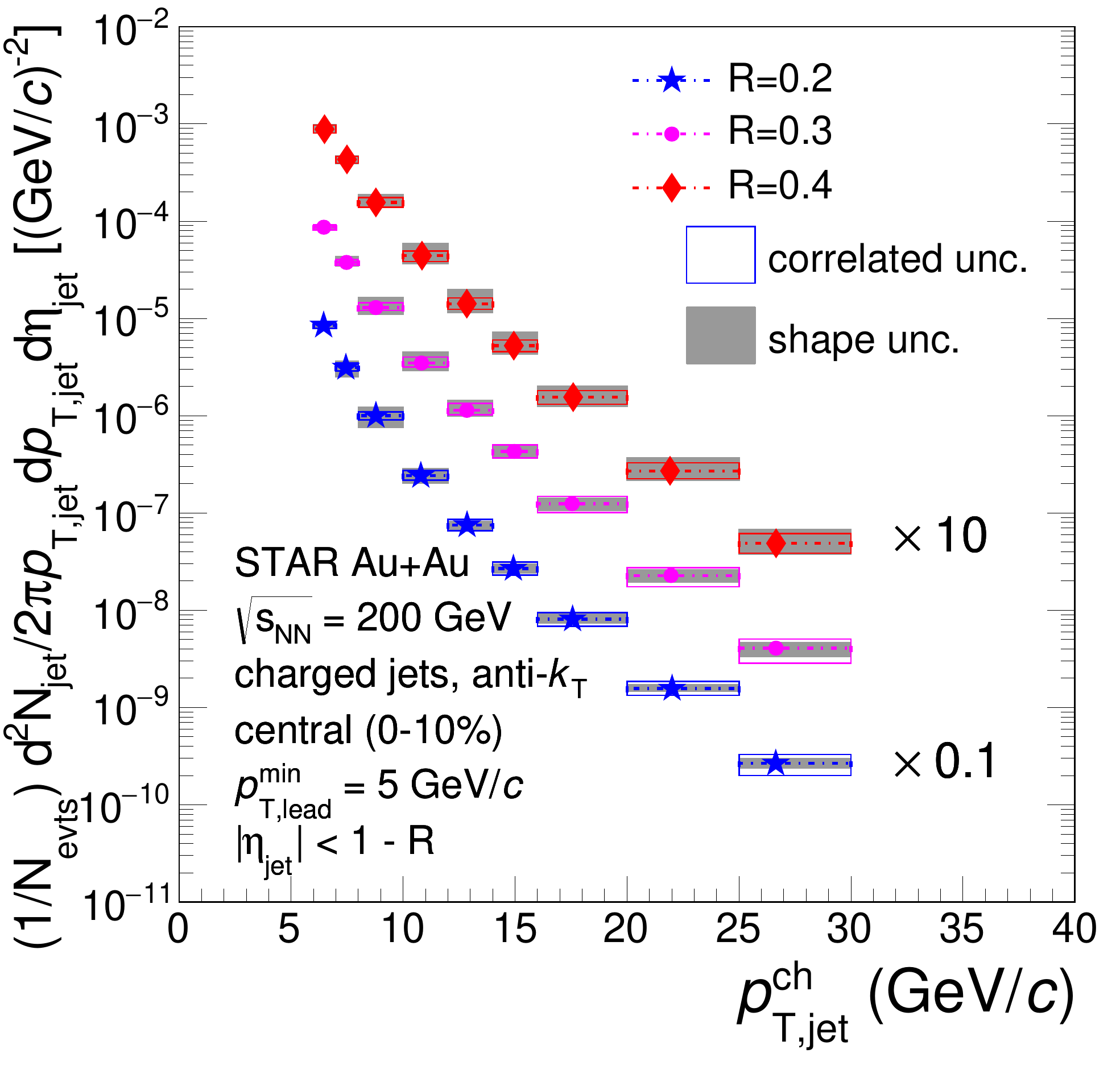}
\includegraphics[width=0.45\textwidth]{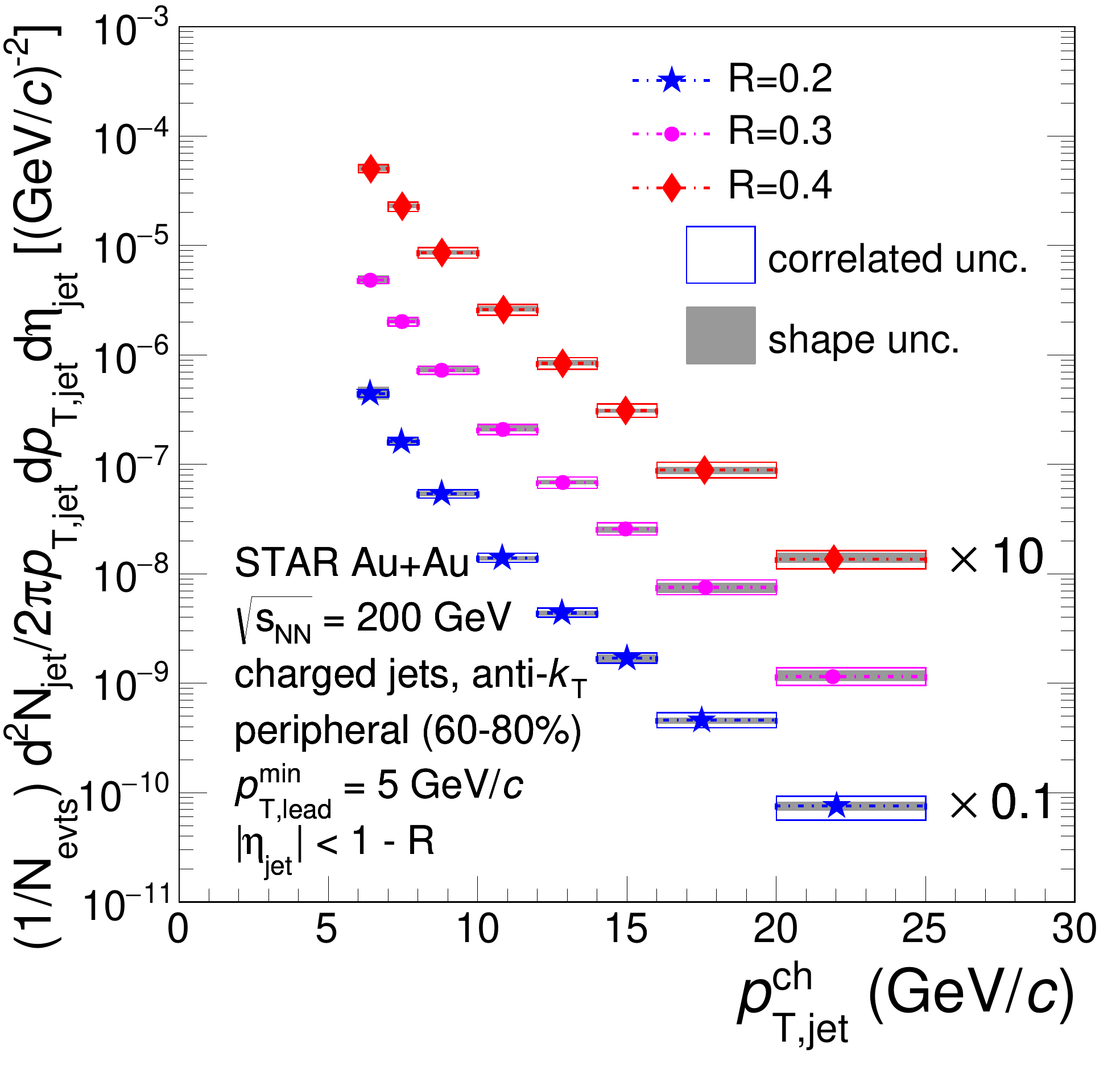}
\includegraphics[width=0.45\textwidth]{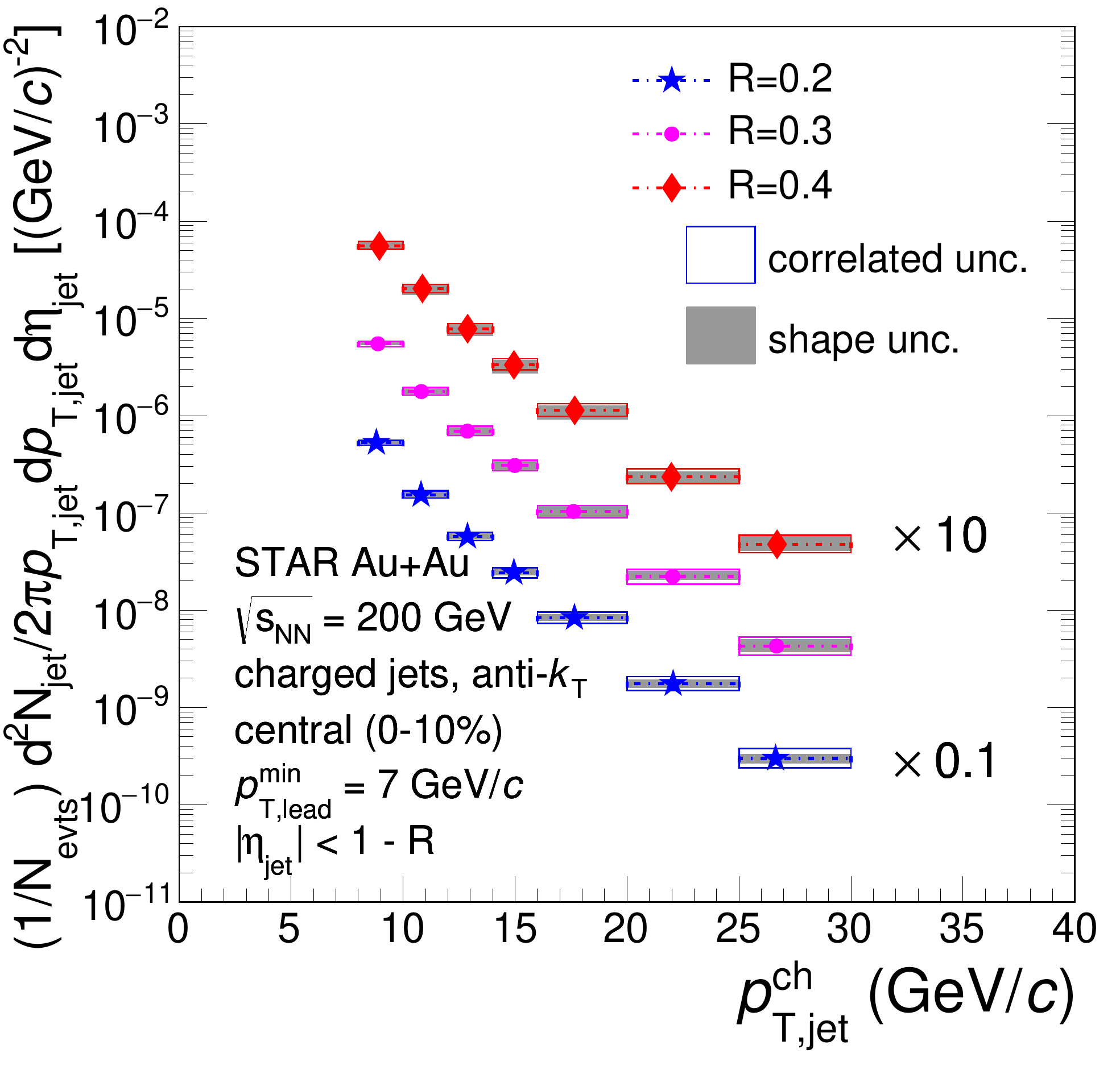}
\includegraphics[width=0.45\textwidth]{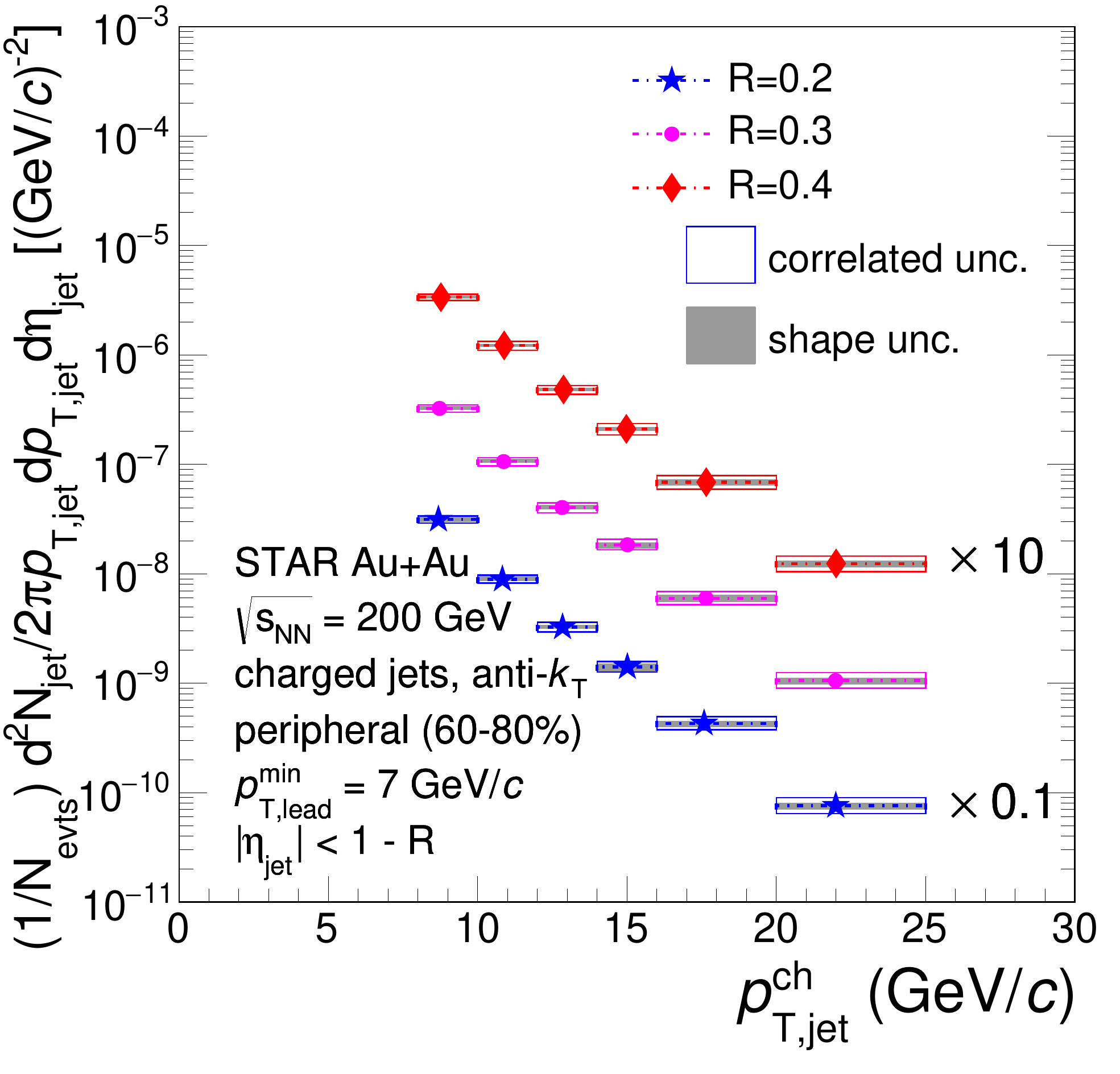}
\caption{(Color online) Corrected quasi-inclusive charged-particle jet distributions in \AuAu\ collisions at \sqrtsNN=200 
GeV for \rr~=~0.2, 0.3, and 0.4. Left: Peripheral; right: 
central \AuAu\ collisions. Upper: \pTleadmin~=~5 \gev; lower: \pTleadmin~=~7 
\gev. Correlated and shape systematic uncertainties are shown separately. The 
value of \pTjetch\ is shifted horizontally within each bin to account for the spectrum shape.}
\label{fig:CorrSpec}
\end{figure*}

Figure~\ref{fig:CorrSpec} shows fully corrected quasi-inclusive charged jet 
distributions in central and peripheral \AuAu\ collisions at 
\sqrtsNN~=~200 GeV, for \rr~=~0.2, 0.3 and 0.4, and for 
\pTleadmin~=~5 and 7 \gev. The entire dataset is used for each distribution, 
which are therefore not statistically independent. 

\begin{figure}[htbp]
\includegraphics[width=0.49\textwidth]{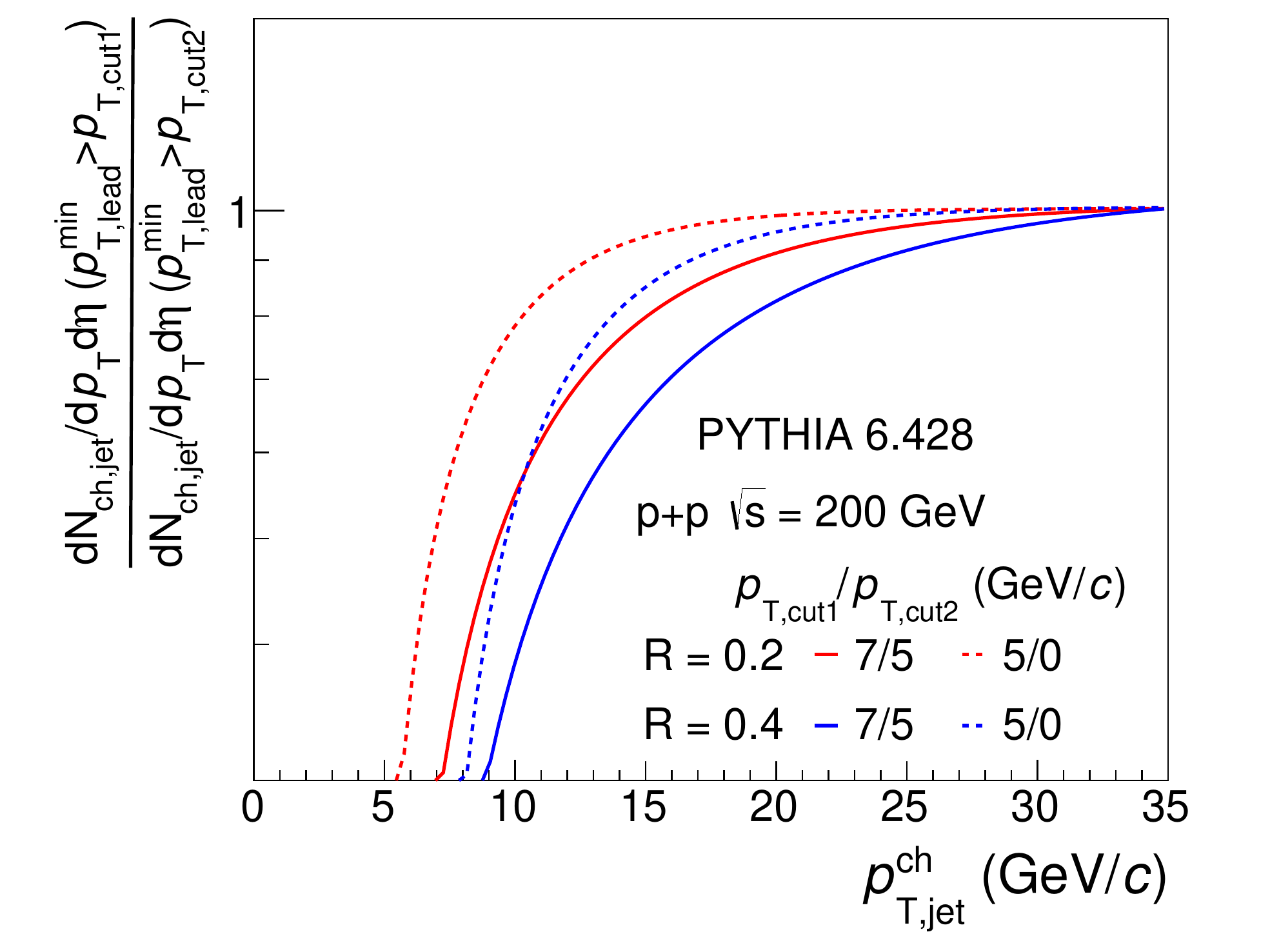}
\caption{(Color online) Ratio of quasi-inclusive charged-particle jet cross sections simulated by PYTHIA for \pp\ collisions at \sqrts~=~200 GeV, $\etajet~=~0$, for \rr~=~0.2 (red) and 0.4 (blue), for \pTleadmin~=~5 \gev\ relative to the unbiased distribution (``5/0,'' dashed) and \pTleadmin~=~7 \gev\ relative to \pTleadmin~=~5 \gev\ (``7/5,'' solid). 
}
\label{fig:ppTrigBias}
\end{figure}

The requirement $\pTlead>\pTleadmin$ imposes 
a bias on the reported jet population. This bias must be quantified in order to 
compare these data to other jet measurements and to theoretical calculations. 
The magnitude of the bias is expected to increase monotonically with 
increasing value of \pTleadmin, and we utilize that expectation to determine the range in \pTjetch\ in which 
the corrected distributions do not 
depend significantly on the value chosen for \pTleadmin. 

We first explore the effect of the bias in \pp\ collisions at \sqrts~=~200 GeV, 
using PYTHIA simulations. Figure~\ref{fig:ppTrigBias} shows the ratios of quasi-inclusive 
charged jet cross sections with \rr~=~0.2 and 0.4 from this simulation for 
\pTleadmin~=~5 \gev\ relative to the unbiased distribution (labeled ``5/0''), and 
\pTleadmin~=~7 \gev\ relative to \pTleadmin~=~5 \gev\ (labeled ``7/5''). The ratio 
rises more rapidly above threshold for \rr~=~0.2 than for \rr~=~0.4, and more 
rapidly for 5/0 than 7/5. The bias due to \pTleadmin~=~5 \gev\ is less 
than 10\% (i.e., the ratio 5/0 is larger than 0.9) for $\pTjetch>13$ 
\gev\ for \rr~=~0.2 and $\pTjetch>17$ \gev\ for \rr~=~0.4. The relative bias due to 
\pTleadmin~=~7 \gev\ relative to \pTleadmin~=~5 \gev\ is less than 10\% (i.e., the 
ratio 7/5 is larger than 0.9) for $\pTjetch>19$ \gev\ for \rr~=~0.2 
and $\pTjetch>24$ \gev\ for \rr~=~0.4. It is evident that measurement of the 
7/5 ratio provides a conservative estimate of the range over which the 
bias due to choosing the value \pTleadmin~=~5 \gev\ is small.

\begin{figure*}[htbp]
\includegraphics[width=0.99\textwidth]{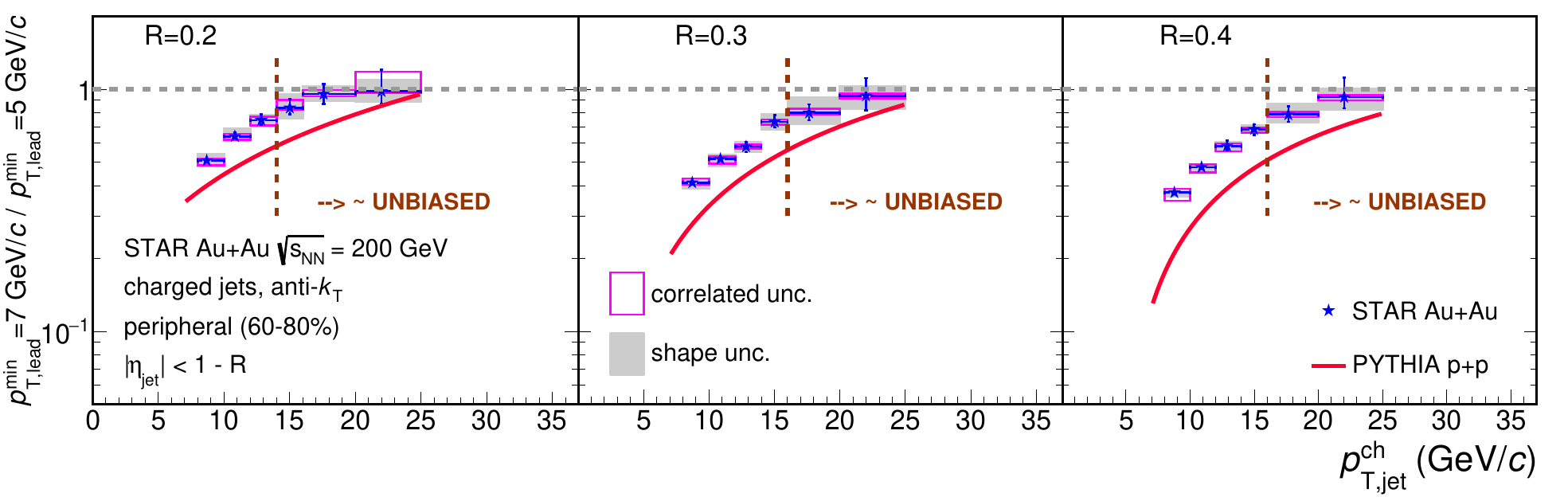}
\includegraphics[width=0.99\textwidth]{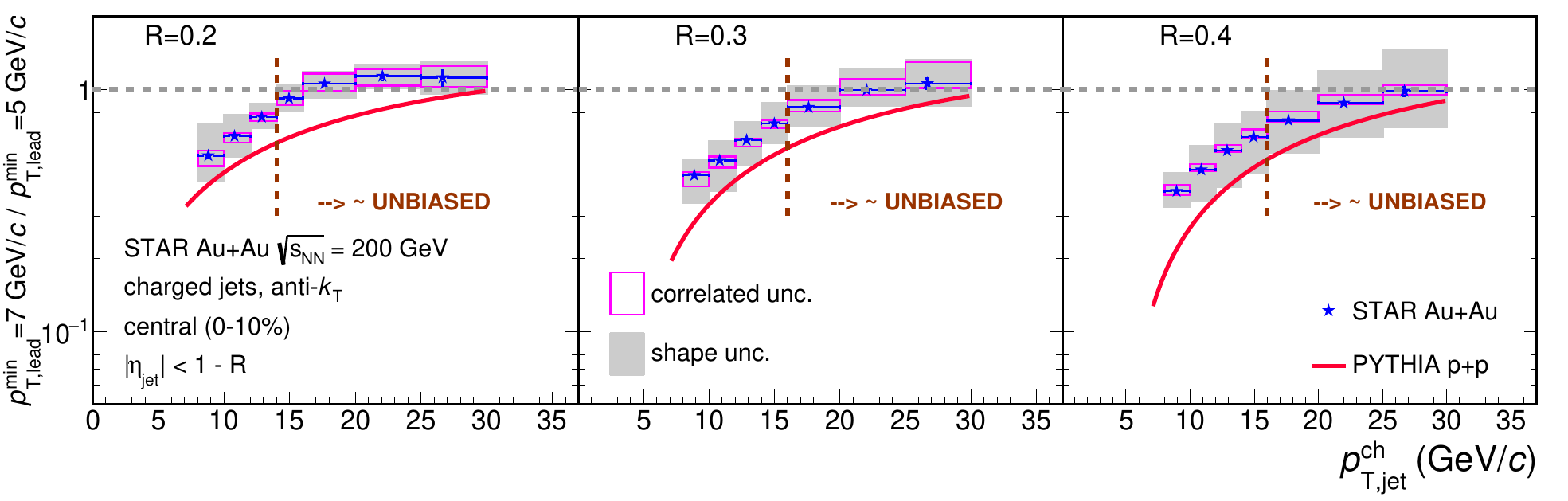}
\caption{(Color online) Ratio of distributions from Fig.~\ref{fig:CorrSpec} for \pTleadmin~=~7 
\gev\ and 5 \gev, for \rr~=~0.2, 0.3, and 0.4 in peripheral (upper) and central (lower) \AuAu\ collisions. The red lines show the corresponding ratios from a PYTHIA simulation of \pp\ collisions (Fig.~\ref{fig:ppTrigBias}).
}
\label{fig:pTleadRat}
\end{figure*}

Figure~\ref{fig:pTleadRat} shows the ratios of distributions from
Fig.~\ref{fig:CorrSpec}
for \pTleadmin~=~7 \gev\ and
\pTleadmin~=~5 \gev\, for 
\rr~=~0.2, 0.3 and 0.4 in peripheral and central \AuAu\ collisions. The 
systematic uncertainty of the ratio accounts for the 
correlated systematic uncertainties of numerator and denominator. For uncorrected 
distributions such a ratio must have value unity or below since the numerator is drawn 
from a subset of the data used in the denominator; however, the figure shows the 
ratio of corrected distributions, and such a constraint has not been imposed. 

Figure~\ref{fig:pTleadRat} also shows the corresponding 7/5 ratios for \pp\ 
collisions simulated by PYTHIA (Fig.~\ref{fig:ppTrigBias}). The ratios for \pp\ 
collisions rise more slowly as a function of \pTjetch\ than those for peripheral 
\AuAu\ collisions and central \AuAu\ collisions, indicating differences in the 
distribution of high-\pT\ jet fragments.

As discussed above for \pp\ collisions, the 7/5 ratio provides a conservative estimate of the region in which the bias due to the choice of value \pTleadmin~=~5 \gev\ is small.
The ratios in Fig.~\ref{fig:pTleadRat} are consistent with or larger than 0.9
in the range $\pTjetch>15$ \gev\ for jets with \rr~=~0.2 and $\pTjetch>17$ \gev\ 
for jets with \rr~=~0.3 and 0.4. In the following figures we indicate these ranges 
by the label ``$\sim$unbiased." 

Jet quenching may induce energy transport to angles larger than \rr\ with respect to the jet axis, effectively suppressing the jet yield at a given value of \pTjetch. In the next sections we discuss measurements of jet yield modification in central \AuAu\ collisions, using both the \RCP\ and \RAA\ observables.

\subsection{Yield suppression: \RCP}
\label{sect:RCP}

\begin{figure*}[htbp]
\includegraphics[width=0.99\textwidth]{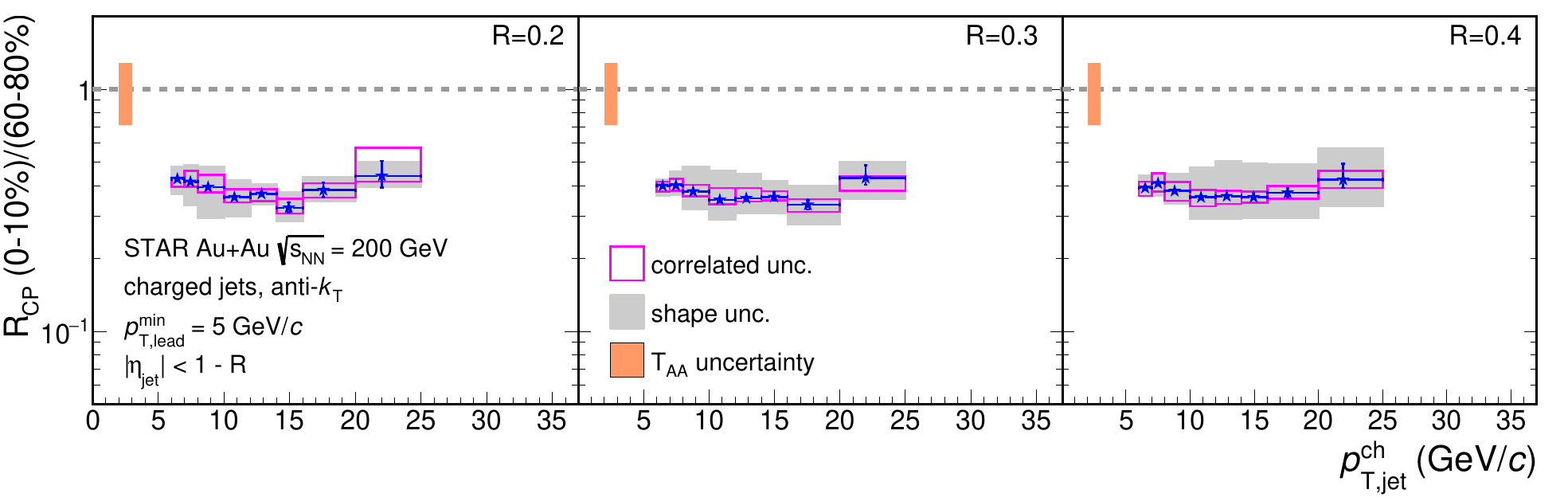}
\caption{(Color online) \RCP\ [(0--10\%)/(60--80\%)] for \AuAu\ collisions at \sqrtsNN~=~200 GeV, for \rr~=~0.2, 0.3 and 0.4. See text for details. 
}
\label{fig:RCP}
\end{figure*}

Figure~\ref{fig:RCP} shows the distribution of \RCP\ from this measurement, for 
\rr~=~0.2, 0.3, and 0.4. Given the close similarity of the 7/5 ratio for central and peripheral \AuAu\ collisions shown 
in Fig.~\ref{fig:pTleadRat}, we show \RCP\ over the full measured range of \pTjetch, without specification of an ``Unbiased" region.
The systematic uncertainty of \RCP\ takes into account the 
correlated uncertainties of numerator and denominator. The uncertainty in the 
ratio due to \TAAavg\ is independent of \pTjetch\ and is dominated by the uncertainty in \TAAavg\
for peripheral collisions.
We observe that $\RCP\approx0.4$ for all \rr, with at most a weak dependence on 
\pTjetch.

\begin{figure*}[htbp]
\includegraphics[width=0.9\textwidth]{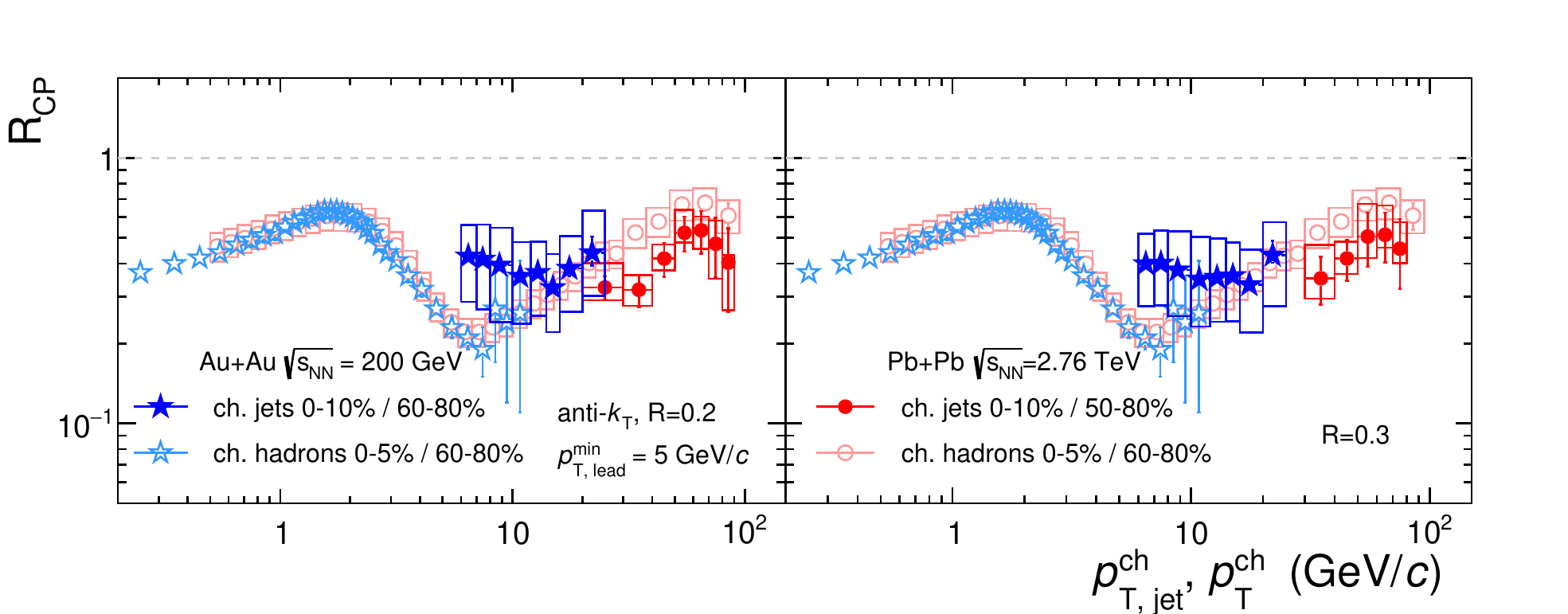}
\caption{(Color online) \RCP\ distributions from Fig.~\ref{fig:RCP} compared to that measured 
in \PbPb\ collisions at \sqrtsNN~=~2.76 
TeV~\cite{Abelev:2013kqa}, for \rr~=~0.2 (left) and \rr~=~0.3 (right). Also shown 
are \RCP\ for inclusive charged hadrons in \AuAu\ collisions at \sqrtsNN~=~200 
GeV~\cite{Adams:2003kv} and in \PbPb\ collisions at \sqrtsNN~=~2.76 
TeV~\cite{Aad:2015wga}. Data from RHIC are in blue; data from the LHC are in red. The charged hadrons \RCP\ distributions are the same in 
the two panels.
The different choices of centrality class are discussed in the text.  
}
\label{fig:RCP_ALICESTAR}
\end{figure*}

Figure~\ref{fig:RCP_ALICESTAR} compares \RCP\ from 
Fig.~\ref{fig:RCP} to that for charged jets with \rr~=~0.2 and 0.3 measured in 
\PbPb\ collisions at \sqrtsNN~=~2.76 TeV~\cite{Abelev:2013kqa}, 
and to \RCP\ for  charged hadrons measured in \AuAu\ collisions at 
\sqrtsNN~=~200 GeV ~\cite{Adams:2003kv} and \PbPb\ collisions at \sqrtsNN~=~2.76 
TeV~\cite{Aad:2015wga}.
Note that for this measurement, central and peripheral collisions correspond 
to the 0--10\% and 60--80\% percentile intervals of 
the \AuAu\ inelastic cross section, respectively, while for the LHC jet measurements in the 
figure the corresponding intervals are 0--10\% and 50--80\%; and for the  
charged hadron measurements at both RHIC 
and the LHC the centrality intervals are 0--5\% and 60--80\%. 

The values of charged-hadron \RCP\ at RHIC and the LHC agree within 
uncertainties over their common range in \pT. The magnitude of 
charged-particle jet \RCP\ is likewise consistent within uncertainties at RHIC and LHC, 
though their \pTjetch\ intervals do not overlap. (Note that the bias due to 
\pTleadmin~=~5 \gev\ is small for $\pTjetch>15$ \gev; 
see Fig.~\ref{fig:pTleadRat}.) The apparent lack of dependence of charged-particle jet 
\RCP\ on \pTjetch\ is in contrast to the significant \pT-dependence of 
charged-hadron \RCP.

The inclusive charged-hadron distribution at high-\pT\ arises
predominantly from the leading hadron of the corresponding jet. The 
correlation between hadron \pT\ and its parent jet \pTjetch\ has a distribution 
that reflects the fragmentation process and which may generate 
different \pT-dependence 
of \RCP\ for hadrons and jets. The comparison of hadron and jet suppression in 
Fig.~\ref{fig:RCP_ALICESTAR} thus provides new constraints on theoretical 
descriptions of jet quenching.

\begin{center}
\begin{table}
\caption{\pT-shift between jet yield distributions in peripheral and central collisions normalized 
by the average number of binary collisions for quasi-inclusive jets (left) and semi-inclusive recoil jets (right). 
}
\label{Tab:pTshift}
\begin{tabular}{ |c|c|c| }
\hline
\multicolumn{3}{|c|}
{\parbox[t]{8cm}{\AuAu\ collisions, \sqrtsNN~=~200 GeV }} 
\\ \hline 
 & \multicolumn{2}{|c|}{\pT-shift peripheral$\rightarrow$central [\gev]} \\ 
\cline{2-3} 
\rr & quasi-inclusive jet (this analysis) & h+jet \cite{Adamczyk:2017yhe} \\ 
 & $15<\pTjetch<25$ \gev & $10<\pTjetch<20$ \gev \\ \hline \hline
0.2 & $-3.2\pm0.3_\mathrm{stat}\pm0.6_\mathrm{sys}$
& $-4.4\pm0.2_\mathrm{stat}\pm1.2_\mathrm{sys}$\\
0.3 & $-3.3\pm0.3_\mathrm{stat}\pm0.6_\mathrm{sys}$
& $-5.0\pm0.5_\mathrm{stat}\pm1.2_\mathrm{sys}$\\
0.4 & $-3.3\pm0.3_\mathrm{stat}\pm0.7_\mathrm{sys}$
& $-5.1\pm0.5_\mathrm{stat}\pm1.2_\mathrm{sys}$\\
\hline
\end{tabular}
\end{table}
\end{center}

The suppression of \RCP\ 
as a function of \pTjetch\ can be expressed equivalently as a 
\pT-shift of the spectrum in central, relative to peripheral, \AuAu\ collisions. 
This representation enables 
direct comparison of different suppression measurements since it removes the 
effect of the spectrum shape. The shift can be interpreted as the 
population-averaged energy transport out of the jet cone due to jet 
quenching~\cite{Adam:2015doa,Adamczyk:2017yhe}.
Table~\ref{Tab:pTshift} shows the \pT-shift values 
corresponding to \RCP\ in Fig.~\ref{fig:RCP} in the range 
$15<\pTjetch<25$ \gev, chosen to minimize the effect of the bias due to the \pTleadmin\ cut. 
The uncertainty in the value of the \pT-shift takes into account the correlated 
uncertainties of the central and peripheral \AuAu\ distributions. 

Table~\ref{Tab:pTshift} compares
the \pT-shift measured in this 
analysis to that for semi-inclusive recoil jet yield suppression measured using
hadron+jet correlations in \AuAu\ collisions at \sqrtsNN~=~200 
GeV~\cite{Adamczyk:2017yhe}. Note that 
the in-medium path-length distribution of jets contributing to the two 
measurements may differ~\cite{Adamczyk:2017yhe}. 
While the central values of the \pT-shift for the inclusive jet distributions 
are consistently smaller than those for recoil jets, no significant difference 
in \pT-shift 
is observed within the uncertainties. 
\subsection{Yield suppression: \RAApyth}
\label{sect:RAA}

This section presents measurements of \RAApyth, in which the reference is the 
inclusive charged-particle jet distribution for \pp\ collisions at \sqrts~=~200 GeV 
calculated by PYTHIA, which was validated by comparing to other STAR 
hadron and jet measurements (Sec.~\ref{sect:ppRef}). No \pTleadmin\ cut is 
imposed on this reference jet population.

\begin{figure*}[htbp]
\includegraphics[width=0.95\textwidth]{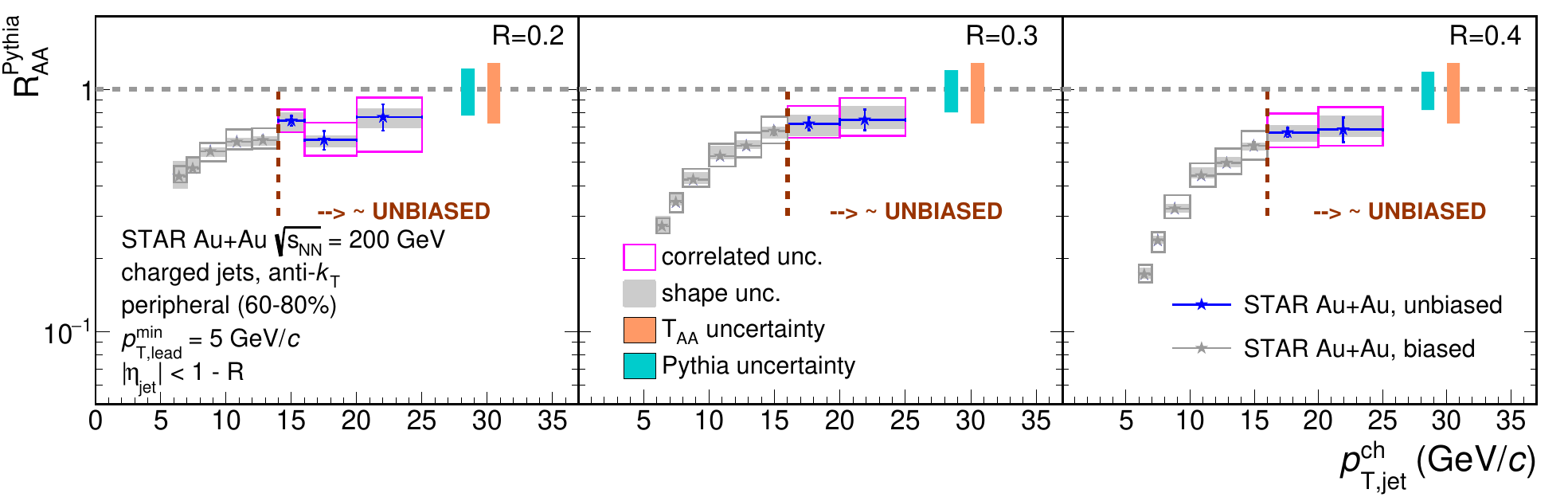}
\includegraphics[width=0.95\textwidth]{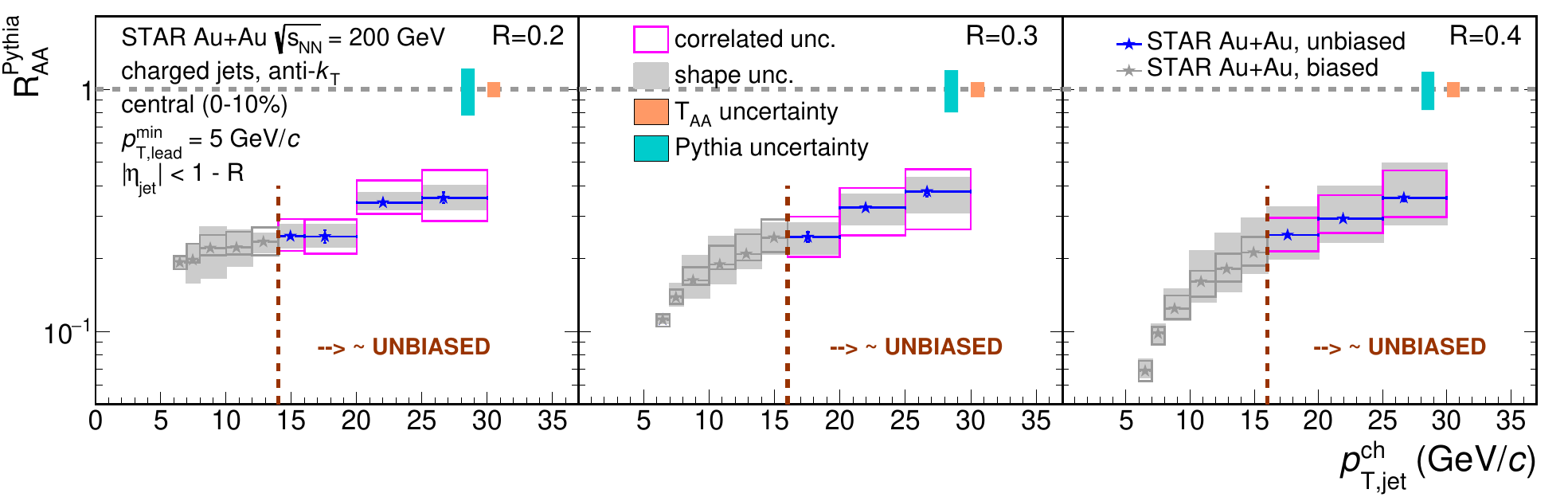}
\caption{(Color online) \RAApyth\ for quasi-inclusive charged jets in peripheral (upper) and central (lower) \AuAu\ collisions at 
\sqrtsNN~=~200 GeV, for \rr~=~0.2, 0.3, and 0.4. 
The reference spectrum for \pp\ collisions at \sqrts~=~200 GeV is generated by 
PYTHIA; see text for details. The region where the bias due to the \pTleadmin\ cut 
is small is indicated by 
the vertical dashed line.}
\label{fig:RAA}
\end{figure*}

Figure~\ref{fig:RAA} shows \RAApyth\ for quasi-inclusive jets in central \AuAu\ 
collisions at 
\sqrtsNN~=~200 GeV, for \rr~=~0.2, 0.3 and 0.4.  The region where the bias due to the \pTleadmin\ cut 
is small for the central \AuAu\ collisions is indicated by 
the vertical dashed line.

\begin{figure*}[htbp]
\includegraphics[width=0.98\textwidth]{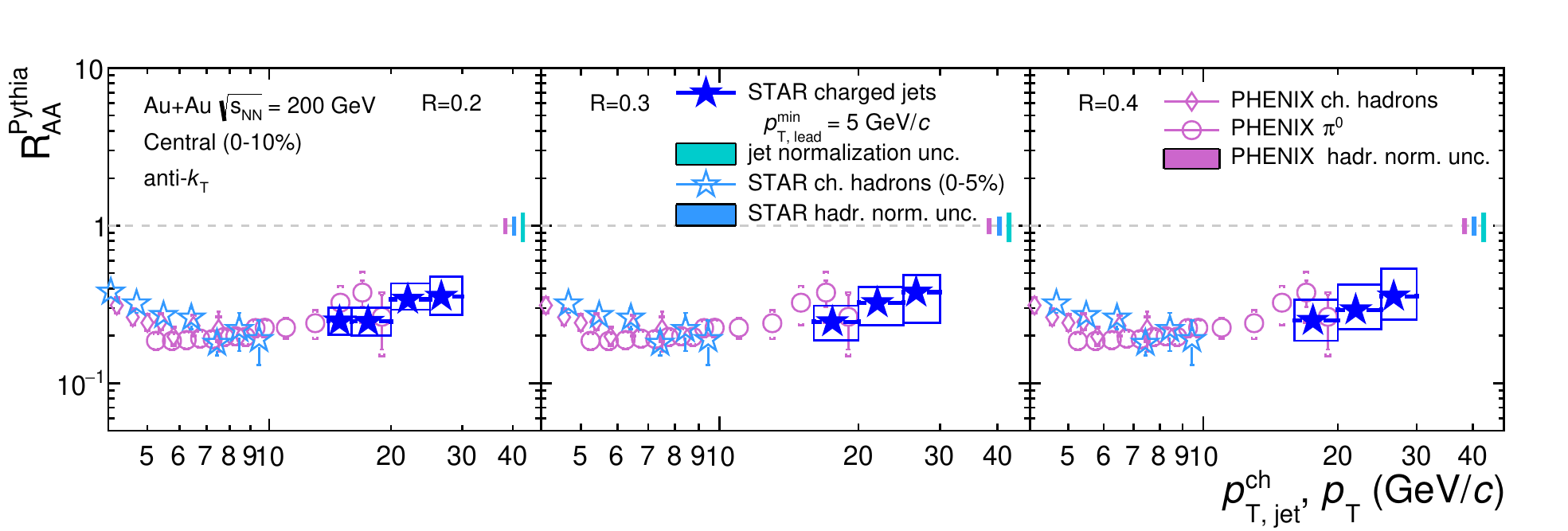}
\caption{(Color online) Comparison of \RAApyth\ from 
Fig.~\ref{fig:RAA} (stars) to charged hadron \cite{Adams:2003kv, Adler:2003au} and \pizero \cite{Adare:2012wg} \RAA\ at \sqrtsNN~=~200 GeV. 
Only points from the region where the bias in the data due to the \pTleadmin\ cut is small are shown.  } 
\label{fig:RAA_hadrons}
\end{figure*}
\begin{figure*}[htbp]
\includegraphics[width=0.98\textwidth]{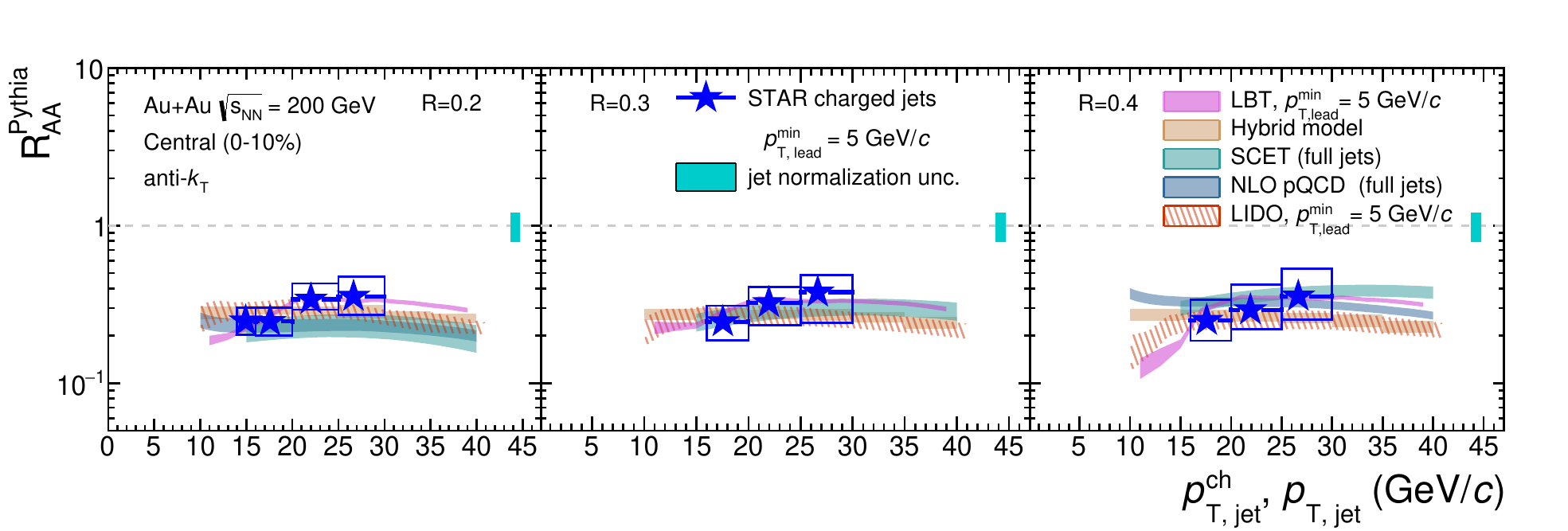}
\caption{(Color online) Comparison of \RAApyth\ from 
Fig.~\ref{fig:RAA} (stars) to the theoretical calculations described in Sect.~\ref{sect:Theory}. Only data in the unbiased region are shown. The Hybrid, LBT and LIDO calculations are for charged-particle jets, while NLO and SCET are for fully reconstructed jets. The LBT and LIDO calculations account for the effect of \pTleadmin~=~5 \gev\ in the \AuAu\ spectrum.} 
\label{fig:RAA_theory}
\end{figure*}

Figure~\ref{fig:RAA_hadrons} compares \RAApyth\ from Fig.~\ref{fig:RAA} to charged-hadron and \pizero\ \RAA\ measured in \AuAu\ collisions at \sqrtsNN~=~200~GeV.  
The values of \pizero\ and jet \RAA\ agree within uncertainties in this region.  

Figure~\ref{fig:RAA_theory} compares measured charged-particle jet \RAA\ to the theoretical calculations presented in  Sect.~\ref{sect:Theory}. The Hybrid, LBT and LIDO calculations are carried out for charged jets, while the SCET and NLO pQCD calculations are for fully reconstructed jets. The \pTjet-dependence of  full jet \RAA\ is weak, however, so that comparison of these calculations with the  charged-particle jet measurement is meaningful. The LBT and LIDO calculations also include a cut on the leading constituent for the \AuAu\ spectrum,  corresponding to \pTleadmin~=~5 \gev\ applied in the data analysis.
All calculations are consistent within uncertainties with the measured inclusive jet \RAA\ in the unbiased region. The largest differences between models is seen for \rr~=~0.4; future measurements of inclusive jet \RAA\ with improved systematic precision may be able to discriminate between these models.

\subsection{Medium-induced jet broadening}
\label{sect:Rratios}

The dependence of the inclusive jet yield on resolution parameter \rr\ is sensitive to the jet 
energy profile transverse to its axis. Ratios of inclusive cross sections 
are of particular interest for measuring the transverse jet energy profile and 
its modification due to jet quenching since there is significant 
cancellation of systematic uncertainties in the ratio, both 
experimenally~\cite{Abelev:2013fn,Chatrchyan:2014gia} and 
theoretically~\cite{Soyez:2011np,Dasgupta:2016bnd,Casalderrey-Solana:2016jvj}.

The ratio of inclusive jet cross sections for small \rr\ (\rr~=~0.2) and large \rr\ (\rr~=~0.4 or 0.5) is found to be 
less than unity in \pp\ collisions at \sqrts~=~2.76 and 7 TeV~\cite{Abelev:2013fn,Chatrchyan:2014gia,ALICE:2014dla}, consistent with pQCD 
calculations at NLO and next-to-next-to-leading-order (NNLO)~\cite{Soyez:2011np,Dasgupta:2016bnd}.  A value of this ratio less than unity is expected qualitatively,  since jets subtend finite area and larger-\rr\ jet reconstruction collects more energy. However, the specific value of the ratio reflects the transverse jet energy profile: The areal energy density in a jet is on average largest near the jet axis, decreasing with increasing distance from the axis. 
The ratio of semi-inclusive recoil jet yields  for different \rr\ 
is likewise measured to be less than unity in
\pp\ collisions at \sqrts~=~7 TeV~\cite{Adam:2015doa}, with the ratio described 
better by PYTHIA than a pQCD calculation at 
NLO~\cite{Adam:2015doa,deFlorian:2009fw}. 

In nucleus-nucleus collisions, broadening of the transverse jet energy profile 
due to quenching has been explored by measuring the ratio of charged-particle jet 
inclusive cross sections with different \rr\ in \PbPb\ collisions at 
\sqrtsNN~=~2.76 TeV~\cite{Abelev:2013kqa}, and the ratio of semi-inclusive recoil 
jet yields with different \rr\ 
in \PbPb\ collisions at \sqrtsNN~=~2.76 TeV~\cite{Adam:2015doa} and in \AuAu\ 
collisions at \sqrtsNN~=~200 GeV~\cite{Adamczyk:2017yhe}. In both measurements, no 
significant medium-induced broadening is observed. Note that this 
observable is 
different from the jet shape observable employed 
in Refs.~\cite{Chatrchyan:2013kwa,Khachatryan:2016tfj}, with different experimental 
and theoretical uncertainties.

\begin{figure*}[htbp]
\includegraphics[width=0.99\textwidth]{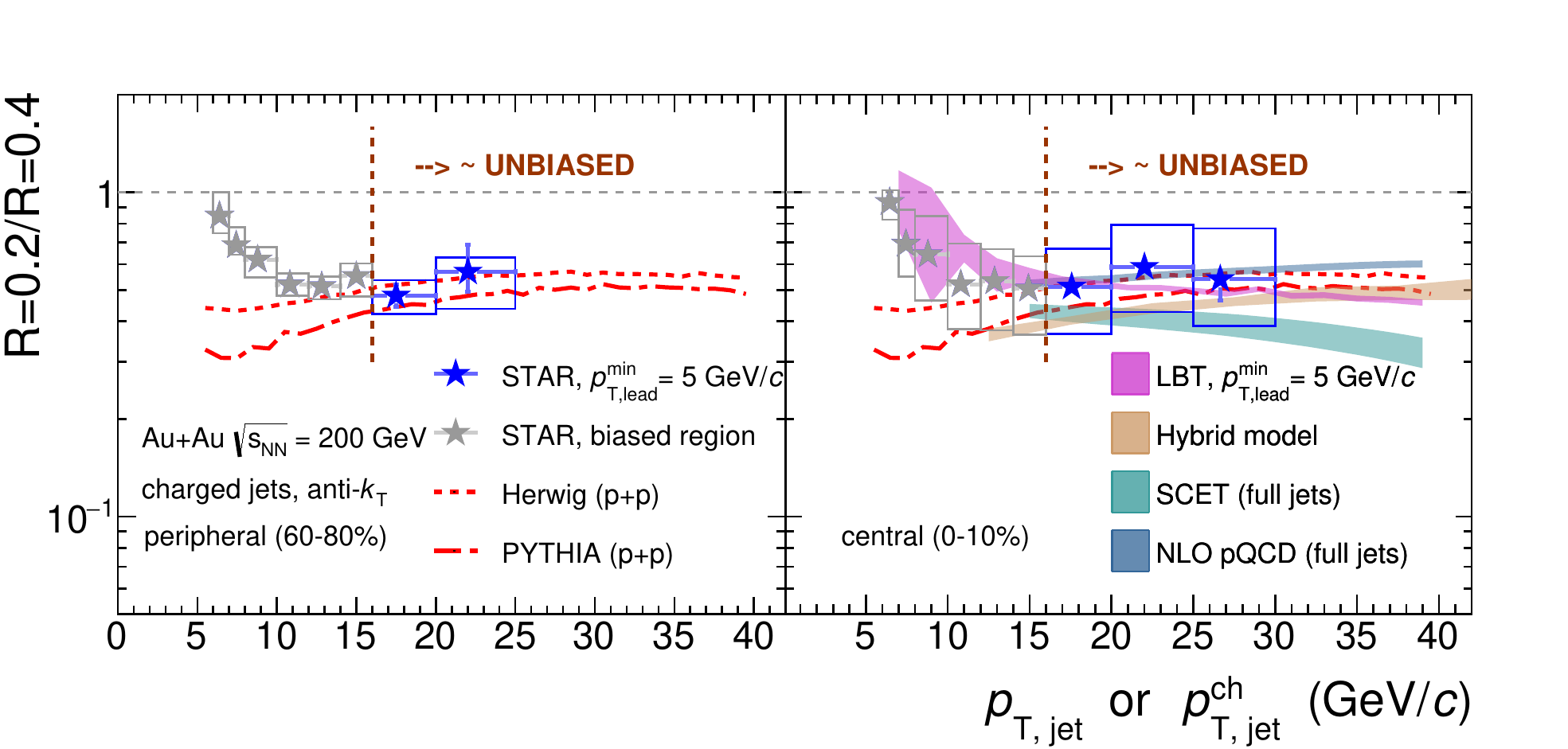}
\caption{(Color online) Ratio of quasi-inclusive yields for charged jets with \rr~=~0.2 and 
\rr~=~0.4, for peripheral (left) and central (right) \AuAu\ collisions at 
\sqrtsNN~=~200 GeV and \pTleadmin~=~5~GeV/$c$. Calculations are 
shown for charged jets in \pp\ collisions at \sqrts~=~200 GeV simulated by 
PYTHIA~\cite{Sjostrand:2006za} and HERWIG~\cite{Bahr:2008pv}; both panels show 
the same distributions from these calculations. Also shown are predictions for 
reconstructed jets in \AuAu\ collisions from the theoretical calculations 
described in Sect.~\ref{sect:Theory}. The region where the bias due to the \pTleadmin\ cut 
is small is indicated by the vertical dashed line.
}
\label{fig:RRratio_theory}
\end{figure*}

Figure~\ref{fig:RRratio_theory} shows the ratio of distributions from
Fig.~\ref{fig:CorrSpec} for \rr~=~0.2 and 0.4, for central 
and peripheral \AuAu\ collisions. The measured ratio is less than unity for both 
centralities, as observed in \pp\ 
collisions~\cite{Abelev:2013fn,Chatrchyan:2014gia,ALICE:2014dla,Adam:2015doa}. The panels also show 
calculations 
for \pp\ collisions at \sqrts~=~200 GeV from PYTHIA and HERWIG, which agree within 
uncertainties with the ratios measured in \AuAu\ collisions. This indicates that 
there is no 
significant modification of the transverse jet profile due to quenching in 
central \AuAu\ collisions at \sqrtsNN~=~200 GeV, 
consistent with related measurements at RHIC and 
LHC~\cite{ALICE:2014dla,Adam:2015doa,Adamczyk:2017yhe}. 

This observation is in contrast to measurements of dijet asymmetry \AJ\ at 
RHIC~\cite{Adamczyk:2016fqm}, which find that energy lost due to quenching for 
jets with \rr~=~0.2 is largely recovered for jets with \rr~=~0.4, indicating a 
significant medium-induced modification of the transverse profile for the 
jet population selected in that analysis. However, that population differs 
significantly from the jet population used in the analysis reported here. Assessment 
of the two analyses and interpretation of their observed differences in terms of 
transverse jet profile modification requires the modeling of both measurements in a 
common theoretical framework (e.g. Ref.~\cite{Kauder:2018cdt}).

Figure~\ref{fig:RRratio_theory} (right panel) also shows theoretical calculations based on the
Hybrid, LBT, NLO, and SCET models  presented in Sec.~\ref{sect:Theory}. These four models predict significantly different values for this ratio, though all calculations agree with the measurement within uncertainties. Future measurements of this observable, with improved systematic precision and at larger \rr, may discriminate between the models.

\section{Summary}
\label{sect:Summary}

We have reported the first measurement of inclusive  charged-particle jet production in central 
and peripheral \AuAu\ collisions at \sqrtsNN~=~200 GeV, over the range 
$5<\pTjetch<30$ \gev.
The large  uncorrelated  background is suppressed by the requirement that the leading hadron in the 
jet satisfies $\pTlead>\pTleadmin$, where \pTleadmin~=~5 \gev. The bias imposed by this requirement
is quantified by comparing distributions for \pTleadmin~=~5 and 7 \gev, and the region of the 
measurement where the bias is small is identified.

A PM is developed, incorporating uncorrelated soft particle emission and a PYTHIA-generated jet distribution,
motivated by the excellent description by such an approach of event-by-event transverse energy fluctuations in \aaa\ collisions over a wide range in \sqrtsNN. The PM describes the uncorrected jet distributions in this analysis well, indicating that the  
background underlying jet measurements in central \AuAu\ 
collisions at RHIC is to a large extent statistically distributed, with 
dynamical correlations playing a much lesser role. This picture is also supported by an earlier mixed event analysis of semi-inclusive hadron-jet 
distributions at RHIC.

Comparison of the charged-particle jet yield in central and peripheral \AuAu\ collisions reveals a suppression for central \AuAu\ collisions, with magnitude of the suppression 
similar to that in central \PbPb\ collisions at the LHC. No significant \pTjetch-dependence of inclusive jet suppression is observed, in contrast to the marked \pT-dependence of inclusive hadron suppression in central \aaa\ collisions at both RHIC and the LHC.

Jet yield suppression at fixed \pTjetch\ can be expressed equivalently as a shift in the yield distribution as a function of \pTjetch, where the magnitude of the shift
corresponds to medium-induced energy transport out of the jet cone. The \pTjetch shift for the inclusive jet population with \rr~=~0.4 is --3.3~$\pm 0.3_\mathrm{stat}\pm 0.7_\mathrm{sys}$~\gev, consistent with
that measured for semi-inclusive recoil jets. We note that in-medium path-length distributions
for these two measurements may differ.

The charged-particle jet yield in central \AuAu\ collisions is also compared to that for
\pp\ collisions generated by PYTHIA, which was validated using a STAR measurement of fully 
reconstructed jets in \pp\ collisions and inclusive single-particle spectra. The magnitude of suppression of inclusive \pizero\ and jet production from this comparison are consistent
within uncertainties. 

Comparison is also made to several theoretical calculations 
of jet quenching (NLO pQCD, SCET, Hybrid model, LIDO), which are consistent with the measurement within uncertainties. Greater
precision is needed to discriminate the models.

Finally, medium-induced broadening of the jet transverse energy distribution is 
explored by measuring the ratio of inclusive yields for \rr~=~0.2 and 0.4. 
No significant medium-induced modification is observed in central \AuAu\ collisions, consistent with similar measurements at the LHC.
In comparison to jet quenching calculations, NLO predicts a larger ratio than that observed, but SCET and the Hybrid model
are consistent with the measurement. The absence of medium-induced broadening in this inclusive jet analysis is in contrast to the broadening observed in di-jet asymmetry measurements at RHIC. Interpretation of this difference requires modeling to carefully assess underlying biases in each of the two analyses.

The results presented here provide new constraints on theoretical models of jet quenching,
and new insights into the nature of the large backgrounds to jet measurements in heavy-ion collisions.

\section{Acknowledgments}
\label{sect:Acknowledgements}

We thank Weiyao Ke, Daniel Pablos, Krishna Rajagopal, Ivan Vitev, and Xin-Nian Wang for providing 
theoretical calculations. We thank the RHIC Operations Group and RCF at BNL, the NERSC Center at LBNL, and the Open Science Grid consortium for providing resources and support.  This work was supported in part by the Office of Nuclear Physics within the U.S. DOE Office of Science; the U.S. National Science Foundation; the Ministry of Education and Science of the Russian Federation; National Natural Science Foundation of China; Chinese Academy of Science; the Ministry of Science and Technology of China and the Chinese Ministry of Education; the Higher Education Sprout Project by Ministry of Education at NCKU; the National Research Foundation of Korea; Czech Science Foundation and Ministry of Education, Youth and Sports of the Czech Republic; Hungarian National Research, Development and Innovation Office; New National Excellency Programme of the Hungarian Ministry of Human Capacities; Department of Atomic Energy and Department of Science and Technology of the Government of India; the National Science Centre of Poland; the Ministry  of Science, Education and Sports of the Republic of Croatia; RosAtom of Russia and German Bundesministerium fur Bildung, Wissenschaft, Forschung and Technologie (BMBF); Helmholtz Association; Ministry of Education, Culture, Sports, Science, and Technology (MEXT); and Japan Society for the Promotion of Science (JSPS).

\clearpage

\bibliographystyle{apsrev4-1}

\bibliography{references.bib}

\begin{thebibliography}{111}%
\makeatletter
\providecommand \@ifxundefined [1]{%
 \@ifx{#1\undefined}
}%
\providecommand \@ifnum [1]{%
 \ifnum #1\expandafter \@firstoftwo
 \else \expandafter \@secondoftwo
 \fi
}%
\providecommand \@ifx [1]{%
 \ifx #1\expandafter \@firstoftwo
 \else \expandafter \@secondoftwo
 \fi
}%
\providecommand \natexlab [1]{#1}%
\providecommand \enquote  [1]{``#1''}%
\providecommand \bibnamefont  [1]{#1}%
\providecommand \bibfnamefont [1]{#1}%
\providecommand \citenamefont [1]{#1}%
\providecommand \href@noop [0]{\@secondoftwo}%
\providecommand \href [0]{\begingroup \@sanitize@url \@href}%
\providecommand \@href[1]{\@@startlink{#1}\@@href}%
\providecommand \@@href[1]{\endgroup#1\@@endlink}%
\providecommand \@sanitize@url [0]{\catcode `\\12\catcode `\$12\catcode
  `\&12\catcode `\#12\catcode `\^12\catcode `\_12\catcode `\%12\relax}%
\providecommand \@@startlink[1]{}%
\providecommand \@@endlink[0]{}%
\providecommand \url  [0]{\begingroup\@sanitize@url \@url }%
\providecommand \@url [1]{\endgroup\@href {#1}{\urlprefix }}%
\providecommand \urlprefix  [0]{URL }%
\providecommand \Eprint [0]{\href }%
\providecommand \doibase [0]{http://dx.doi.org/}%
\providecommand \selectlanguage [0]{\@gobble}%
\providecommand \bibinfo  [0]{\@secondoftwo}%
\providecommand \bibfield  [0]{\@secondoftwo}%
\providecommand \translation [1]{[#1]}%
\providecommand \BibitemOpen [0]{}%
\providecommand \bibitemStop [0]{}%
\providecommand \bibitemNoStop [0]{.\EOS\space}%
\providecommand \EOS [0]{\spacefactor3000\relax}%
\providecommand \BibitemShut  [1]{\csname bibitem#1\endcsname}%
\let\auto@bib@innerbib\@empty
\bibitem [{\citenamefont {Busza}\ \emph {et~al.}(2018)\citenamefont {Busza},
  \citenamefont {Rajagopal},\ and\ \citenamefont {van~der
  Schee}}]{Busza:2018rrf}%
  \BibitemOpen
  \bibfield  {author} {\bibinfo {author} {\bibfnamefont {W.}~\bibnamefont
  {Busza}}, \bibinfo {author} {\bibfnamefont {K.}~\bibnamefont {Rajagopal}}, \
  and\ \bibinfo {author} {\bibfnamefont {W.}~\bibnamefont {van~der Schee}},\
  }\href {\doibase 10.1146/annurev-nucl-101917-020852} {\bibfield  {journal}
  {\bibinfo  {journal} {Annu. Rev. Nucl. Part. Sci.}\ }\textbf {\bibinfo
  {volume} {68}},\ \bibinfo {pages} {339} (\bibinfo {year} {2018})},\ \Eprint
  {http://arxiv.org/abs/1802.04801} {arXiv:1802.04801 [hep-ph]} \BibitemShut
  {NoStop}%
\bibitem [{\citenamefont {Heinz}\ and\ \citenamefont
  {Snellings}(2013)}]{Heinz:2013th}%
  \BibitemOpen
  \bibfield  {author} {\bibinfo {author} {\bibfnamefont {U.}~\bibnamefont
  {Heinz}}\ and\ \bibinfo {author} {\bibfnamefont {R.}~\bibnamefont
  {Snellings}},\ }\href {\doibase 10.1146/annurev-nucl-102212-170540}
  {\bibfield  {journal} {\bibinfo  {journal} {Annu. Rev. Nucl. Part. Sci.}\
  }\textbf {\bibinfo {volume} {63}},\ \bibinfo {pages} {123} (\bibinfo {year}
  {2013})},\ \Eprint {http://arxiv.org/abs/1301.2826} {arXiv:1301.2826
  [nucl-th]} \BibitemShut {NoStop}%
\bibitem [{\citenamefont {Burke}\ \emph {et~al.}(2014)\citenamefont {Burke}
  \emph {et~al.}}]{Burke:2013yra}%
  \BibitemOpen
  \bibfield  {author} {\bibinfo {author} {\bibfnamefont {K.~M.}\ \bibnamefont
  {Burke}} \emph {et~al.},\ }\href {\doibase 10.1103/PhysRevC.90.014909}
  {\bibfield  {journal} {\bibinfo  {journal} {Phys. Rev. C}\ }\textbf {\bibinfo
  {volume} {90}},\ \bibinfo {pages} {014909} (\bibinfo {year} {2014})},\
  \Eprint {http://arxiv.org/abs/1312.5003} {arXiv:1312.5003 [nucl-th]}
  \BibitemShut {NoStop}%
\bibitem [{\citenamefont {Abelev}\ \emph
  {et~al.}(2006{\natexlab{a}})\citenamefont {Abelev} \emph
  {et~al.}}]{Abelev:2006uq}%
  \BibitemOpen
  \bibfield  {author} {\bibinfo {author} {\bibfnamefont {B.~I.}\ \bibnamefont
  {Abelev}} \emph {et~al.} (\bibinfo {collaboration} {STAR}),\ }\href {\doibase
  10.1103/PhysRevLett.97.252001} {\bibfield  {journal} {\bibinfo  {journal}
  {Phys. Rev. Lett.}\ }\textbf {\bibinfo {volume} {97}},\ \bibinfo {pages}
  {252001} (\bibinfo {year} {2006}{\natexlab{a}})},\ \Eprint
  {http://arxiv.org/abs/hep-ex/0608030} {arXiv:hep-ex/0608030} \BibitemShut
  {NoStop}%
\bibitem [{\citenamefont {Adamczyk}\ \emph
  {et~al.}(2017{\natexlab{a}})\citenamefont {Adamczyk} \emph
  {et~al.}}]{Adamczyk:2016okk}%
  \BibitemOpen
  \bibfield  {author} {\bibinfo {author} {\bibfnamefont {L.}~\bibnamefont
  {Adamczyk}} \emph {et~al.} (\bibinfo {collaboration} {STAR}),\ }\href
  {\doibase 10.1103/PhysRevD.95.071103} {\bibfield  {journal} {\bibinfo
  {journal} {Phys. Rev. D}\ }\textbf {\bibinfo {volume} {95}},\ \bibinfo
  {pages} {071103} (\bibinfo {year} {2017}{\natexlab{a}})},\ \Eprint
  {http://arxiv.org/abs/1610.06616} {arXiv:1610.06616 [hep-ex]} \BibitemShut
  {NoStop}%
\bibitem [{\citenamefont {Abelev}\ \emph
  {et~al.}(2013{\natexlab{a}})\citenamefont {Abelev} \emph
  {et~al.}}]{Abelev:2013fn}%
  \BibitemOpen
  \bibfield  {author} {\bibinfo {author} {\bibfnamefont {B.}~\bibnamefont
  {Abelev}} \emph {et~al.} (\bibinfo {collaboration} {ALICE}),\ }\href
  {\doibase 10.1016/j.physletb.2013.04.026} {\bibfield  {journal} {\bibinfo
  {journal} {Phys. Lett. B}\ }\textbf {\bibinfo {volume} {722}},\ \bibinfo
  {pages} {262} (\bibinfo {year} {2013}{\natexlab{a}})},\ \Eprint
  {http://arxiv.org/abs/1301.3475} {arXiv:1301.3475 [nucl-ex]} \BibitemShut
  {NoStop}%
\bibitem [{\citenamefont {Aad}\ \emph {et~al.}(2015{\natexlab{a}})\citenamefont
  {Aad} \emph {et~al.}}]{Aad:2014vwa}%
  \BibitemOpen
  \bibfield  {author} {\bibinfo {author} {\bibfnamefont {G.}~\bibnamefont
  {Aad}} \emph {et~al.} (\bibinfo {collaboration} {ATLAS}),\ }\href {\doibase
  10.1007/JHEP02(2015)153, 10.1007/JHEP09(2015)141} {\bibfield  {journal}
  {\bibinfo  {journal} {J. High Energy Phys.}\ }\textbf {\bibinfo {volume}
  {02}},\ \bibinfo {pages} {153} (\bibinfo {year} {2015}{\natexlab{a}})},\
  \bibinfo {note} {[Erratum: J. High Energy Phys. {\bf 09}, 141 (2015)]},\
  \Eprint {http://arxiv.org/abs/1410.8857} {arXiv:1410.8857 [hep-ex]}
  \BibitemShut {NoStop}%
\bibitem [{\citenamefont {Khachatryan}\ \emph
  {et~al.}(2017{\natexlab{a}})\citenamefont {Khachatryan} \emph
  {et~al.}}]{Khachatryan:2016mlc}%
  \BibitemOpen
  \bibfield  {author} {\bibinfo {author} {\bibfnamefont {V.}~\bibnamefont
  {Khachatryan}} \emph {et~al.} (\bibinfo {collaboration} {CMS}),\ }\href
  {\doibase 10.1007/JHEP03(2017)156} {\bibfield  {journal} {\bibinfo  {journal}
  {J. High Energy Phys.}\ }\textbf {\bibinfo {volume} {03}},\ \bibinfo {pages}
  {156} (\bibinfo {year} {2017}{\natexlab{a}})},\ \Eprint
  {http://arxiv.org/abs/1609.05331} {arXiv:1609.05331 [hep-ex]} \BibitemShut
  {NoStop}%
\bibitem [{\citenamefont {Majumder}\ and\ \citenamefont
  {Van~Leeuwen}(2011)}]{Majumder:2010qh}%
  \BibitemOpen
  \bibfield  {author} {\bibinfo {author} {\bibfnamefont {A.}~\bibnamefont
  {Majumder}}\ and\ \bibinfo {author} {\bibfnamefont {M.}~\bibnamefont
  {Van~Leeuwen}},\ }\href {\doibase 10.1016/j.ppnp.2010.09.001} {\bibfield
  {journal} {\bibinfo  {journal} {Prog. Part. Nucl. Phys.}\ }\textbf {\bibinfo
  {volume} {66}},\ \bibinfo {pages} {41} (\bibinfo {year} {2011})},\ \Eprint
  {http://arxiv.org/abs/1002.2206} {arXiv:1002.2206 [hep-ph]} \BibitemShut
  {NoStop}%
\bibitem [{\citenamefont {Cao}\ and\ \citenamefont {Wang}(2020)}]{Cao:2020wlm}%
  \BibitemOpen
  \bibfield  {author} {\bibinfo {author} {\bibfnamefont {S.}~\bibnamefont
  {Cao}}\ and\ \bibinfo {author} {\bibfnamefont {X.-N.}\ \bibnamefont {Wang}},\
  }\href@noop {} {\  (\bibinfo {year} {2020})},\ \Eprint
  {http://arxiv.org/abs/2002.04028} {arXiv:2002.04028 [hep-ph]} \BibitemShut
  {NoStop}%
\bibitem [{\citenamefont {Adams}\ \emph
  {et~al.}(2005{\natexlab{a}})\citenamefont {Adams} \emph
  {et~al.}}]{Adams:2005dq}%
  \BibitemOpen
  \bibfield  {author} {\bibinfo {author} {\bibfnamefont {J.}~\bibnamefont
  {Adams}} \emph {et~al.} (\bibinfo {collaboration} {STAR}),\ }\href {\doibase
  10.1016/j.nuclphysa.2005.03.085} {\bibfield  {journal} {\bibinfo  {journal}
  {Nucl. Phys. A}\ }\textbf {\bibinfo {volume} {757}},\ \bibinfo {pages} {102}
  (\bibinfo {year} {2005}{\natexlab{a}})},\ \Eprint
  {http://arxiv.org/abs/nucl-ex/0501009} {arXiv:nucl-ex/0501009} \BibitemShut
  {NoStop}%
\bibitem [{\citenamefont {Wang}\ and\ \citenamefont
  {Gyulassy}(1992)}]{Wang:1991xy}%
  \BibitemOpen
  \bibfield  {author} {\bibinfo {author} {\bibfnamefont {X.-N.}\ \bibnamefont
  {Wang}}\ and\ \bibinfo {author} {\bibfnamefont {M.}~\bibnamefont
  {Gyulassy}},\ }\href {\doibase 10.1103/PhysRevLett.68.1480} {\bibfield
  {journal} {\bibinfo  {journal} {Phys. Rev. Lett.}\ }\textbf {\bibinfo
  {volume} {68}},\ \bibinfo {pages} {1480} (\bibinfo {year}
  {1992})}\BibitemShut {NoStop}%
\bibitem [{\citenamefont {Adler}\ \emph {et~al.}(2002)\citenamefont {Adler}
  \emph {et~al.}}]{Adler:2002xw}%
  \BibitemOpen
  \bibfield  {author} {\bibinfo {author} {\bibfnamefont {C.}~\bibnamefont
  {Adler}} \emph {et~al.} (\bibinfo {collaboration} {STAR}),\ }\href {\doibase
  10.1103/PhysRevLett.89.202301} {\bibfield  {journal} {\bibinfo  {journal}
  {Phys. Rev. Lett.}\ }\textbf {\bibinfo {volume} {89}},\ \bibinfo {pages}
  {202301} (\bibinfo {year} {2002})},\ \Eprint
  {http://arxiv.org/abs/nucl-ex/0206011} {arXiv:nucl-ex/0206011 [nucl-ex]}
  \BibitemShut {NoStop}%
\bibitem [{\citenamefont {Adler}\ \emph
  {et~al.}(2003{\natexlab{a}})\citenamefont {Adler} \emph
  {et~al.}}]{Adler:2002tq}%
  \BibitemOpen
  \bibfield  {author} {\bibinfo {author} {\bibfnamefont {C.}~\bibnamefont
  {Adler}} \emph {et~al.} (\bibinfo {collaboration} {STAR}),\ }\href {\doibase
  10.1103/PhysRevLett.90.082302} {\bibfield  {journal} {\bibinfo  {journal}
  {Phys. Rev. Lett.}\ }\textbf {\bibinfo {volume} {90}},\ \bibinfo {pages}
  {082302} (\bibinfo {year} {2003}{\natexlab{a}})},\ \Eprint
  {http://arxiv.org/abs/nucl-ex/0210033} {arXiv:nucl-ex/0210033 [nucl-ex]}
  \BibitemShut {NoStop}%
\bibitem [{\citenamefont {Adams}\ \emph {et~al.}(2003)\citenamefont {Adams}
  \emph {et~al.}}]{Adams:2003kv}%
  \BibitemOpen
  \bibfield  {author} {\bibinfo {author} {\bibfnamefont {J.}~\bibnamefont
  {Adams}} \emph {et~al.} (\bibinfo {collaboration} {STAR}),\ }\href {\doibase
  10.1103/PhysRevLett.91.172302} {\bibfield  {journal} {\bibinfo  {journal}
  {Phys. Rev. Lett.}\ }\textbf {\bibinfo {volume} {91}},\ \bibinfo {pages}
  {172302} (\bibinfo {year} {2003})},\ \Eprint
  {http://arxiv.org/abs/nucl-ex/0305015} {arXiv:nucl-ex/0305015 [nucl-ex]}
  \BibitemShut {NoStop}%
\bibitem [{\citenamefont {Adams}\ \emph {et~al.}(2006)\citenamefont {Adams}
  \emph {et~al.}}]{Adams:2006yt}%
  \BibitemOpen
  \bibfield  {author} {\bibinfo {author} {\bibfnamefont {J.}~\bibnamefont
  {Adams}} \emph {et~al.} (\bibinfo {collaboration} {STAR}),\ }\href {\doibase
  10.1103/PhysRevLett.97.162301} {\bibfield  {journal} {\bibinfo  {journal}
  {Phys. Rev. Lett.}\ }\textbf {\bibinfo {volume} {97}},\ \bibinfo {pages}
  {162301} (\bibinfo {year} {2006})},\ \Eprint
  {http://arxiv.org/abs/nucl-ex/0604018} {arXiv:nucl-ex/0604018} \BibitemShut
  {NoStop}%
\bibitem [{\citenamefont {Adamczyk}\ \emph {et~al.}(2014)\citenamefont
  {Adamczyk} \emph {et~al.}}]{Adamczyk:2013jei}%
  \BibitemOpen
  \bibfield  {author} {\bibinfo {author} {\bibfnamefont {L.}~\bibnamefont
  {Adamczyk}} \emph {et~al.} (\bibinfo {collaboration} {STAR}),\ }\href
  {\doibase 10.1103/PhysRevLett.112.122301} {\bibfield  {journal} {\bibinfo
  {journal} {Phys. Rev. Lett.}\ }\textbf {\bibinfo {volume} {112}},\ \bibinfo
  {pages} {122301} (\bibinfo {year} {2014})},\ \Eprint
  {http://arxiv.org/abs/1302.6184} {arXiv:1302.6184 [nucl-ex]} \BibitemShut
  {NoStop}%
\bibitem [{\citenamefont {Adcox}\ \emph
  {et~al.}(2002{\natexlab{a}})\citenamefont {Adcox} \emph
  {et~al.}}]{Adcox:2001jp}%
  \BibitemOpen
  \bibfield  {author} {\bibinfo {author} {\bibfnamefont {K.}~\bibnamefont
  {Adcox}} \emph {et~al.} (\bibinfo {collaboration} {PHENIX}),\ }\href
  {\doibase 10.1103/PhysRevLett.88.022301} {\bibfield  {journal} {\bibinfo
  {journal} {Phys. Rev. Lett.}\ }\textbf {\bibinfo {volume} {88}},\ \bibinfo
  {pages} {022301} (\bibinfo {year} {2002}{\natexlab{a}})},\ \Eprint
  {http://arxiv.org/abs/nucl-ex/0109003} {arXiv:nucl-ex/0109003 [nucl-ex]}
  \BibitemShut {NoStop}%
\bibitem [{\citenamefont {Adare}\ \emph {et~al.}(2013)\citenamefont {Adare}
  \emph {et~al.}}]{Adare:2012wg}%
  \BibitemOpen
  \bibfield  {author} {\bibinfo {author} {\bibfnamefont {A.}~\bibnamefont
  {Adare}} \emph {et~al.} (\bibinfo {collaboration} {PHENIX}),\ }\href
  {\doibase 10.1103/PhysRevC.87.034911} {\bibfield  {journal} {\bibinfo
  {journal} {Phys. Rev. C}\ }\textbf {\bibinfo {volume} {87}},\ \bibinfo
  {pages} {034911} (\bibinfo {year} {2013})},\ \Eprint
  {http://arxiv.org/abs/1208.2254} {arXiv:1208.2254 [nucl-ex]} \BibitemShut
  {NoStop}%
\bibitem [{\citenamefont {Adare}\ \emph {et~al.}(2010)\citenamefont {Adare}
  \emph {et~al.}}]{Adare:2010ry}%
  \BibitemOpen
  \bibfield  {author} {\bibinfo {author} {\bibfnamefont {A.}~\bibnamefont
  {Adare}} \emph {et~al.} (\bibinfo {collaboration} {PHENIX}),\ }\href
  {\doibase 10.1103/PhysRevLett.104.252301} {\bibfield  {journal} {\bibinfo
  {journal} {Phys. Rev. Lett.}\ }\textbf {\bibinfo {volume} {104}},\ \bibinfo
  {pages} {252301} (\bibinfo {year} {2010})},\ \Eprint
  {http://arxiv.org/abs/1002.1077} {arXiv:1002.1077 [nucl-ex]} \BibitemShut
  {NoStop}%
\bibitem [{\citenamefont {Aamodt}\ \emph {et~al.}(2012)\citenamefont {Aamodt}
  \emph {et~al.}}]{Aamodt:2011vg}%
  \BibitemOpen
  \bibfield  {author} {\bibinfo {author} {\bibfnamefont {K.}~\bibnamefont
  {Aamodt}} \emph {et~al.} (\bibinfo {collaboration} {ALICE}),\ }\href
  {\doibase 10.1103/PhysRevLett.108.092301} {\bibfield  {journal} {\bibinfo
  {journal} {Phys. Rev. Lett.}\ }\textbf {\bibinfo {volume} {108}},\ \bibinfo
  {pages} {092301} (\bibinfo {year} {2012})},\ \Eprint
  {http://arxiv.org/abs/1110.0121} {arXiv:1110.0121 [nucl-ex]} \BibitemShut
  {NoStop}%
\bibitem [{\citenamefont {Abelev}\ \emph
  {et~al.}(2013{\natexlab{b}})\citenamefont {Abelev} \emph
  {et~al.}}]{Abelev:2012hxa}%
  \BibitemOpen
  \bibfield  {author} {\bibinfo {author} {\bibfnamefont {B.}~\bibnamefont
  {Abelev}} \emph {et~al.} (\bibinfo {collaboration} {ALICE}),\ }\href
  {\doibase 10.1016/j.physletb.2013.01.051} {\bibfield  {journal} {\bibinfo
  {journal} {Phys. Lett. B}\ }\textbf {\bibinfo {volume} {720}},\ \bibinfo
  {pages} {52} (\bibinfo {year} {2013}{\natexlab{b}})},\ \Eprint
  {http://arxiv.org/abs/1208.2711} {arXiv:1208.2711 [hep-ex]} \BibitemShut
  {NoStop}%
\bibitem [{\citenamefont {Adam}\ \emph {et~al.}(2016)\citenamefont {Adam} \emph
  {et~al.}}]{Adam:2016jp}%
  \BibitemOpen
  \bibfield  {author} {\bibinfo {author} {\bibfnamefont {J.}~\bibnamefont
  {Adam}} \emph {et~al.} (\bibinfo {collaboration} {ALICE}),\ }\href {\doibase
  10.1016/j.physletb.2016.10.048} {\bibfield  {journal} {\bibinfo  {journal}
  {Phys. Lett. B}\ }\textbf {\bibinfo {volume} {763}},\ \bibinfo {pages} {238}
  (\bibinfo {year} {2016})},\ \Eprint {http://arxiv.org/abs/1608.07201}
  {arXiv:1608.07201 [nucl-ex]} \BibitemShut {NoStop}%
\bibitem [{\citenamefont {Chatrchyan}\ \emph
  {et~al.}(2012{\natexlab{a}})\citenamefont {Chatrchyan} \emph
  {et~al.}}]{CMS:2012aa}%
  \BibitemOpen
  \bibfield  {author} {\bibinfo {author} {\bibfnamefont {S.}~\bibnamefont
  {Chatrchyan}} \emph {et~al.} (\bibinfo {collaboration} {CMS}),\ }\href
  {\doibase 10.1140/epjc/s10052-012-1945-x} {\bibfield  {journal} {\bibinfo
  {journal} {Eur. Phys. J. C}\ }\textbf {\bibinfo {volume} {72}},\ \bibinfo
  {pages} {1945} (\bibinfo {year} {2012}{\natexlab{a}})},\ \Eprint
  {http://arxiv.org/abs/1202.2554} {arXiv:1202.2554 [nucl-ex]} \BibitemShut
  {NoStop}%
\bibitem [{\citenamefont {Chatrchyan}\ \emph
  {et~al.}(2012{\natexlab{b}})\citenamefont {Chatrchyan} \emph
  {et~al.}}]{Chatrchyan:2012wg}%
  \BibitemOpen
  \bibfield  {author} {\bibinfo {author} {\bibfnamefont {S.}~\bibnamefont
  {Chatrchyan}} \emph {et~al.} (\bibinfo {collaboration} {CMS}),\ }\href
  {\doibase 10.1140/epjc/s10052-012-2012-3} {\bibfield  {journal} {\bibinfo
  {journal} {Eur. Phys. J. C}\ }\textbf {\bibinfo {volume} {72}},\ \bibinfo
  {pages} {2012} (\bibinfo {year} {2012}{\natexlab{b}})},\ \Eprint
  {http://arxiv.org/abs/1201.3158} {arXiv:1201.3158 [nucl-ex]} \BibitemShut
  {NoStop}%
\bibitem [{\citenamefont {Baier}(2003)}]{Baier:2002tc}%
  \BibitemOpen
  \bibfield  {author} {\bibinfo {author} {\bibfnamefont {R.}~\bibnamefont
  {Baier}},\ }\href {\doibase 10.1016/S0375-9474(02)01429-X} {\bibfield
  {journal} {\bibinfo  {journal} {Nucl. Phys. A}\ }\textbf {\bibinfo {volume}
  {715}},\ \bibinfo {pages} {209} (\bibinfo {year} {2003})},\ \Eprint
  {http://arxiv.org/abs/hep-ph/0209038} {arXiv:hep-ph/0209038 [hep-ph]}
  \BibitemShut {NoStop}%
\bibitem [{\citenamefont {Drees}\ \emph {et~al.}(2005)\citenamefont {Drees},
  \citenamefont {Feng},\ and\ \citenamefont {Jia}}]{Drees:2003zh}%
  \BibitemOpen
  \bibfield  {author} {\bibinfo {author} {\bibfnamefont {A.}~\bibnamefont
  {Drees}}, \bibinfo {author} {\bibfnamefont {H.}~\bibnamefont {Feng}}, \ and\
  \bibinfo {author} {\bibfnamefont {J.}~\bibnamefont {Jia}},\ }\href {\doibase
  10.1103/PhysRevC.71.034909} {\bibfield  {journal} {\bibinfo  {journal} {Phys.
  Rev. C}\ }\textbf {\bibinfo {volume} {71}},\ \bibinfo {pages} {034909}
  (\bibinfo {year} {2005})},\ \Eprint {http://arxiv.org/abs/nucl-th/0310044}
  {arXiv:nucl-th/0310044 [nucl-th]} \BibitemShut {NoStop}%
\bibitem [{\citenamefont {Dainese}\ \emph {et~al.}(2005)\citenamefont
  {Dainese}, \citenamefont {Loizides},\ and\ \citenamefont
  {Paic}}]{Dainese:2004te}%
  \BibitemOpen
  \bibfield  {author} {\bibinfo {author} {\bibfnamefont {A.}~\bibnamefont
  {Dainese}}, \bibinfo {author} {\bibfnamefont {C.}~\bibnamefont {Loizides}}, \
  and\ \bibinfo {author} {\bibfnamefont {G.}~\bibnamefont {Paic}},\ }\href
  {\doibase 10.1140/epjc/s2004-02077-x} {\bibfield  {journal} {\bibinfo
  {journal} {Eur. Phys. J. C}\ }\textbf {\bibinfo {volume} {38}},\ \bibinfo
  {pages} {461} (\bibinfo {year} {2005})},\ \Eprint
  {http://arxiv.org/abs/hep-ph/0406201} {arXiv:hep-ph/0406201 [hep-ph]}
  \BibitemShut {NoStop}%
\bibitem [{\citenamefont {Eskola}\ \emph {et~al.}(2005)\citenamefont {Eskola},
  \citenamefont {Honkanen}, \citenamefont {Salgado},\ and\ \citenamefont
  {Wiedemann}}]{Eskola:2004cr}%
  \BibitemOpen
  \bibfield  {author} {\bibinfo {author} {\bibfnamefont {K.}~\bibnamefont
  {Eskola}}, \bibinfo {author} {\bibfnamefont {H.}~\bibnamefont {Honkanen}},
  \bibinfo {author} {\bibfnamefont {C.}~\bibnamefont {Salgado}}, \ and\
  \bibinfo {author} {\bibfnamefont {U.}~\bibnamefont {Wiedemann}},\ }\href
  {\doibase 10.1016/j.nuclphysa.2004.09.070} {\bibfield  {journal} {\bibinfo
  {journal} {Nucl. Phys. A}\ }\textbf {\bibinfo {volume} {747}},\ \bibinfo
  {pages} {511} (\bibinfo {year} {2005})},\ \Eprint
  {http://arxiv.org/abs/hep-ph/0406319} {arXiv:hep-ph/0406319 [hep-ph]}
  \BibitemShut {NoStop}%
\bibitem [{\citenamefont {Renk}(2006)}]{Renk:2006nd}%
  \BibitemOpen
  \bibfield  {author} {\bibinfo {author} {\bibfnamefont {T.}~\bibnamefont
  {Renk}},\ }\href {\doibase 10.1103/PhysRevC.74.024903} {\bibfield  {journal}
  {\bibinfo  {journal} {Phys. Rev. C}\ }\textbf {\bibinfo {volume} {74}},\
  \bibinfo {pages} {024903} (\bibinfo {year} {2006})},\ \Eprint
  {http://arxiv.org/abs/hep-ph/0602045} {arXiv:hep-ph/0602045 [hep-ph]}
  \BibitemShut {NoStop}%
\bibitem [{\citenamefont {Loizides}(2007)}]{Loizides:2006cs}%
  \BibitemOpen
  \bibfield  {author} {\bibinfo {author} {\bibfnamefont {C.}~\bibnamefont
  {Loizides}},\ }\href {\doibase 10.1140/epjc/s10052-006-0059-8} {\bibfield
  {journal} {\bibinfo  {journal} {Eur. Phys. J. C}\ }\textbf {\bibinfo {volume}
  {49}},\ \bibinfo {pages} {339} (\bibinfo {year} {2007})},\ \Eprint
  {http://arxiv.org/abs/hep-ph/0608133} {arXiv:hep-ph/0608133 [hep-ph]}
  \BibitemShut {NoStop}%
\bibitem [{\citenamefont {Zhang}\ \emph {et~al.}(2007)\citenamefont {Zhang},
  \citenamefont {Owens}, \citenamefont {Wang},\ and\ \citenamefont
  {Wang}}]{Zhang:2007ja}%
  \BibitemOpen
  \bibfield  {author} {\bibinfo {author} {\bibfnamefont {H.}~\bibnamefont
  {Zhang}}, \bibinfo {author} {\bibfnamefont {J.}~\bibnamefont {Owens}},
  \bibinfo {author} {\bibfnamefont {E.}~\bibnamefont {Wang}}, \ and\ \bibinfo
  {author} {\bibfnamefont {X.-N.}\ \bibnamefont {Wang}},\ }\href {\doibase
  10.1103/PhysRevLett.98.212301} {\bibfield  {journal} {\bibinfo  {journal}
  {Phys. Rev. Lett.}\ }\textbf {\bibinfo {volume} {98}},\ \bibinfo {pages}
  {212301} (\bibinfo {year} {2007})},\ \Eprint
  {http://arxiv.org/abs/nucl-th/0701045} {arXiv:nucl-th/0701045 [nucl-th]}
  \BibitemShut {NoStop}%
\bibitem [{\citenamefont {Renk}(2012)}]{Renk:2012cb}%
  \BibitemOpen
  \bibfield  {author} {\bibinfo {author} {\bibfnamefont {T.}~\bibnamefont
  {Renk}},\ }\href {\doibase 10.1103/PhysRevC.86.061901} {\bibfield  {journal}
  {\bibinfo  {journal} {Phys. Rev. C}\ }\textbf {\bibinfo {volume} {86}},\
  \bibinfo {pages} {061901} (\bibinfo {year} {2012})},\ \Eprint
  {http://arxiv.org/abs/1204.5572} {arXiv:1204.5572 [hep-ph]} \BibitemShut
  {NoStop}%
\bibitem [{\citenamefont {Abelev}\ \emph
  {et~al.}(2014{\natexlab{a}})\citenamefont {Abelev} \emph
  {et~al.}}]{Abelev:2013kqa}%
  \BibitemOpen
  \bibfield  {author} {\bibinfo {author} {\bibfnamefont {B.}~\bibnamefont
  {Abelev}} \emph {et~al.} (\bibinfo {collaboration} {ALICE}),\ }\href
  {\doibase 10.1007/JHEP03(2014)013} {\bibfield  {journal} {\bibinfo  {journal}
  {J. High Energy Phys.}\ }\textbf {\bibinfo {volume} {03}},\ \bibinfo {pages}
  {013} (\bibinfo {year} {2014}{\natexlab{a}})},\ \Eprint
  {http://arxiv.org/abs/1311.0633} {arXiv:1311.0633 [nucl-ex]} \BibitemShut
  {NoStop}%
\bibitem [{\citenamefont {Adam}\ \emph
  {et~al.}(2015{\natexlab{a}})\citenamefont {Adam} \emph
  {et~al.}}]{Adam:2015ewa}%
  \BibitemOpen
  \bibfield  {author} {\bibinfo {author} {\bibfnamefont {J.}~\bibnamefont
  {Adam}} \emph {et~al.} (\bibinfo {collaboration} {ALICE}),\ }\href {\doibase
  10.1016/j.physletb.2015.04.039} {\bibfield  {journal} {\bibinfo  {journal}
  {Phys. Lett. B}\ }\textbf {\bibinfo {volume} {746}},\ \bibinfo {pages} {1}
  (\bibinfo {year} {2015}{\natexlab{a}})},\ \Eprint
  {http://arxiv.org/abs/1502.01689} {arXiv:1502.01689 [nucl-ex]} \BibitemShut
  {NoStop}%
\bibitem [{\citenamefont {Acharya}\ \emph
  {et~al.}(2020{\natexlab{a}})\citenamefont {Acharya} \emph
  {et~al.}}]{Acharya:2019jyg}%
  \BibitemOpen
  \bibfield  {author} {\bibinfo {author} {\bibfnamefont {S.}~\bibnamefont
  {Acharya}} \emph {et~al.} (\bibinfo {collaboration} {ALICE}),\ }\href
  {\doibase 10.1103/PhysRevC.101.034911} {\bibfield  {journal} {\bibinfo
  {journal} {Phys. Rev. C}\ }\textbf {\bibinfo {volume} {101}},\ \bibinfo
  {pages} {034911} (\bibinfo {year} {2020}{\natexlab{a}})},\ \Eprint
  {http://arxiv.org/abs/1909.09718} {arXiv:1909.09718 [nucl-ex]} \BibitemShut
  {NoStop}%
\bibitem [{\citenamefont {Aad}\ \emph {et~al.}(2015{\natexlab{b}})\citenamefont
  {Aad} \emph {et~al.}}]{Aad:2014bxa}%
  \BibitemOpen
  \bibfield  {author} {\bibinfo {author} {\bibfnamefont {G.}~\bibnamefont
  {Aad}} \emph {et~al.} (\bibinfo {collaboration} {ATLAS}),\ }\href {\doibase
  10.1103/PhysRevLett.114.072302} {\bibfield  {journal} {\bibinfo  {journal}
  {Phys. Rev. Lett.}\ }\textbf {\bibinfo {volume} {114}},\ \bibinfo {pages}
  {072302} (\bibinfo {year} {2015}{\natexlab{b}})},\ \Eprint
  {http://arxiv.org/abs/1411.2357} {arXiv:1411.2357 [hep-ex]} \BibitemShut
  {NoStop}%
\bibitem [{\citenamefont {Khachatryan}\ \emph
  {et~al.}(2017{\natexlab{b}})\citenamefont {Khachatryan} \emph
  {et~al.}}]{Khachatryan:2016jfl}%
  \BibitemOpen
  \bibfield  {author} {\bibinfo {author} {\bibfnamefont {V.}~\bibnamefont
  {Khachatryan}} \emph {et~al.} (\bibinfo {collaboration} {CMS}),\ }\href
  {\doibase 10.1103/PhysRevC.96.015202} {\bibfield  {journal} {\bibinfo
  {journal} {Phys. Rev. C}\ }\textbf {\bibinfo {volume} {96}},\ \bibinfo
  {pages} {015202} (\bibinfo {year} {2017}{\natexlab{b}})},\ \Eprint
  {http://arxiv.org/abs/1609.05383} {arXiv:1609.05383 [nucl-ex]} \BibitemShut
  {NoStop}%
\bibitem [{\citenamefont {Adam}\ \emph
  {et~al.}(2015{\natexlab{b}})\citenamefont {Adam} \emph
  {et~al.}}]{Adam:2015doa}%
  \BibitemOpen
  \bibfield  {author} {\bibinfo {author} {\bibfnamefont {J.}~\bibnamefont
  {Adam}} \emph {et~al.} (\bibinfo {collaboration} {ALICE}),\ }\href {\doibase
  10.1007/JHEP09(2015)170} {\bibfield  {journal} {\bibinfo  {journal} {J. High
  Energy Phys.}\ }\textbf {\bibinfo {volume} {09}},\ \bibinfo {pages} {170}
  (\bibinfo {year} {2015}{\natexlab{b}})},\ \Eprint
  {http://arxiv.org/abs/1506.03984} {arXiv:1506.03984 [nucl-ex]} \BibitemShut
  {NoStop}%
\bibitem [{\citenamefont {Aad}\ \emph {et~al.}(2010)\citenamefont {Aad} \emph
  {et~al.}}]{Aad:2010bu}%
  \BibitemOpen
  \bibfield  {author} {\bibinfo {author} {\bibfnamefont {G.}~\bibnamefont
  {Aad}} \emph {et~al.} (\bibinfo {collaboration} {ATLAS}),\ }\href {\doibase
  10.1103/PhysRevLett.105.252303} {\bibfield  {journal} {\bibinfo  {journal}
  {Phys. Rev. Lett.}\ }\textbf {\bibinfo {volume} {105}},\ \bibinfo {pages}
  {252303} (\bibinfo {year} {2010})},\ \Eprint {http://arxiv.org/abs/1011.6182}
  {arXiv:1011.6182 [hep-ex]} \BibitemShut {NoStop}%
\bibitem [{\citenamefont {Chatrchyan}\ \emph
  {et~al.}(2012{\natexlab{c}})\citenamefont {Chatrchyan} \emph
  {et~al.}}]{Chatrchyan:2012nia}%
  \BibitemOpen
  \bibfield  {author} {\bibinfo {author} {\bibfnamefont {S.}~\bibnamefont
  {Chatrchyan}} \emph {et~al.} (\bibinfo {collaboration} {CMS}),\ }\href
  {\doibase 10.1016/j.physletb.2012.04.058} {\bibfield  {journal} {\bibinfo
  {journal} {Phys. Lett. B}\ }\textbf {\bibinfo {volume} {712}},\ \bibinfo
  {pages} {176} (\bibinfo {year} {2012}{\natexlab{c}})},\ \Eprint
  {http://arxiv.org/abs/1202.5022} {arXiv:1202.5022 [nucl-ex]} \BibitemShut
  {NoStop}%
\bibitem [{\citenamefont {Chatrchyan}\ \emph {et~al.}(2013)\citenamefont
  {Chatrchyan} \emph {et~al.}}]{Chatrchyan:2012gt}%
  \BibitemOpen
  \bibfield  {author} {\bibinfo {author} {\bibfnamefont {S.}~\bibnamefont
  {Chatrchyan}} \emph {et~al.} (\bibinfo {collaboration} {CMS}),\ }\href
  {\doibase 10.1016/j.physletb.2012.11.003} {\bibfield  {journal} {\bibinfo
  {journal} {Phys. Lett. B}\ }\textbf {\bibinfo {volume} {718}},\ \bibinfo
  {pages} {773} (\bibinfo {year} {2013})},\ \Eprint
  {http://arxiv.org/abs/1205.0206} {arXiv:1205.0206 [nucl-ex]} \BibitemShut
  {NoStop}%
\bibitem [{\citenamefont {Sirunyan}\ \emph
  {et~al.}(2018{\natexlab{a}})\citenamefont {Sirunyan} \emph
  {et~al.}}]{Sirunyan:2018qec}%
  \BibitemOpen
  \bibfield  {author} {\bibinfo {author} {\bibfnamefont {A.~M.}\ \bibnamefont
  {Sirunyan}} \emph {et~al.} (\bibinfo {collaboration} {CMS}),\ }\href
  {\doibase 10.1103/PhysRevLett.121.242301} {\bibfield  {journal} {\bibinfo
  {journal} {Phys. Rev. Lett.}\ }\textbf {\bibinfo {volume} {121}},\ \bibinfo
  {pages} {242301} (\bibinfo {year} {2018}{\natexlab{a}})},\ \Eprint
  {http://arxiv.org/abs/1801.04895} {arXiv:1801.04895 [hep-ex]} \BibitemShut
  {NoStop}%
\bibitem [{\citenamefont {Acharya}\ \emph {et~al.}(2018)\citenamefont {Acharya}
  \emph {et~al.}}]{Acharya:2017goa}%
  \BibitemOpen
  \bibfield  {author} {\bibinfo {author} {\bibfnamefont {S.}~\bibnamefont
  {Acharya}} \emph {et~al.} (\bibinfo {collaboration} {ALICE}),\ }\href
  {\doibase 10.1016/j.physletb.2017.11.044} {\bibfield  {journal} {\bibinfo
  {journal} {Phys. Lett. B}\ }\textbf {\bibinfo {volume} {776}},\ \bibinfo
  {pages} {249} (\bibinfo {year} {2018})},\ \Eprint
  {http://arxiv.org/abs/1702.00804} {arXiv:1702.00804 [nucl-ex]} \BibitemShut
  {NoStop}%
\bibitem [{\citenamefont {Acharya}\ \emph
  {et~al.}(2020{\natexlab{b}})\citenamefont {Acharya} \emph
  {et~al.}}]{Acharya:2019djg}%
  \BibitemOpen
  \bibfield  {author} {\bibinfo {author} {\bibfnamefont {S.}~\bibnamefont
  {Acharya}} \emph {et~al.} (\bibinfo {collaboration} {ALICE}),\ }\href
  {\doibase 10.1016/j.physletb.2020.135227} {\bibfield  {journal} {\bibinfo
  {journal} {Phys. Lett. B}\ }\textbf {\bibinfo {volume} {802}},\ \bibinfo
  {pages} {135227} (\bibinfo {year} {2020}{\natexlab{b}})},\ \Eprint
  {http://arxiv.org/abs/1905.02512} {arXiv:1905.02512 [nucl-ex]} \BibitemShut
  {NoStop}%
\bibitem [{\citenamefont {Sirunyan}\ \emph
  {et~al.}(2018{\natexlab{b}})\citenamefont {Sirunyan} \emph
  {et~al.}}]{Sirunyan:2017bsd}%
  \BibitemOpen
  \bibfield  {author} {\bibinfo {author} {\bibfnamefont {A.~M.}\ \bibnamefont
  {Sirunyan}} \emph {et~al.} (\bibinfo {collaboration} {CMS}),\ }\href
  {\doibase 10.1103/PhysRevLett.120.142302} {\bibfield  {journal} {\bibinfo
  {journal} {Phys. Rev. Lett.}\ }\textbf {\bibinfo {volume} {120}},\ \bibinfo
  {pages} {142302} (\bibinfo {year} {2018}{\natexlab{b}})},\ \Eprint
  {http://arxiv.org/abs/1708.09429} {arXiv:1708.09429 [nucl-ex]} \BibitemShut
  {NoStop}%
\bibitem [{\citenamefont {Adamczyk}\ \emph
  {et~al.}(2017{\natexlab{b}})\citenamefont {Adamczyk} \emph
  {et~al.}}]{Adamczyk:2016fqm}%
  \BibitemOpen
  \bibfield  {author} {\bibinfo {author} {\bibfnamefont {L.}~\bibnamefont
  {Adamczyk}} \emph {et~al.} (\bibinfo {collaboration} {STAR}),\ }\href
  {\doibase 10.1103/PhysRevLett.119.062301} {\bibfield  {journal} {\bibinfo
  {journal} {Phys. Rev. Lett.}\ }\textbf {\bibinfo {volume} {119}},\ \bibinfo
  {pages} {062301} (\bibinfo {year} {2017}{\natexlab{b}})},\ \Eprint
  {http://arxiv.org/abs/1609.03878} {arXiv:1609.03878 [nucl-ex]} \BibitemShut
  {NoStop}%
\bibitem [{\citenamefont {Adamczyk}\ \emph
  {et~al.}(2017{\natexlab{c}})\citenamefont {Adamczyk} \emph
  {et~al.}}]{Adamczyk:2017yhe}%
  \BibitemOpen
  \bibfield  {author} {\bibinfo {author} {\bibfnamefont {L.}~\bibnamefont
  {Adamczyk}} \emph {et~al.} (\bibinfo {collaboration} {STAR}),\ }\href
  {\doibase 10.1103/PhysRevC.96.024905} {\bibfield  {journal} {\bibinfo
  {journal} {Phys. Rev. C}\ }\textbf {\bibinfo {volume} {96}},\ \bibinfo
  {pages} {024905} (\bibinfo {year} {2017}{\natexlab{c}})},\ \Eprint
  {http://arxiv.org/abs/1702.01108} {arXiv:1702.01108 [nucl-ex]} \BibitemShut
  {NoStop}%
\bibitem [{\citenamefont {Cacciari}\ \emph
  {et~al.}(2008{\natexlab{a}})\citenamefont {Cacciari}, \citenamefont {Salam},\
  and\ \citenamefont {Soyez}}]{FastJetAntikt}%
  \BibitemOpen
  \bibfield  {author} {\bibinfo {author} {\bibfnamefont {M.}~\bibnamefont
  {Cacciari}}, \bibinfo {author} {\bibfnamefont {G.~P.}\ \bibnamefont {Salam}},
  \ and\ \bibinfo {author} {\bibfnamefont {G.}~\bibnamefont {Soyez}},\ }\href
  {\doibase 10.1088/1126-6708/2008/04/063} {\bibfield  {journal} {\bibinfo
  {journal} {J. High Energy Phys.}\ }\textbf {\bibinfo {volume} {04}},\
  \bibinfo {pages} {063} (\bibinfo {year} {2008}{\natexlab{a}})},\ \Eprint
  {http://arxiv.org/abs/0802.1189} {arXiv:0802.1189 [hep-ph]} \BibitemShut
  {NoStop}%
\bibitem [{\citenamefont {Skands}(2010)}]{Skands:2010ak}%
  \BibitemOpen
  \bibfield  {author} {\bibinfo {author} {\bibfnamefont {P.~Z.}\ \bibnamefont
  {Skands}},\ }\href {\doibase 10.1103/PhysRevD.82.074018} {\bibfield
  {journal} {\bibinfo  {journal} {Phys. Rev. D}\ }\textbf {\bibinfo {volume}
  {82}},\ \bibinfo {pages} {074018} (\bibinfo {year} {2010})},\ \Eprint
  {http://arxiv.org/abs/1005.3457} {arXiv:1005.3457 [hep-ph]} \BibitemShut
  {NoStop}%
\bibitem [{\citenamefont {Adam}\ \emph {et~al.}(2019)\citenamefont {Adam} \emph
  {et~al.}}]{Adam:2019aml}%
  \BibitemOpen
  \bibfield  {author} {\bibinfo {author} {\bibfnamefont {J.}~\bibnamefont
  {Adam}} \emph {et~al.} (\bibinfo {collaboration} {STAR}),\ }\href {\doibase
  10.1103/PhysRevD.100.052005} {\bibfield  {journal} {\bibinfo  {journal}
  {Phys. Rev. D}\ }\textbf {\bibinfo {volume} {100}},\ \bibinfo {pages}
  {052005} (\bibinfo {year} {2019})},\ \Eprint
  {http://arxiv.org/abs/1906.02740} {arXiv:1906.02740 [hep-ex]} \BibitemShut
  {NoStop}%
\bibitem [{\citenamefont {Ackermann}\ \emph {et~al.}(2003)\citenamefont
  {Ackermann} \emph {et~al.}}]{Ackermann:2002ad}%
  \BibitemOpen
  \bibfield  {author} {\bibinfo {author} {\bibfnamefont {K.~H.}\ \bibnamefont
  {Ackermann}} \emph {et~al.} (\bibinfo {collaboration} {STAR}),\ }\href
  {\doibase 10.1016/S0168-9002(02)01960-5} {\bibfield  {journal} {\bibinfo
  {journal} {Nucl. Instrum. Meth. A}\ }\textbf {\bibinfo {volume} {499}},\
  \bibinfo {pages} {624} (\bibinfo {year} {2003})}\BibitemShut {NoStop}%
\bibitem [{\citenamefont {Anderson}\ \emph {et~al.}(2003)\citenamefont
  {Anderson} \emph {et~al.}}]{Anderson:2003ur}%
  \BibitemOpen
  \bibfield  {author} {\bibinfo {author} {\bibfnamefont {M.}~\bibnamefont
  {Anderson}} \emph {et~al.},\ }\href {\doibase 10.1016/S0168-9002(02)01964-2}
  {\bibfield  {journal} {\bibinfo  {journal} {Nucl. Instrum. Meth. A}\ }\textbf
  {\bibinfo {volume} {499}},\ \bibinfo {pages} {659} (\bibinfo {year}
  {2003})},\ \Eprint {http://arxiv.org/abs/nucl-ex/0301015}
  {arXiv:nucl-ex/0301015 [nucl-ex]} \BibitemShut {NoStop}%
\bibitem [{\citenamefont {Adams}\ \emph
  {et~al.}(2005{\natexlab{b}})\citenamefont {Adams} \emph
  {et~al.}}]{Adams:2003qm}%
  \BibitemOpen
  \bibfield  {author} {\bibinfo {author} {\bibfnamefont {J.}~\bibnamefont
  {Adams}} \emph {et~al.} (\bibinfo {collaboration} {STAR}),\ }\href {\doibase
  10.1016/j.physletb.2005.04.041} {\bibfield  {journal} {\bibinfo  {journal}
  {Phys. Lett. B}\ }\textbf {\bibinfo {volume} {616}},\ \bibinfo {pages} {8}
  (\bibinfo {year} {2005}{\natexlab{b}})},\ \Eprint
  {http://arxiv.org/abs/nucl-ex/0309012} {arXiv:nucl-ex/0309012 [nucl-ex]}
  \BibitemShut {NoStop}%
\bibitem [{\citenamefont {Agakishiev}\ \emph
  {et~al.}(2012{\natexlab{a}})\citenamefont {Agakishiev} \emph
  {et~al.}}]{Agakishiev:2011dc}%
  \BibitemOpen
  \bibfield  {author} {\bibinfo {author} {\bibfnamefont {G.}~\bibnamefont
  {Agakishiev}} \emph {et~al.} (\bibinfo {collaboration} {STAR}),\ }\href
  {\doibase 10.1103/PhysRevLett.108.072302} {\bibfield  {journal} {\bibinfo
  {journal} {Phys. Rev. Lett.}\ }\textbf {\bibinfo {volume} {108}},\ \bibinfo
  {pages} {072302} (\bibinfo {year} {2012}{\natexlab{a}})},\ \Eprint
  {http://arxiv.org/abs/1110.0579} {arXiv:1110.0579 [nucl-ex]} \BibitemShut
  {NoStop}%
\bibitem [{\citenamefont {Abelev}\ \emph
  {et~al.}(2006{\natexlab{b}})\citenamefont {Abelev} \emph
  {et~al.}}]{Abelev:2006jr}%
  \BibitemOpen
  \bibfield  {author} {\bibinfo {author} {\bibfnamefont {B.~I.}\ \bibnamefont
  {Abelev}} \emph {et~al.} (\bibinfo {collaboration} {STAR}),\ }\href {\doibase
  10.1103/PhysRevLett.97.152301} {\bibfield  {journal} {\bibinfo  {journal}
  {Phys. Rev. Lett.}\ }\textbf {\bibinfo {volume} {97}},\ \bibinfo {pages}
  {152301} (\bibinfo {year} {2006}{\natexlab{b}})},\ \Eprint
  {http://arxiv.org/abs/nucl-ex/0606003} {arXiv:nucl-ex/0606003 [nucl-ex]}
  \BibitemShut {NoStop}%
\bibitem [{\citenamefont {Agakishiev}\ \emph
  {et~al.}(2012{\natexlab{b}})\citenamefont {Agakishiev} \emph
  {et~al.}}]{Agakishiev:2011ar}%
  \BibitemOpen
  \bibfield  {author} {\bibinfo {author} {\bibfnamefont {G.}~\bibnamefont
  {Agakishiev}} \emph {et~al.} (\bibinfo {collaboration} {STAR}),\ }\href
  {\doibase 10.1103/PhysRevLett.108.072301} {\bibfield  {journal} {\bibinfo
  {journal} {Phys. Rev. Lett.}\ }\textbf {\bibinfo {volume} {108}},\ \bibinfo
  {pages} {072301} (\bibinfo {year} {2012}{\natexlab{b}})},\ \Eprint
  {http://arxiv.org/abs/1107.2955} {arXiv:1107.2955 [nucl-ex]} \BibitemShut
  {NoStop}%
\bibitem [{\citenamefont {Miller}\ \emph {et~al.}(2007)\citenamefont {Miller},
  \citenamefont {Reygers}, \citenamefont {Sanders},\ and\ \citenamefont
  {Steinberg}}]{Miller:2007ri}%
  \BibitemOpen
  \bibfield  {author} {\bibinfo {author} {\bibfnamefont {M.~L.}\ \bibnamefont
  {Miller}}, \bibinfo {author} {\bibfnamefont {K.}~\bibnamefont {Reygers}},
  \bibinfo {author} {\bibfnamefont {S.~J.}\ \bibnamefont {Sanders}}, \ and\
  \bibinfo {author} {\bibfnamefont {P.}~\bibnamefont {Steinberg}},\ }\href
  {\doibase 10.1146/annurev.nucl.57.090506.123020} {\bibfield  {journal}
  {\bibinfo  {journal} {Annu. Rev. Nucl. Part. Sci.}\ }\textbf {\bibinfo
  {volume} {57}},\ \bibinfo {pages} {205} (\bibinfo {year} {2007})},\ \Eprint
  {http://arxiv.org/abs/nucl-ex/0701025} {arXiv:nucl-ex/0701025 [nucl-ex]}
  \BibitemShut {NoStop}%
\bibitem [{\citenamefont {de~Barros}\ \emph {et~al.}(2013)\citenamefont
  {de~Barros}, \citenamefont {Fenton-Olsen}, \citenamefont {Jacobs},\ and\
  \citenamefont {Ploskon}}]{deBarros:2012ws}%
  \BibitemOpen
  \bibfield  {author} {\bibinfo {author} {\bibfnamefont {G.}~\bibnamefont
  {de~Barros}}, \bibinfo {author} {\bibfnamefont {B.}~\bibnamefont
  {Fenton-Olsen}}, \bibinfo {author} {\bibfnamefont {P.}~\bibnamefont
  {Jacobs}}, \ and\ \bibinfo {author} {\bibfnamefont {M.}~\bibnamefont
  {Ploskon}},\ }\href {\doibase 10.1016/j.nuclphysa.2012.12.019} {\bibfield
  {journal} {\bibinfo  {journal} {Nucl. Phys. A}\ }\textbf {\bibinfo {volume}
  {910-911}},\ \bibinfo {pages} {314} (\bibinfo {year} {2013})},\ \Eprint
  {http://arxiv.org/abs/1208.1518} {arXiv:1208.1518 [hep-ex]} \BibitemShut
  {NoStop}%
\bibitem [{\citenamefont {Cacciari}\ \emph {et~al.}(2012)\citenamefont
  {Cacciari}, \citenamefont {Salam},\ and\ \citenamefont
  {Soyez}}]{Cacciari:2011ma}%
  \BibitemOpen
  \bibfield  {author} {\bibinfo {author} {\bibfnamefont {M.}~\bibnamefont
  {Cacciari}}, \bibinfo {author} {\bibfnamefont {G.~P.}\ \bibnamefont {Salam}},
  \ and\ \bibinfo {author} {\bibfnamefont {G.}~\bibnamefont {Soyez}},\ }\href
  {\doibase 10.1140/epjc/s10052-012-1896-2} {\bibfield  {journal} {\bibinfo
  {journal} {Eur. Phys. J. C}\ }\textbf {\bibinfo {volume} {72}},\ \bibinfo
  {pages} {1896} (\bibinfo {year} {2012})},\ \Eprint
  {http://arxiv.org/abs/1111.6097} {arXiv:1111.6097 [hep-ph]} \BibitemShut
  {NoStop}%
\bibitem [{\citenamefont {Cacciari}\ \emph
  {et~al.}(2008{\natexlab{b}})\citenamefont {Cacciari}, \citenamefont {Salam},\
  and\ \citenamefont {Soyez}}]{FastJetArea}%
  \BibitemOpen
  \bibfield  {author} {\bibinfo {author} {\bibfnamefont {M.}~\bibnamefont
  {Cacciari}}, \bibinfo {author} {\bibfnamefont {G.~P.}\ \bibnamefont {Salam}},
  \ and\ \bibinfo {author} {\bibfnamefont {G.}~\bibnamefont {Soyez}},\ }\href
  {\doibase 10.1088/1126-6708/2008/04/005} {\bibfield  {journal} {\bibinfo
  {journal} {J. High Energy Phys.}\ }\textbf {\bibinfo {volume} {04}},\
  \bibinfo {pages} {005} (\bibinfo {year} {2008}{\natexlab{b}})},\ \Eprint
  {http://arxiv.org/abs/0802.1188} {arXiv:0802.1188 [hep-ph]} \BibitemShut
  {NoStop}%
\bibitem [{\citenamefont {Cacciari}\ and\ \citenamefont
  {Salam}(2008)}]{FastJetPileup}%
  \BibitemOpen
  \bibfield  {author} {\bibinfo {author} {\bibfnamefont {M.}~\bibnamefont
  {Cacciari}}\ and\ \bibinfo {author} {\bibfnamefont {G.~P.}\ \bibnamefont
  {Salam}},\ }\href {\doibase 10.1016/j.physletb.2007.09.077} {\bibfield
  {journal} {\bibinfo  {journal} {Phys. Lett.}\ }\textbf {\bibinfo {volume}
  {B659}},\ \bibinfo {pages} {119} (\bibinfo {year} {2008})},\ \Eprint
  {http://arxiv.org/abs/0707.1378} {arXiv:0707.1378 [hep-ph]} \BibitemShut
  {NoStop}%
\bibitem [{\citenamefont {Cowan}(2002)}]{Cowan:2002in}%
  \BibitemOpen
  \bibfield  {author} {\bibinfo {author} {\bibfnamefont {G.}~\bibnamefont
  {Cowan}},\ }\href@noop {} {\bibfield  {journal} {\bibinfo  {journal} {Conf.
  Proc. C}\ }\textbf {\bibinfo {volume} {0203181}},\ \bibinfo {pages} {248}
  (\bibinfo {year} {2002})}\BibitemShut {NoStop}%
\bibitem [{\citenamefont {H{\"o}cker}\ and\ \citenamefont
  {Kartvelishvili}(1996)}]{Hocker:1995kb}%
  \BibitemOpen
  \bibfield  {author} {\bibinfo {author} {\bibfnamefont {A.}~\bibnamefont
  {H{\"o}cker}}\ and\ \bibinfo {author} {\bibfnamefont {V.}~\bibnamefont
  {Kartvelishvili}},\ }\href {\doibase 10.1016/0168-9002(95)01478-0} {\bibfield
   {journal} {\bibinfo  {journal} {Nucl. Instrum. Meth. A}\ }\textbf {\bibinfo
  {volume} {372}},\ \bibinfo {pages} {469} (\bibinfo {year} {1996})},\ \Eprint
  {http://arxiv.org/abs/hep-ph/9509307} {arXiv:hep-ph/9509307 [hep-ph]}
  \BibitemShut {NoStop}%
\bibitem [{\citenamefont {{D'Agostini}}(2010)}]{dAgostini2010}%
  \BibitemOpen
  \bibfield  {author} {\bibinfo {author} {\bibfnamefont {G.}~\bibnamefont
  {{D'Agostini}}},\ }\href@noop {} {\bibfield  {journal} {\bibinfo  {journal}
  {ArXiv e-prints}\ } (\bibinfo {year} {2010})},\ \Eprint
  {http://arxiv.org/abs/1010.0632} {arXiv:1010.0632 [physics.data-an]}
  \BibitemShut {NoStop}%
\bibitem [{\citenamefont {Jacobs}(2011)}]{Jacobs:2010wq}%
  \BibitemOpen
  \bibfield  {author} {\bibinfo {author} {\bibfnamefont {P.}~\bibnamefont
  {Jacobs}} (\bibinfo {collaboration} {STAR}),\ }\href {\doibase
  10.1016/j.nuclphysa.2011.02.064} {\bibfield  {journal} {\bibinfo  {journal}
  {Nucl. Phys. A}\ }\textbf {\bibinfo {volume} {855}},\ \bibinfo {pages} {299}
  (\bibinfo {year} {2011})},\ \Eprint {http://arxiv.org/abs/1012.2406}
  {arXiv:1012.2406 [nucl-ex]} \BibitemShut {NoStop}%
\bibitem [{\citenamefont {Adare}\ \emph {et~al.}(2015)\citenamefont {Adare}
  \emph {et~al.}}]{Adare:2014bga}%
  \BibitemOpen
  \bibfield  {author} {\bibinfo {author} {\bibfnamefont {A.}~\bibnamefont
  {Adare}} \emph {et~al.} (\bibinfo {collaboration} {PHENIX}),\ }\href
  {\doibase 10.1103/PhysRevC.92.034913} {\bibfield  {journal} {\bibinfo
  {journal} {Phys. Rev. C}\ }\textbf {\bibinfo {volume} {92}},\ \bibinfo
  {pages} {034913} (\bibinfo {year} {2015})},\ \Eprint
  {http://arxiv.org/abs/1412.1043} {arXiv:1412.1043 [nucl-ex]} \BibitemShut
  {NoStop}%
\bibitem [{\citenamefont {D'Agostini}(1995)}]{D'Agostini:1994zf}%
  \BibitemOpen
  \bibfield  {author} {\bibinfo {author} {\bibfnamefont {G.}~\bibnamefont
  {D'Agostini}},\ }\href {\doibase 10.1016/0168-9002(95)00274-X} {\bibfield
  {journal} {\bibinfo  {journal} {Nucl. Instrum. Meth. A}\ }\textbf {\bibinfo
  {volume} {362}},\ \bibinfo {pages} {487} (\bibinfo {year}
  {1995})}\BibitemShut {NoStop}%
\bibitem [{\citenamefont {Adare}\ \emph {et~al.}(2011)\citenamefont {Adare}
  \emph {et~al.}}]{Adare:2010fe}%
  \BibitemOpen
  \bibfield  {author} {\bibinfo {author} {\bibfnamefont {A.}~\bibnamefont
  {Adare}} \emph {et~al.} (\bibinfo {collaboration} {PHENIX}),\ }\href
  {\doibase 10.1103/PhysRevD.83.052004} {\bibfield  {journal} {\bibinfo
  {journal} {Phys. Rev. D}\ }\textbf {\bibinfo {volume} {83}},\ \bibinfo
  {pages} {052004} (\bibinfo {year} {2011})},\ \Eprint
  {http://arxiv.org/abs/1005.3674} {arXiv:1005.3674 [hep-ex]} \BibitemShut
  {NoStop}%
\bibitem [{\citenamefont {Adamczyk}\ \emph {et~al.}(2015)\citenamefont
  {Adamczyk} \emph {et~al.}}]{Adamczyk:2015lme}%
  \BibitemOpen
  \bibfield  {author} {\bibinfo {author} {\bibfnamefont {L.}~\bibnamefont
  {Adamczyk}} \emph {et~al.} (\bibinfo {collaboration} {STAR}),\ }\href
  {\doibase 10.1103/PhysRevC.92.024912} {\bibfield  {journal} {\bibinfo
  {journal} {Phys. Rev. C}\ }\textbf {\bibinfo {volume} {92}},\ \bibinfo
  {pages} {024912} (\bibinfo {year} {2015})},\ \Eprint
  {http://arxiv.org/abs/1504.01317} {arXiv:1504.01317 [hep-ex]} \BibitemShut
  {NoStop}%
\bibitem [{\citenamefont {Adcox}\ \emph
  {et~al.}(2002{\natexlab{b}})\citenamefont {Adcox} \emph
  {et~al.}}]{Adcox:2002pa}%
  \BibitemOpen
  \bibfield  {author} {\bibinfo {author} {\bibfnamefont {K.}~\bibnamefont
  {Adcox}} \emph {et~al.} (\bibinfo {collaboration} {PHENIX}),\ }\href
  {\doibase 10.1103/PhysRevC.66.024901} {\bibfield  {journal} {\bibinfo
  {journal} {Phys. Rev. C}\ }\textbf {\bibinfo {volume} {66}},\ \bibinfo
  {pages} {024901} (\bibinfo {year} {2002}{\natexlab{b}})},\ \Eprint
  {http://arxiv.org/abs/nucl-ex/0203015} {arXiv:nucl-ex/0203015 [nucl-ex]}
  \BibitemShut {NoStop}%
\bibitem [{\citenamefont {Adams}\ \emph
  {et~al.}(2005{\natexlab{c}})\citenamefont {Adams} \emph
  {et~al.}}]{Adams:2003uw}%
  \BibitemOpen
  \bibfield  {author} {\bibinfo {author} {\bibfnamefont {J.}~\bibnamefont
  {Adams}} \emph {et~al.} (\bibinfo {collaboration} {STAR}),\ }\href {\doibase
  10.1103/PhysRevC.71.064906} {\bibfield  {journal} {\bibinfo  {journal} {Phys.
  Rev. C}\ }\textbf {\bibinfo {volume} {71}},\ \bibinfo {pages} {064906}
  (\bibinfo {year} {2005}{\natexlab{c}})},\ \Eprint
  {http://arxiv.org/abs/nucl-ex/0308033} {arXiv:nucl-ex/0308033 [nucl-ex]}
  \BibitemShut {NoStop}%
\bibitem [{\citenamefont {Adamczyk}\ \emph {et~al.}(2013)\citenamefont
  {Adamczyk} \emph {et~al.}}]{Adamczyk:2013up}%
  \BibitemOpen
  \bibfield  {author} {\bibinfo {author} {\bibfnamefont {L.}~\bibnamefont
  {Adamczyk}} \emph {et~al.} (\bibinfo {collaboration} {STAR}),\ }\href
  {\doibase 10.1103/PhysRevC.87.064902} {\bibfield  {journal} {\bibinfo
  {journal} {Phys. Rev. C}\ }\textbf {\bibinfo {volume} {87}},\ \bibinfo
  {pages} {064902} (\bibinfo {year} {2013})},\ \Eprint
  {http://arxiv.org/abs/1301.6633} {arXiv:1301.6633 [nucl-ex]} \BibitemShut
  {NoStop}%
\bibitem [{\citenamefont {Appelshauser}\ \emph {et~al.}(1999)\citenamefont
  {Appelshauser} \emph {et~al.}}]{Appelshauser:1999ft}%
  \BibitemOpen
  \bibfield  {author} {\bibinfo {author} {\bibfnamefont {H.}~\bibnamefont
  {Appelshauser}} \emph {et~al.} (\bibinfo {collaboration} {NA49}),\ }\href
  {\doibase 10.1016/S0370-2693(99)00673-5} {\bibfield  {journal} {\bibinfo
  {journal} {Phys. Lett. B}\ }\textbf {\bibinfo {volume} {459}},\ \bibinfo
  {pages} {679} (\bibinfo {year} {1999})},\ \Eprint
  {http://arxiv.org/abs/hep-ex/9904014} {arXiv:hep-ex/9904014 [hep-ex]}
  \BibitemShut {NoStop}%
\bibitem [{\citenamefont {Adamova}\ \emph {et~al.}(2003)\citenamefont {Adamova}
  \emph {et~al.}}]{Adamova:2003pz}%
  \BibitemOpen
  \bibfield  {author} {\bibinfo {author} {\bibfnamefont {D.}~\bibnamefont
  {Adamova}} \emph {et~al.} (\bibinfo {collaboration} {CERES}),\ }\href
  {\doibase 10.1016/j.nuclphysa.2003.07.018} {\bibfield  {journal} {\bibinfo
  {journal} {Nucl. Phys. A}\ }\textbf {\bibinfo {volume} {727}},\ \bibinfo
  {pages} {97} (\bibinfo {year} {2003})},\ \Eprint
  {http://arxiv.org/abs/nucl-ex/0305002} {arXiv:nucl-ex/0305002 [nucl-ex]}
  \BibitemShut {NoStop}%
\bibitem [{\citenamefont {Adler}\ \emph
  {et~al.}(2004{\natexlab{a}})\citenamefont {Adler} \emph
  {et~al.}}]{Adler:2003xq}%
  \BibitemOpen
  \bibfield  {author} {\bibinfo {author} {\bibfnamefont {S.~S.}\ \bibnamefont
  {Adler}} \emph {et~al.} (\bibinfo {collaboration} {PHENIX}),\ }\href
  {\doibase 10.1103/PhysRevLett.93.092301} {\bibfield  {journal} {\bibinfo
  {journal} {Phys. Rev. Lett.}\ }\textbf {\bibinfo {volume} {93}},\ \bibinfo
  {pages} {092301} (\bibinfo {year} {2004}{\natexlab{a}})},\ \Eprint
  {http://arxiv.org/abs/nucl-ex/0310005} {arXiv:nucl-ex/0310005 [nucl-ex]}
  \BibitemShut {NoStop}%
\bibitem [{\citenamefont {Abelev}\ \emph
  {et~al.}(2014{\natexlab{b}})\citenamefont {Abelev} \emph
  {et~al.}}]{Abelev:2014ckr}%
  \BibitemOpen
  \bibfield  {author} {\bibinfo {author} {\bibfnamefont {B.~B.}\ \bibnamefont
  {Abelev}} \emph {et~al.} (\bibinfo {collaboration} {ALICE}),\ }\href
  {\doibase 10.1140/epjc/s10052-014-3077-y} {\bibfield  {journal} {\bibinfo
  {journal} {Eur. Phys. J. C}\ }\textbf {\bibinfo {volume} {74}},\ \bibinfo
  {pages} {3077} (\bibinfo {year} {2014}{\natexlab{b}})},\ \Eprint
  {http://arxiv.org/abs/1407.5530} {arXiv:1407.5530 [nucl-ex]} \BibitemShut
  {NoStop}%
\bibitem [{\citenamefont {Wang}\ and\ \citenamefont
  {Gyulassy}(1991)}]{Wang:1991hta}%
  \BibitemOpen
  \bibfield  {author} {\bibinfo {author} {\bibfnamefont {X.-N.}\ \bibnamefont
  {Wang}}\ and\ \bibinfo {author} {\bibfnamefont {M.}~\bibnamefont
  {Gyulassy}},\ }\href {\doibase 10.1103/PhysRevD.44.3501} {\bibfield
  {journal} {\bibinfo  {journal} {Phys. Rev. D}\ }\textbf {\bibinfo {volume}
  {44}},\ \bibinfo {pages} {3501} (\bibinfo {year} {1991})}\BibitemShut
  {NoStop}%
\bibitem [{\citenamefont {Lokhtin}\ \emph {et~al.}(2011)\citenamefont
  {Lokhtin}, \citenamefont {Belyaev},\ and\ \citenamefont
  {Snigirev}}]{Lokhtin:2011qq}%
  \BibitemOpen
  \bibfield  {author} {\bibinfo {author} {\bibfnamefont {I.}~\bibnamefont
  {Lokhtin}}, \bibinfo {author} {\bibfnamefont {A.}~\bibnamefont {Belyaev}}, \
  and\ \bibinfo {author} {\bibfnamefont {A.}~\bibnamefont {Snigirev}},\ }\href
  {\doibase 10.1140/epjc/s10052-011-1650-1} {\bibfield  {journal} {\bibinfo
  {journal} {Eur. Phys. J. C}\ }\textbf {\bibinfo {volume} {71}},\ \bibinfo
  {pages} {1650} (\bibinfo {year} {2011})},\ \Eprint
  {http://arxiv.org/abs/1103.1853} {arXiv:1103.1853 [hep-ph]} \BibitemShut
  {NoStop}%
\bibitem [{\citenamefont {Tannenbaum}(2001)}]{Tannenbaum:2001gs}%
  \BibitemOpen
  \bibfield  {author} {\bibinfo {author} {\bibfnamefont {M.}~\bibnamefont
  {Tannenbaum}},\ }\href {\doibase 10.1016/S0370-2693(00)01325-3} {\bibfield
  {journal} {\bibinfo  {journal} {Phys. Lett. B}\ }\textbf {\bibinfo {volume}
  {498}},\ \bibinfo {pages} {29} (\bibinfo {year} {2001})}\BibitemShut
  {NoStop}%
\bibitem [{\citenamefont {Abelev}\ \emph {et~al.}(2009)\citenamefont {Abelev}
  \emph {et~al.}}]{Abelev:2008ab}%
  \BibitemOpen
  \bibfield  {author} {\bibinfo {author} {\bibfnamefont {B.}~\bibnamefont
  {Abelev}} \emph {et~al.} (\bibinfo {collaboration} {STAR}),\ }\href {\doibase
  10.1103/PhysRevC.79.034909} {\bibfield  {journal} {\bibinfo  {journal} {Phys.
  Rev. C}\ }\textbf {\bibinfo {volume} {79}},\ \bibinfo {pages} {034909}
  (\bibinfo {year} {2009})},\ \Eprint {http://arxiv.org/abs/0808.2041}
  {arXiv:0808.2041 [nucl-ex]} \BibitemShut {NoStop}%
\bibitem [{\citenamefont {Adler}\ \emph
  {et~al.}(2003{\natexlab{b}})\citenamefont {Adler} \emph
  {et~al.}}]{Adler:2003pb}%
  \BibitemOpen
  \bibfield  {author} {\bibinfo {author} {\bibfnamefont {S.~S.}\ \bibnamefont
  {Adler}} \emph {et~al.} (\bibinfo {collaboration} {PHENIX}),\ }\href
  {\doibase 10.1103/PhysRevLett.91.241803} {\bibfield  {journal} {\bibinfo
  {journal} {Phys. Rev. Lett.}\ }\textbf {\bibinfo {volume} {91}},\ \bibinfo
  {pages} {241803} (\bibinfo {year} {2003}{\natexlab{b}})},\ \Eprint
  {http://arxiv.org/abs/hep-ex/0304038} {arXiv:hep-ex/0304038 [hep-ex]}
  \BibitemShut {NoStop}%
\bibitem [{\citenamefont {Sj{\"o}strand}\ \emph {et~al.}(2006)\citenamefont
  {Sj{\"o}strand}, \citenamefont {Mrenna},\ and\ \citenamefont
  {Skands}}]{Sjostrand:2006za}%
  \BibitemOpen
  \bibfield  {author} {\bibinfo {author} {\bibfnamefont {T.}~\bibnamefont
  {Sj{\"o}strand}}, \bibinfo {author} {\bibfnamefont {S.}~\bibnamefont
  {Mrenna}}, \ and\ \bibinfo {author} {\bibfnamefont {P.~Z.}\ \bibnamefont
  {Skands}},\ }\href {\doibase 10.1088/1126-6708/2006/05/026} {\bibfield
  {journal} {\bibinfo  {journal} {J. High Energy Phys.}\ }\textbf {\bibinfo
  {volume} {05}},\ \bibinfo {pages} {026} (\bibinfo {year} {2006})},\ \Eprint
  {http://arxiv.org/abs/hep-ph/0603175} {arXiv:hep-ph/0603175 [hep-ph]}
  \BibitemShut {NoStop}%
\bibitem [{\citenamefont {Adam}\ \emph {et~al.}(2020)\citenamefont {Adam} \emph
  {et~al.}}]{Adam:2019xpp}%
  \BibitemOpen
  \bibfield  {author} {\bibinfo {author} {\bibfnamefont {J.}~\bibnamefont
  {Adam}} \emph {et~al.} (\bibinfo {collaboration} {STAR}),\ }\href {\doibase
  10.1103/PhysRevD.101.052004} {\bibfield  {journal} {\bibinfo  {journal}
  {Phys. Rev. D}\ }\textbf {\bibinfo {volume} {101}},\ \bibinfo {pages}
  {052004} (\bibinfo {year} {2020})},\ \Eprint
  {http://arxiv.org/abs/1912.08187} {arXiv:1912.08187 [nucl-ex]} \BibitemShut
  {NoStop}%
\bibitem [{\citenamefont {Vitev}\ and\ \citenamefont
  {Zhang}(2010)}]{Vitev:2009rd}%
  \BibitemOpen
  \bibfield  {author} {\bibinfo {author} {\bibfnamefont {I.}~\bibnamefont
  {Vitev}}\ and\ \bibinfo {author} {\bibfnamefont {B.-W.}\ \bibnamefont
  {Zhang}},\ }\href {\doibase 10.1103/PhysRevLett.104.132001} {\bibfield
  {journal} {\bibinfo  {journal} {Phys. Rev. Lett.}\ }\textbf {\bibinfo
  {volume} {104}},\ \bibinfo {pages} {132001} (\bibinfo {year} {2010})},\
  \Eprint {http://arxiv.org/abs/0910.1090} {arXiv:0910.1090 [hep-ph]}
  \BibitemShut {NoStop}%
\bibitem [{\citenamefont {Vitev}\ and\ \citenamefont
  {Zhang}(2008)}]{Vitev:2008vk}%
  \BibitemOpen
  \bibfield  {author} {\bibinfo {author} {\bibfnamefont {I.}~\bibnamefont
  {Vitev}}\ and\ \bibinfo {author} {\bibfnamefont {B.-W.}\ \bibnamefont
  {Zhang}},\ }\href {\doibase 10.1016/j.physletb.2008.10.019} {\bibfield
  {journal} {\bibinfo  {journal} {Phys. Lett. B}\ }\textbf {\bibinfo {volume}
  {669}},\ \bibinfo {pages} {337} (\bibinfo {year} {2008})},\ \Eprint
  {http://arxiv.org/abs/0804.3805} {arXiv:0804.3805 [hep-ph]} \BibitemShut
  {NoStop}%
\bibitem [{\citenamefont {Sharma}\ \emph {et~al.}(2009)\citenamefont {Sharma},
  \citenamefont {Vitev},\ and\ \citenamefont {Zhang}}]{Sharma:2009hn}%
  \BibitemOpen
  \bibfield  {author} {\bibinfo {author} {\bibfnamefont {R.}~\bibnamefont
  {Sharma}}, \bibinfo {author} {\bibfnamefont {I.}~\bibnamefont {Vitev}}, \
  and\ \bibinfo {author} {\bibfnamefont {B.-W.}\ \bibnamefont {Zhang}},\ }\href
  {\doibase 10.1103/PhysRevC.80.054902} {\bibfield  {journal} {\bibinfo
  {journal} {Phys. Rev. C}\ }\textbf {\bibinfo {volume} {80}},\ \bibinfo
  {pages} {054902} (\bibinfo {year} {2009})},\ \Eprint
  {http://arxiv.org/abs/0904.0032} {arXiv:0904.0032 [hep-ph]} \BibitemShut
  {NoStop}%
\bibitem [{\citenamefont {Chien}\ and\ \citenamefont
  {Vitev}(2016)}]{Chien:2015hda}%
  \BibitemOpen
  \bibfield  {author} {\bibinfo {author} {\bibfnamefont {Y.-T.}\ \bibnamefont
  {Chien}}\ and\ \bibinfo {author} {\bibfnamefont {I.}~\bibnamefont {Vitev}},\
  }\href {\doibase 10.1007/JHEP05(2016)023} {\bibfield  {journal} {\bibinfo
  {journal} {J. High Energy Phys.}\ }\textbf {\bibinfo {volume} {05}},\
  \bibinfo {pages} {023} (\bibinfo {year} {2016})},\ \Eprint
  {http://arxiv.org/abs/1509.07257} {arXiv:1509.07257 [hep-ph]} \BibitemShut
  {NoStop}%
\bibitem [{\citenamefont {Chien}\ \emph {et~al.}(2016)\citenamefont {Chien},
  \citenamefont {Emerman}, \citenamefont {Kang}, \citenamefont {Ovanesyan},\
  and\ \citenamefont {Vitev}}]{Chien:2015vja}%
  \BibitemOpen
  \bibfield  {author} {\bibinfo {author} {\bibfnamefont {Y.-T.}\ \bibnamefont
  {Chien}}, \bibinfo {author} {\bibfnamefont {A.}~\bibnamefont {Emerman}},
  \bibinfo {author} {\bibfnamefont {Z.-B.}\ \bibnamefont {Kang}}, \bibinfo
  {author} {\bibfnamefont {G.}~\bibnamefont {Ovanesyan}}, \ and\ \bibinfo
  {author} {\bibfnamefont {I.}~\bibnamefont {Vitev}},\ }\href {\doibase
  10.1103/PhysRevD.93.074030} {\bibfield  {journal} {\bibinfo  {journal} {Phys.
  Rev. D}\ }\textbf {\bibinfo {volume} {93}},\ \bibinfo {pages} {074030}
  (\bibinfo {year} {2016})},\ \Eprint {http://arxiv.org/abs/1509.02936}
  {arXiv:1509.02936 [hep-ph]} \BibitemShut {NoStop}%
\bibitem [{\citenamefont {Idilbi}\ and\ \citenamefont
  {Majumder}(2009)}]{Idilbi:2008vm}%
  \BibitemOpen
  \bibfield  {author} {\bibinfo {author} {\bibfnamefont {A.}~\bibnamefont
  {Idilbi}}\ and\ \bibinfo {author} {\bibfnamefont {A.}~\bibnamefont
  {Majumder}},\ }\href {\doibase 10.1103/PhysRevD.80.054022} {\bibfield
  {journal} {\bibinfo  {journal} {Phys. Rev. D}\ }\textbf {\bibinfo {volume}
  {80}},\ \bibinfo {pages} {054022} (\bibinfo {year} {2009})},\ \Eprint
  {http://arxiv.org/abs/0808.1087} {arXiv:0808.1087 [hep-ph]} \BibitemShut
  {NoStop}%
\bibitem [{\citenamefont {D'Eramo}\ \emph {et~al.}(2011)\citenamefont
  {D'Eramo}, \citenamefont {Liu},\ and\ \citenamefont
  {Rajagopal}}]{DEramo:2010wup}%
  \BibitemOpen
  \bibfield  {author} {\bibinfo {author} {\bibfnamefont {F.}~\bibnamefont
  {D'Eramo}}, \bibinfo {author} {\bibfnamefont {H.}~\bibnamefont {Liu}}, \ and\
  \bibinfo {author} {\bibfnamefont {K.}~\bibnamefont {Rajagopal}},\ }\href
  {\doibase 10.1103/PhysRevD.84.065015} {\bibfield  {journal} {\bibinfo
  {journal} {Phys. Rev. D}\ }\textbf {\bibinfo {volume} {84}},\ \bibinfo
  {pages} {065015} (\bibinfo {year} {2011})},\ \Eprint
  {http://arxiv.org/abs/1006.1367} {arXiv:1006.1367 [hep-ph]} \BibitemShut
  {NoStop}%
\bibitem [{\citenamefont {Ovanesyan}\ and\ \citenamefont
  {Vitev}(2011)}]{Ovanesyan:2011xy}%
  \BibitemOpen
  \bibfield  {author} {\bibinfo {author} {\bibfnamefont {G.}~\bibnamefont
  {Ovanesyan}}\ and\ \bibinfo {author} {\bibfnamefont {I.}~\bibnamefont
  {Vitev}},\ }\href {\doibase 10.1007/JHEP06(2011)080} {\bibfield  {journal}
  {\bibinfo  {journal} {J. High Energy Phys.}\ }\textbf {\bibinfo {volume}
  {06}},\ \bibinfo {pages} {080} (\bibinfo {year} {2011})},\ \Eprint
  {http://arxiv.org/abs/1103.1074} {arXiv:1103.1074 [hep-ph]} \BibitemShut
  {NoStop}%
\bibitem [{\citenamefont {Abelev}\ \emph
  {et~al.}(2014{\natexlab{c}})\citenamefont {Abelev} \emph
  {et~al.}}]{Abelev:2014ypa}%
  \BibitemOpen
  \bibfield  {author} {\bibinfo {author} {\bibfnamefont {B.~B.}\ \bibnamefont
  {Abelev}} \emph {et~al.} (\bibinfo {collaboration} {ALICE}),\ }\href
  {\doibase 10.1140/epjc/s10052-014-3108-8} {\bibfield  {journal} {\bibinfo
  {journal} {Eur. Phys. J. C}\ }\textbf {\bibinfo {volume} {74}},\ \bibinfo
  {pages} {3108} (\bibinfo {year} {2014}{\natexlab{c}})},\ \Eprint
  {http://arxiv.org/abs/1405.3794} {arXiv:1405.3794 [nucl-ex]} \BibitemShut
  {NoStop}%
\bibitem [{\citenamefont {Aad}\ \emph {et~al.}(2015{\natexlab{c}})\citenamefont
  {Aad} \emph {et~al.}}]{Aad:2015wga}%
  \BibitemOpen
  \bibfield  {author} {\bibinfo {author} {\bibfnamefont {G.}~\bibnamefont
  {Aad}} \emph {et~al.} (\bibinfo {collaboration} {ATLAS}),\ }\href {\doibase
  10.1007/JHEP09(2015)050} {\bibfield  {journal} {\bibinfo  {journal} {J. High
  Energy Phys.}\ }\textbf {\bibinfo {volume} {09}},\ \bibinfo {pages} {050}
  (\bibinfo {year} {2015}{\natexlab{c}})},\ \Eprint
  {http://arxiv.org/abs/1504.04337} {arXiv:1504.04337 [hep-ex]} \BibitemShut
  {NoStop}%
\bibitem [{\citenamefont {Casalderrey-Solana}\ \emph
  {et~al.}(2017)\citenamefont {Casalderrey-Solana}, \citenamefont {Gulhan},
  \citenamefont {Milhano}, \citenamefont {Pablos},\ and\ \citenamefont
  {Rajagopal}}]{Casalderrey-Solana:2016jvj}%
  \BibitemOpen
  \bibfield  {author} {\bibinfo {author} {\bibfnamefont {J.}~\bibnamefont
  {Casalderrey-Solana}}, \bibinfo {author} {\bibfnamefont {D.}~\bibnamefont
  {Gulhan}}, \bibinfo {author} {\bibfnamefont {G.}~\bibnamefont {Milhano}},
  \bibinfo {author} {\bibfnamefont {D.}~\bibnamefont {Pablos}}, \ and\ \bibinfo
  {author} {\bibfnamefont {K.}~\bibnamefont {Rajagopal}},\ }\href {\doibase
  10.1007/JHEP03(2017)135} {\bibfield  {journal} {\bibinfo  {journal} {J. High
  Energy Phys.}\ }\textbf {\bibinfo {volume} {03}},\ \bibinfo {pages} {135}
  (\bibinfo {year} {2017})},\ \Eprint {http://arxiv.org/abs/1609.05842}
  {arXiv:1609.05842 [hep-ph]} \BibitemShut {NoStop}%
\bibitem [{\citenamefont {Casalderrey-Solana}\ \emph
  {et~al.}(2019)\citenamefont {Casalderrey-Solana}, \citenamefont {Hulcher},
  \citenamefont {Milhano}, \citenamefont {Pablos},\ and\ \citenamefont
  {Rajagopal}}]{Casalderrey-Solana:2018wrw}%
  \BibitemOpen
  \bibfield  {author} {\bibinfo {author} {\bibfnamefont {J.}~\bibnamefont
  {Casalderrey-Solana}}, \bibinfo {author} {\bibfnamefont {Z.}~\bibnamefont
  {Hulcher}}, \bibinfo {author} {\bibfnamefont {G.}~\bibnamefont {Milhano}},
  \bibinfo {author} {\bibfnamefont {D.}~\bibnamefont {Pablos}}, \ and\ \bibinfo
  {author} {\bibfnamefont {K.}~\bibnamefont {Rajagopal}},\ }\href {\doibase
  10.1103/PhysRevC.99.051901} {\bibfield  {journal} {\bibinfo  {journal} {Phys.
  Rev. C}\ }\textbf {\bibinfo {volume} {99}},\ \bibinfo {pages} {051901}
  (\bibinfo {year} {2019})},\ \Eprint {http://arxiv.org/abs/1808.07386}
  {arXiv:1808.07386 [hep-ph]} \BibitemShut {NoStop}%
\bibitem [{\citenamefont {He}\ \emph {et~al.}(2015)\citenamefont {He},
  \citenamefont {Luo}, \citenamefont {Wang},\ and\ \citenamefont
  {Zhu}}]{He:2015pra}%
  \BibitemOpen
  \bibfield  {author} {\bibinfo {author} {\bibfnamefont {Y.}~\bibnamefont
  {He}}, \bibinfo {author} {\bibfnamefont {T.}~\bibnamefont {Luo}}, \bibinfo
  {author} {\bibfnamefont {X.-N.}\ \bibnamefont {Wang}}, \ and\ \bibinfo
  {author} {\bibfnamefont {Y.}~\bibnamefont {Zhu}},\ }\href {\doibase
  10.1103/PhysRevC.91.054908} {\bibfield  {journal} {\bibinfo  {journal} {Phys.
  Rev. C}\ }\textbf {\bibinfo {volume} {91}},\ \bibinfo {pages} {054908}
  (\bibinfo {year} {2015})},\ \bibinfo {note} {[Erratum: Phys.Rev.C 97, 019902
  (2018)]},\ \Eprint {http://arxiv.org/abs/1503.03313} {arXiv:1503.03313
  [nucl-th]} \BibitemShut {NoStop}%
\bibitem [{\citenamefont {He}\ \emph {et~al.}(2019)\citenamefont {He},
  \citenamefont {Cao}, \citenamefont {Chen}, \citenamefont {Luo}, \citenamefont
  {Pang},\ and\ \citenamefont {Wang}}]{He:2018xjv}%
  \BibitemOpen
  \bibfield  {author} {\bibinfo {author} {\bibfnamefont {Y.}~\bibnamefont
  {He}}, \bibinfo {author} {\bibfnamefont {S.}~\bibnamefont {Cao}}, \bibinfo
  {author} {\bibfnamefont {W.}~\bibnamefont {Chen}}, \bibinfo {author}
  {\bibfnamefont {T.}~\bibnamefont {Luo}}, \bibinfo {author} {\bibfnamefont
  {L.-G.}\ \bibnamefont {Pang}}, \ and\ \bibinfo {author} {\bibfnamefont
  {X.-N.}\ \bibnamefont {Wang}},\ }\href {\doibase 10.1103/PhysRevC.99.054911}
  {\bibfield  {journal} {\bibinfo  {journal} {Phys. Rev. C}\ }\textbf {\bibinfo
  {volume} {99}},\ \bibinfo {pages} {054911} (\bibinfo {year} {2019})},\
  \Eprint {http://arxiv.org/abs/1809.02525} {arXiv:1809.02525 [nucl-th]}
  \BibitemShut {NoStop}%
\bibitem [{\citenamefont {Pang}\ \emph {et~al.}(2015)\citenamefont {Pang},
  \citenamefont {Hatta}, \citenamefont {Wang},\ and\ \citenamefont
  {Xiao}}]{Pang:2014ipa}%
  \BibitemOpen
  \bibfield  {author} {\bibinfo {author} {\bibfnamefont {L.-G.}\ \bibnamefont
  {Pang}}, \bibinfo {author} {\bibfnamefont {Y.}~\bibnamefont {Hatta}},
  \bibinfo {author} {\bibfnamefont {X.-N.}\ \bibnamefont {Wang}}, \ and\
  \bibinfo {author} {\bibfnamefont {B.-W.}\ \bibnamefont {Xiao}},\ }\href
  {\doibase 10.1103/PhysRevD.91.074027} {\bibfield  {journal} {\bibinfo
  {journal} {Phys. Rev. D}\ }\textbf {\bibinfo {volume} {91}},\ \bibinfo
  {pages} {074027} (\bibinfo {year} {2015})},\ \Eprint
  {http://arxiv.org/abs/1411.7767} {arXiv:1411.7767 [hep-ph]} \BibitemShut
  {NoStop}%
\bibitem [{\citenamefont {Ke}\ \emph {et~al.}(2019)\citenamefont {Ke},
  \citenamefont {Xu},\ and\ \citenamefont {Bass}}]{Ke:2018jem}%
  \BibitemOpen
  \bibfield  {author} {\bibinfo {author} {\bibfnamefont {W.}~\bibnamefont
  {Ke}}, \bibinfo {author} {\bibfnamefont {Y.}~\bibnamefont {Xu}}, \ and\
  \bibinfo {author} {\bibfnamefont {S.~A.}\ \bibnamefont {Bass}},\ }\href
  {\doibase 10.1103/PhysRevC.100.064911} {\bibfield  {journal} {\bibinfo
  {journal} {Phys. Rev. C}\ }\textbf {\bibinfo {volume} {100}},\ \bibinfo
  {pages} {064911} (\bibinfo {year} {2019})},\ \Eprint
  {http://arxiv.org/abs/1810.08177} {arXiv:1810.08177 [nucl-th]} \BibitemShut
  {NoStop}%
\bibitem [{\citenamefont {Ke}\ \emph {et~al.}(2020)\citenamefont {Ke},
  \citenamefont {Wang}, \citenamefont {Fan},\ and\ \citenamefont
  {Bass}}]{Ke:2020nsm}%
  \BibitemOpen
  \bibfield  {author} {\bibinfo {author} {\bibfnamefont {W.}~\bibnamefont
  {Ke}}, \bibinfo {author} {\bibfnamefont {X.-N.}\ \bibnamefont {Wang}},
  \bibinfo {author} {\bibfnamefont {W.}~\bibnamefont {Fan}}, \ and\ \bibinfo
  {author} {\bibfnamefont {S.}~\bibnamefont {Bass}}\ }(\bibinfo {year} {2020})\
  \Eprint {http://arxiv.org/abs/2008.07622} {arXiv:2008.07622 [nucl-th]}
  \BibitemShut {NoStop}%
\bibitem [{\citenamefont {Adler}\ \emph
  {et~al.}(2004{\natexlab{b}})\citenamefont {Adler} \emph
  {et~al.}}]{Adler:2003au}%
  \BibitemOpen
  \bibfield  {author} {\bibinfo {author} {\bibfnamefont {S.~S.}\ \bibnamefont
  {Adler}} \emph {et~al.} (\bibinfo {collaboration} {PHENIX}),\ }\href
  {\doibase 10.1103/PhysRevC.69.034910} {\bibfield  {journal} {\bibinfo
  {journal} {Phys. Rev. C}\ }\textbf {\bibinfo {volume} {69}},\ \bibinfo
  {pages} {034910} (\bibinfo {year} {2004}{\natexlab{b}})},\ \Eprint
  {http://arxiv.org/abs/nucl-ex/0308006} {arXiv:nucl-ex/0308006 [nucl-ex]}
  \BibitemShut {NoStop}%
\bibitem [{\citenamefont {Chatrchyan}\ \emph
  {et~al.}(2014{\natexlab{a}})\citenamefont {Chatrchyan} \emph
  {et~al.}}]{Chatrchyan:2014gia}%
  \BibitemOpen
  \bibfield  {author} {\bibinfo {author} {\bibfnamefont {S.}~\bibnamefont
  {Chatrchyan}} \emph {et~al.} (\bibinfo {collaboration} {CMS}),\ }\href
  {\doibase 10.1103/PhysRevD.90.072006} {\bibfield  {journal} {\bibinfo
  {journal} {Phys. Rev. D}\ }\textbf {\bibinfo {volume} {90}},\ \bibinfo
  {pages} {072006} (\bibinfo {year} {2014}{\natexlab{a}})},\ \Eprint
  {http://arxiv.org/abs/1406.0324} {arXiv:1406.0324 [hep-ex]} \BibitemShut
  {NoStop}%
\bibitem [{\citenamefont {Soyez}(2011)}]{Soyez:2011np}%
  \BibitemOpen
  \bibfield  {author} {\bibinfo {author} {\bibfnamefont {G.}~\bibnamefont
  {Soyez}},\ }\href {\doibase 10.1016/j.physletb.2011.02.061} {\bibfield
  {journal} {\bibinfo  {journal} {Phys. Lett. B}\ }\textbf {\bibinfo {volume}
  {698}},\ \bibinfo {pages} {59} (\bibinfo {year} {2011})},\ \Eprint
  {http://arxiv.org/abs/1101.2665} {arXiv:1101.2665 [hep-ph]} \BibitemShut
  {NoStop}%
\bibitem [{\citenamefont {Dasgupta}\ \emph {et~al.}(2016)\citenamefont
  {Dasgupta}, \citenamefont {Dreyer}, \citenamefont {Salam},\ and\
  \citenamefont {Soyez}}]{Dasgupta:2016bnd}%
  \BibitemOpen
  \bibfield  {author} {\bibinfo {author} {\bibfnamefont {M.}~\bibnamefont
  {Dasgupta}}, \bibinfo {author} {\bibfnamefont {F.~A.}\ \bibnamefont
  {Dreyer}}, \bibinfo {author} {\bibfnamefont {G.~P.}\ \bibnamefont {Salam}}, \
  and\ \bibinfo {author} {\bibfnamefont {G.}~\bibnamefont {Soyez}},\ }\href
  {\doibase 10.1007/JHEP06(2016)057} {\bibfield  {journal} {\bibinfo  {journal}
  {J. High Energy Phys.}\ }\textbf {\bibinfo {volume} {06}},\ \bibinfo {pages}
  {057} (\bibinfo {year} {2016})},\ \Eprint {http://arxiv.org/abs/1602.01110}
  {arXiv:1602.01110 [hep-ph]} \BibitemShut {NoStop}%
\bibitem [{\citenamefont {Abelev}\ \emph {et~al.}(2015)\citenamefont {Abelev}
  \emph {et~al.}}]{ALICE:2014dla}%
  \BibitemOpen
  \bibfield  {author} {\bibinfo {author} {\bibfnamefont {B.~B.}\ \bibnamefont
  {Abelev}} \emph {et~al.} (\bibinfo {collaboration} {ALICE}),\ }\href
  {\doibase 10.1103/PhysRevD.91.112012} {\bibfield  {journal} {\bibinfo
  {journal} {Phys. Rev. D}\ }\textbf {\bibinfo {volume} {91}},\ \bibinfo
  {pages} {112012} (\bibinfo {year} {2015})},\ \Eprint
  {http://arxiv.org/abs/1411.4969} {arXiv:1411.4969 [nucl-ex]} \BibitemShut
  {NoStop}%
\bibitem [{\citenamefont {de~Florian}(2009)}]{deFlorian:2009fw}%
  \BibitemOpen
  \bibfield  {author} {\bibinfo {author} {\bibfnamefont {D.}~\bibnamefont
  {de~Florian}},\ }\href {\doibase 10.1103/PhysRevD.79.114014} {\bibfield
  {journal} {\bibinfo  {journal} {Phys. Rev. D}\ }\textbf {\bibinfo {volume}
  {79}},\ \bibinfo {pages} {114014} (\bibinfo {year} {2009})},\ \Eprint
  {http://arxiv.org/abs/0904.4402} {arXiv:0904.4402 [hep-ph]} \BibitemShut
  {NoStop}%
\bibitem [{\citenamefont {Chatrchyan}\ \emph
  {et~al.}(2014{\natexlab{b}})\citenamefont {Chatrchyan} \emph
  {et~al.}}]{Chatrchyan:2013kwa}%
  \BibitemOpen
  \bibfield  {author} {\bibinfo {author} {\bibfnamefont {S.}~\bibnamefont
  {Chatrchyan}} \emph {et~al.} (\bibinfo {collaboration} {CMS}),\ }\href
  {\doibase 10.1016/j.physletb.2014.01.042} {\bibfield  {journal} {\bibinfo
  {journal} {Phys. Lett. B}\ }\textbf {\bibinfo {volume} {730}},\ \bibinfo
  {pages} {243} (\bibinfo {year} {2014}{\natexlab{b}})},\ \Eprint
  {http://arxiv.org/abs/1310.0878} {arXiv:1310.0878 [nucl-ex]} \BibitemShut
  {NoStop}%
\bibitem [{\citenamefont {Khachatryan}\ \emph {et~al.}(2016)\citenamefont
  {Khachatryan} \emph {et~al.}}]{Khachatryan:2016tfj}%
  \BibitemOpen
  \bibfield  {author} {\bibinfo {author} {\bibfnamefont {V.}~\bibnamefont
  {Khachatryan}} \emph {et~al.} (\bibinfo {collaboration} {CMS}),\ }\href
  {\doibase 10.1007/JHEP11(2016)055} {\bibfield  {journal} {\bibinfo  {journal}
  {J. High Energy Phys.}\ }\textbf {\bibinfo {volume} {11}},\ \bibinfo {pages}
  {055} (\bibinfo {year} {2016})},\ \Eprint {http://arxiv.org/abs/1609.02466}
  {arXiv:1609.02466 [nucl-ex]} \BibitemShut {NoStop}%
\bibitem [{\citenamefont {Bahr}\ \emph {et~al.}(2008)\citenamefont {Bahr} \emph
  {et~al.}}]{Bahr:2008pv}%
  \BibitemOpen
  \bibfield  {author} {\bibinfo {author} {\bibfnamefont {M.}~\bibnamefont
  {Bahr}} \emph {et~al.},\ }\href {\doibase 10.1140/epjc/s10052-008-0798-9}
  {\bibfield  {journal} {\bibinfo  {journal} {Eur. Phys. J. C}\ }\textbf
  {\bibinfo {volume} {58}},\ \bibinfo {pages} {639} (\bibinfo {year} {2008})},\
  \Eprint {http://arxiv.org/abs/0803.0883} {arXiv:0803.0883 [hep-ph]}
  \BibitemShut {NoStop}%
\bibitem [{\citenamefont {Kauder}(2019)}]{Kauder:2018cdt}%
  \BibitemOpen
  \bibfield  {author} {\bibinfo {author} {\bibfnamefont {K.}~\bibnamefont
  {Kauder}} (\bibinfo {collaboration} {JETSCAPE}),\ }\href {\doibase
  10.1016/j.nuclphysa.2018.09.033} {\bibfield  {journal} {\bibinfo  {journal}
  {Nucl. Phys. A}\ }\textbf {\bibinfo {volume} {982}},\ \bibinfo {pages} {615}
  (\bibinfo {year} {2019})},\ \Eprint {http://arxiv.org/abs/1807.09615}
  {arXiv:1807.09615 [hep-ph]} \BibitemShut {NoStop}%
\end{thebibliography}%

\end{document}